\documentclass[reqno]{amsart}

\usepackage[style=alphabetic,sorting=none,backend=biber]{biblatex}
\bibliography{cwa}
\usepackage{tiz-note}
\usepackage[hyphenbreaks]{breakurl}

\begin{document}

\title{Type theories in category theory}

\author{Tesla Zhang}
\address{The Pennsylvania State University}
\email{yqz5714@psu.edu}
\urladdr{https://personal.psu.edu/yqz5714}
\date\today

\begin{abstract}
We introduce basic notions in category theory to type theorists,
including comprehension categories, categories with attributes,
contextual categories, type categories, and categories with families
along with additional discussions that are not very closely related to type theories
by listing definitions, lemmata (with detailed, diagram-rich proofs), and remarks.
By doing so, this introduction becomes more friendly as a referential material to be
read in random order (instead of from the beginning to the end).
In the end, we list some mistakes made in the early versions of this introduction.

The interpretation of common type formers in dependent type theories are discussed based on existing
categorical constructions instead of mechanically derived from their type theoretical definition.
Non-dependent type formers include unit, products (as fiber products), and functions (as \textit{fiber exponents}),
and dependent ones include sigma types, extensional equalities (as equalizers), pi types,
and the universe of (all) propositions (as the subobject classifier).
\end{abstract}
\maketitle
% % https://q.uiver.app/?q=WzAsMTAsWzEsMF0sWzQsMV0sWzMsMF0sWzIsMCwiXFxibGFja3RyaWFuZ2xlZG93biJdLFs0LDBdLFszLDFdLFsxLDFdLFswLDBdLFsyLDJdLFsyLDFdLFs1LDQsIiIsMCx7InN0eWxlIjp7ImJvZHkiOnsibmFtZSI6Im5vbmUifSwiaGVhZCI6eyJuYW1lIjoibm9uZSJ9fX1dLFs3LDYsIiIsMCx7InN0eWxlIjp7ImJvZHkiOnsibmFtZSI6Im5vbmUifSwiaGVhZCI6eyJuYW1lIjoibm9uZSJ9fX1dLFs5LDMsIiIsMCx7InN0eWxlIjp7InRhaWwiOnsibmFtZSI6Im1vbm8ifSwiaGVhZCI6eyJuYW1lIjoibm9uZSJ9fX1dLFszLDEwLCIiLDIseyJsZXZlbCI6MSwic3R5bGUiOnsiaGVhZCI6eyJuYW1lIjoibm9uZSJ9fX1dLFszLDExLCIiLDAseyJsZXZlbCI6MSwic3R5bGUiOnsiaGVhZCI6eyJuYW1lIjoibm9uZSJ9fX1dLFszLDEwLCIiLDIseyJvZmZzZXQiOi0yLCJsZXZlbCI6MSwic3R5bGUiOnsiaGVhZCI6eyJuYW1lIjoibm9uZSJ9fX1dLFszLDExLCIiLDAseyJvZmZzZXQiOjIsImxldmVsIjoxLCJzdHlsZSI6eyJoZWFkIjp7Im5hbWUiOiJub25lIn19fV0sWzMsMTEsIiIsMCx7Im9mZnNldCI6LTIsImxldmVsIjoxLCJzdHlsZSI6eyJoZWFkIjp7Im5hbWUiOiJub25lIn19fV0sWzMsMTAsIiIsMix7Im9mZnNldCI6MiwibGV2ZWwiOjEsInN0eWxlIjp7ImhlYWQiOnsibmFtZSI6Im5vbmUifX19XV0=
% \[\begin{tikzcd}
% 	{} & {} & \blacktriangledown & {} & {} \\
% 	& {} & {} & {} & {} \\
% 	&& {}
% 	\arrow[""{name=0, anchor=center, inner sep=0}, draw=none, from=2-4, to=1-5]
% 	\arrow[""{name=1, anchor=center, inner sep=0}, draw=none, from=1-1, to=2-2]
% 	\arrow[tail, no head, from=2-3, to=1-3]
% 	\arrow[no head, from=1-3, to=0]
% 	\arrow[no head, from=1-3, to=1]
% 	\arrow[shift left=2, no head, from=1-3, to=0]
% 	\arrow[shift right=2, no head, from=1-3, to=1]
% 	\arrow[shift left=2, no head, from=1-3, to=1]
% 	\arrow[shift right=2, no head, from=1-3, to=0]
% \end{tikzcd}\]
\tableofcontents
\newcommand{\apply}[2]{\textsf{ap}(#1,#2)}
\newcommand{\EE}{\mathcal E}
\newcommand{\DD}{\mathcal D}
\newcommand{\VV}{\mathcal V}
\newcommand{\FF}{\mathcal F}
\newcommand{\truecon}{\textsf{true}}
\newcommand{\falsecon}{\textsf{false}}
\newcommand{\eqlzSimp}{\mathrm{eq}}
\newcommand{\eqlz}[2]{\eqlzSimp(#1,#2)}
\newcommand{\Eqlz}[2]{\mathrm{Eq}(#1,#2)}
\newcommand{\Idty}[3]{\mathrm{Id}_{#1}~#2~#3}
\newcommand{\reflcon}[1]{\textsf{refl}_{#1}}
\newcommand{\Idprop}[3]{\mathrm{Id'}_{#1}~#2~#3}
\newcommand{\sfelim}[1]{#1\textsf{-elim}}
\newcommand{\ctfragment}[1]{\ExecuteMetaData[tt]{#1}}
\newcommand{\ttfragment}[1]{\begin{mathparpagebreakable}
\ctfragment{#1}
\end{mathparpagebreakable}}

\section{Preface}\label{sec:pre}
\begin{warning}[Prerequisites]
We assume the familiarity of the following concepts:
\begin{enumerate}
\item Set theory: sets, subsets, elements, functions, image, fiber, injectivity, and cartesian products.
\item Type theory: dependent types, lambda calculus, (telescopic) contexts, typing judgments, and substitutions.
\item Category theory: categories, objects, morphisms (isomorphisms), commutative diagrams,
functors (endofunctors), and natural transformations (natural isomorphisms).
\end{enumerate}
Typing judgments and substitutions are slightly discussed (in~\cref{sec:tt}),
but not in an introductory way.
\end{warning}

\begin{notation}\label{not:ct}
We introduce some notational conventions in category theory.
\begin{itemize}
\item
We write $A\in\CC$ to say that ``$A$ is an object in the category $\CC$'', and $f\in\CC(A,B)$ to say that
``$f$ is a morphism in the category $\CC$ from $A$ towards $B$''. In the literature, the former
is also written as $A\in\text{Ob}(\CC)$ (so that the category $\CC$ and the set of objects $\text{Ob}(\CC)$ are distinguished)
and the latter is also written as $f\in\text{Hom}_\CC(A,B)$.
The $\text{Hom}_\CC(A,B)$ notation is also (though rarely) used when the category $\CC$ is a complicated expression.
\item We write $\id_A\in\CC(A,A)$ for the identity morphism, $dom(f)$ and $cod(f)$ for the domain projection
and the codomain projection of morphisms.
We also say a morphism $f$ is a morphism \textit{from} $dom(f)$ and \textit{towards} $cod(f)$.
\item We write $f\circ g\in\CC(dom(g),cod(f))$ for composition of morphisms.
\item We use dashed lines in commutative diagrams for unique morphisms.
\item We write $A\cong B\in\CC$ for objects $A$ and $B$ in the category $\CC$ are equivalent up to unique isomorphism.
\item We write $F\simeq G$ for naturally isomorphic functors $F$ and $G$.
\end{itemize}
\lessSpace{-1}
\end{notation}

\begin{notation}\label{not:set}
We introduce some notational conventions and terminologies about sets and functions.
\begin{itemize}
\item For two sets $B\subseteq A$, the ``identity'' function $\iota:B\to A$ is called a \textit{natural inclusion}.
\item We use $-$ for a \textit{hole} in any formula to denote the place where a function puts its argument in,
and the formula itself is the rest of the function body. For example,
$(-\circ f)$ is a function that maps $g$ to $g\circ f$ (in the literature, people also write $g\mapsto g\circ f$).
\item We write $A\simeq B$ for isomorphic sets $A$ and $B$.
\end{itemize}
\lessSpace{-1}
\end{notation}

\begin{convention}[Sterling~\cite{OkTT}]\label{conv:gray}
In the typing rules, some judgments might be \textit{entailed} by another.
We write the entailed judgment(s) out explicitly in gray if they worth mentioning.
\end{convention}

\begin{convention}[Cockx~\cite{DepPM}]\label{conv:box}
We use \fbox{boxes} to distinguish type theory or category theory notations from natural language text,
e.g. ``we combine a term \fbox{$a$} with a term \fbox{$b$} to get a term \fbox{$a~b$}''.
\end{convention}

\section{Background and motivation}\label{sec:tt}
\begin{remark}
The notion of \textit{substitution} in type theory is the process of replacing a variable
with a term. We sometimes refer to the action of replacement with the verb \textit{apply},
like ``apply this substitution to that term''.
We generalize this notion to lists, so that one substitution deals with multiple replacements.
\end{remark}

\begin{defn}\label{def:substobj}
The data carried by a substitution is called a \textit{substitution object} --
a list of mapping from variables to terms.
Substitution objects can be \textit{applied} on terms.
\end{defn}

\begin{example}
Consider a lambda calculus term \fbox{$(\lambda x.\lambda y.u)~a~b$}, we denote its reduced form as $u[a/x,b/y]$
(a notation similar to the one used by Curry~\cite{SubstNotation}),
which means the term $u$ applied by a substitution object $[a/x,b/y]$ (\cref{def:substobj}).
\end{example}

\begin{remark}[Nameless]
Contexts, types, terms, and substitution objects are considered up to $\alpha$-renaming throughout this document.
That is to say, there are no ``names'' in these structures -- $\alpha$-equivalent terms are considered the same.
We can imagine that we are using a nameless representation of binding structures, such as de Bruijn indices~\cite{DBI}.
\end{remark}

\begin{notation}
We introduce some notational conventions in type theory.

We use $\Gamma, \Delta$ to denote \textit{contexts}, \fbox{$\sigma,\gamma$} (recall~\cref{conv:box}) for
substitution objects (\cref{def:substobj}), lowercase letters for terms (preferably $u,v$),
and uppercase letters for types (preferably $A,B$).

We write $u\sigma$ for ``applying the substitution object $\sigma$ to term $u$''.
\end{notation}

\begin{remark}\label{rem:basic-tt}
In a simple type theory, terms are typed in a context and types are canonical mathematical objects.
However, since we are dealing with a dependent type theory, types will be formed in a context too.

We will have two basic judgments: \textit{context formation} $\Gvdash$
(the types in $\Gamma$ are well-typed) and \textit{type formation} \fbox{$\Gvdash\isType A$}
($A$ is a well-typed type in $\Gamma$). Based on these judgments, we define the inference rules for terms
(using~\cref{conv:gray}) using the following syntax
(these are not actual rules, but syntax demonstrations):
\ttfragment{term}
They read ``term $u$ has type $A$ in context $\Gamma$, presupposing \fbox{$\Gvdash\isType A$}''
and ``term $u$ and $v$ are the same term of type $A$ in context $\Gamma$, presupposing $\Gvdash u:A,\Gvdash v:A$''.

We also have the judgment for substitution objects: \ttfragment{subst-obj}
It reads ``in the context $\Gamma$, the $i$-th term in $\sigma$ have the $i$-th type in $\Delta$,
presupposing $\Gvdash,\Delta\vdash$''. It is assumed that $\sigma$ and $\Delta$ have the same length.
\end{remark}

\begin{remark}[Substitutions]\label{rem:subst-ctx}
Although we are going to treat substitution informally,
we would like to emphasize some important insights about substitutions.
\begin{itemize}
\item $\Gvdash\sigma:\Delta$ (a judgment already described in~\cref{rem:basic-tt}):
means ``terms in $\sigma$ can refer to bindings in $\Gamma$'',
and we say ``$\sigma$ \textit{instantiates} the context $\Delta$''.
\item Consider a term $\Delta\vdash u:A$, both $u$ and $A$ can refer to bindings in $\Delta$.
\end{itemize}
We can turn $u$ into another term, typed in the context $\Gamma$ by applying the substitution $\sigma$
since the open references to $\Delta$ can be replaced by the terms in $\sigma$, which can refer to bindings in $\Gamma$.
This process can be described by the following deduction: \ttfragment{subst-ty}
Similarly, for contexts $\Gamma,\Gamma',\Gamma''$ and substitution objects $\sigma,\gamma$, there is:
\ttfragment{subst-ctx}
The above deduction tree induces the definition of a category (\cref{def:cat-ctx}).
\end{remark}

\subsection{Pullbacks and products}
\begin{notation}\label{not:sets}
We refer to the category of small sets\footnote{If you do not know what are \textit{small} sets,
think of them as sets.} as $\SET$.
\end{notation}

\begin{defn}[Terminal]\label{def:termobj}
For a category $\CC$, an object $\textit1\in\CC$ is called \textit{terminal} if for every $A\in\CC$
there is a unique morphism $\textit1_A\in\CC(A,\textit1)$. The visualization is trivial:
% https://q.uiver.app/?q=WzAsNixbMSwxLCJcXHRleHRpdCAxIl0sWzAsMCwiXFxidWxsZXQiXSxbMSwwLCJcXGJ1bGxldCJdLFsyLDAsIlxcYnVsbGV0Il0sWzAsMSwiXFxidWxsZXQiXSxbMiwxLCJcXGJ1bGxldCJdLFsxLDAsIiIsMCx7InN0eWxlIjp7ImJvZHkiOnsibmFtZSI6ImRhc2hlZCJ9fX1dLFsyLDAsIiIsMix7InN0eWxlIjp7ImJvZHkiOnsibmFtZSI6ImRhc2hlZCJ9fX1dLFszLDAsIiIsMix7InN0eWxlIjp7ImJvZHkiOnsibmFtZSI6ImRhc2hlZCJ9fX1dLFs0LDAsIiIsMix7InN0eWxlIjp7ImJvZHkiOnsibmFtZSI6ImRhc2hlZCJ9fX1dLFs1LDAsIiIsMix7InN0eWxlIjp7ImJvZHkiOnsibmFtZSI6ImRhc2hlZCJ9fX1dXQ==
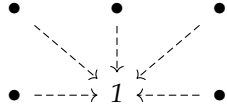
\[\begin{tikzcd}
	\bullet & \bullet & \bullet \\
	\bullet & {\textit 1} & \bullet
	\arrow[dashed, from=1-1, to=2-2]
	\arrow[dashed, from=1-2, to=2-2]
	\arrow[dashed, from=1-3, to=2-2]
	\arrow[dashed, from=2-1, to=2-2]
	\arrow[dashed, from=2-3, to=2-2]
\end{tikzcd}\]
\lessSpace{-1}
\end{defn}

\begin{lem}[Uniqueness]\label{lem:unique-term}
If a category $\CC$ has several terminal objects, they are (uniquely) isomorphic to each other.
\end{lem}
\begin{proof}
By~\cref{def:termobj}, consider two terminal objects $X,Y\in\CC$,
there are unique morphisms $X_Y\in\CC(Y,X)$ and $Y_X\in\CC(X,Y)$
whose compositions cannot be any morphism other than the identity morphism.
Thus an isomorphism.
\end{proof}

\begin{glorious-defn}[Product]\label{def:prodobj}
For a category $\CC$ and $A,B\in\CC$,
the \textit{product object} of $A$ and $B$, denoted $(A\times B)\in\CC$, is an object in $\CC$ equipped with two morphisms
$\pi_1\in\CC(A\times B,A)$ and $\pi_2\in\CC(A\times B,B)$ such that for every object $D\in\CC$ with two morphisms $f_1\in\CC(D,A)$
and $f_2\in\CC(D,B)$, there is a unique morphism in $\CC(D, A\times B)$ (called the \textit{product morphism} of $f_1$ and $f_2$) that commutes the following diagram:
% https://q.uiver.app/?q=WzAsNCxbMSwxLCJBXFx0aW1lcyBCIl0sWzAsMSwiQSJdLFsyLDEsIkIiXSxbMSwwLCJEIl0sWzAsMSwiXFxwaV8xIl0sWzAsMiwiXFxwaV8yIiwyXSxbMywwLCIiLDIseyJzdHlsZSI6eyJib2R5Ijp7Im5hbWUiOiJkYXNoZWQifX19XSxbMywxLCJmXzEiLDFdLFszLDIsImZfMiIsMV1d
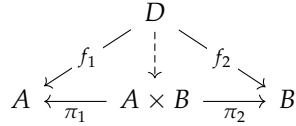
\[\begin{tikzcd}
	& D \\
	A & {A\times B} & B
	\arrow["{\pi_1}", from=2-2, to=2-1]
	\arrow["{\pi_2}"', from=2-2, to=2-3]
	\arrow[dashed, from=1-2, to=2-2]
	\arrow["{f_1}"{description}, from=1-2, to=2-1]
	\arrow["{f_2}"{description}, from=1-2, to=2-3]
\end{tikzcd}\]
\lessSpace{-1}
\end{glorious-defn}

\begin{lem}[Uniqueness]\label{lem:prod-unique}
In a category $\CC$ and $A,B\in\CC$, the product object is unique up to unique isomorphism.
\end{lem}
\begin{proof}
Consider two product objects $P_1,P_2\in\CC$ of $A$ and $B$, then there exists unique morphisms $h_1\in\CC(P_1, P_2)$ and $h_2\in\CC(P_2, P_1)$
commuting the product diagrams. Diagrammatically:
% https://q.uiver.app/?q=WzAsNCxbMCwxLCJBIl0sWzQsMCwiQiJdLFsyLDAsIlBfMSJdLFsyLDEsIlBfMiJdLFsyLDMsImhfMSIsMCx7ImN1cnZlIjotMSwic3R5bGUiOnsiYm9keSI6eyJuYW1lIjoiZGFzaGVkIn19fV0sWzIsMSwiXFxwaV8yIiwxXSxbMywxLCJcXHBpXzIiLDFdLFsyLDAsIlxccGlfMSIsMV0sWzMsMCwiXFxwaV8xIiwxXSxbMywyLCJoXzIiLDAseyJjdXJ2ZSI6LTEsInN0eWxlIjp7ImJvZHkiOnsibmFtZSI6ImRhc2hlZCJ9fX1dXQ==
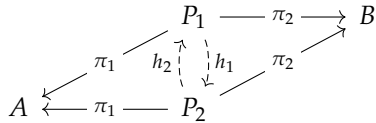
\[\begin{tikzcd}
	&& {P_1} && B \\
	A && {P_2}
	\arrow["{h_1}", curve={height=-6pt}, dashed, from=1-3, to=2-3]
	\arrow["{\pi_2}"{description}, from=1-3, to=1-5]
	\arrow["{\pi_2}"{description}, from=2-3, to=1-5]
	\arrow["{\pi_1}"{description}, from=1-3, to=2-1]
	\arrow["{\pi_1}"{description}, from=2-3, to=2-1]
	\arrow["{h_2}", curve={height=-6pt}, dashed, from=2-3, to=1-3]
\end{tikzcd}\]
The uniqueness of these morphisms implies that $\forall f\in\CC(P_1, P_2), f=h_1$,
let $f=h_1\circ h_2\circ h_1$, so $f=h_1$, so $h_2\circ h_1=\id_{P_1}$, similarly $h_1\circ h_2=\id_{P_2}$.
So, there is a unique isomorphism between $P_1$ and $P_2$.
\end{proof}

\begin{remark}\label{rem:prod-intuition}
In usually categories (like $\SET$), a product object (\cref{def:prodobj}) $A\times B$
is the \textit{simplest} object that \textit{contains} the information of both $A$ and $B$,
because every other object that contains information about both $A$ and $B$ contains information about $A\times B$.
\end{remark}

\begin{exercise}
Prove that in $\SET$ the cartesian product of sets $A,B\in\SET$ is the product object.
\end{exercise}

\begin{exercise}
For readers who are familiar with the notion of \textit{groups} and \textit{direct product of groups},
prove that in the category of groups and group homomorphisms the direct product of groups is the product object.
\end{exercise}

\begin{exercise}
For readers who are familiar with the notion of \textit{topological spaces} and \textit{product space},
prove that in the category of topological spaces and continuous functions the product space of topological spaces is the product object.
\end{exercise}

\begin{glorious-defn}[Pullback]\label{def:pullback}
For a category $\CC$, objects $A,B,C\in\CC$, and morphisms $f\in\CC(A,C)$, $g\in\CC(B,C)$,
the \textit{pullback} of $f$ and $g$ is an object $D\in\CC$ equipped with two morphisms
$a\in\CC(D,A)$ and $b\in\CC(D,B)$ where $f\circ a=g\circ b\in\CC(D,C)$,
such that every object $E\in\CC$ with $a'\in\CC(E,A)$ and $b'\in\CC(E,B)$ where $f\circ a'=g\circ b'\in\CC(E,C)$,
there is a unique morphism in $\CC(E,D)$ (also called the product morphism) commuting the following diagram:
% https://q.uiver.app/?q=WzAsNSxbMSwxLCJBIl0sWzIsMCwiQiJdLFsyLDEsIkMiXSxbMCwwLCJFIl0sWzEsMCwiRCJdLFswLDIsImYiLDJdLFsxLDIsImciXSxbMywwLCJhJyIsMl0sWzMsMSwiYiciLDAseyJvZmZzZXQiOi0xLCJjdXJ2ZSI6LTF9XSxbNCwwLCJhIl0sWzQsMSwiYiIsMl0sWzMsNCwiIiwwLHsic3R5bGUiOnsiYm9keSI6eyJuYW1lIjoiZGFzaGVkIn19fV0sWzQsMiwiIiwyLHsic3R5bGUiOnsibmFtZSI6ImNvcm5lciJ9fV1d
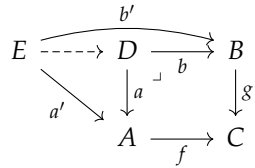
\[\begin{tikzcd}
	E & D & B \\
	& A & C
	\arrow["f"', from=2-2, to=2-3]
	\arrow["g", from=1-3, to=2-3]
	\arrow["{a'}"', from=1-1, to=2-2]
	\arrow["{b'}", shift left=1, curve={height=-6pt}, from=1-1, to=1-3]
	\arrow["a", from=1-2, to=2-2]
	\arrow["b"', from=1-2, to=1-3]
	\arrow[dashed, from=1-1, to=1-2]
	\arrow["\lrcorner"{anchor=center, pos=0.125}, draw=none, from=1-2, to=2-3]
\end{tikzcd}\]
If in a category $\CC$, for all objects $A,B,C\in\CC$ and pairs of morphisms
$f\in\CC(A,C)$, $g\in\CC(B,C)$ the pullback of $f$ and $g$ exists, we say that $\CC$ \textit{has all pullbacks}.
\end{glorious-defn}

\begin{remark}
Note how similar the diagrams in~\cref{def:prodobj,def:pullback} are.
We sometimes denote the pullback $D$ in~\cref{def:pullback} as $A\times_C B$
and refer to $D$ as the \textit{fiber product} of $A$ and $B$.
\end{remark}

\begin{notation}
In~\cref{def:pullback}, the objects $A,B,C,D\in\CC$ constitute a commutative square.
If a commutative square characterizes a pullback, we mark it with a $\lrcorner$
and refer to the square as a \textit{pullback square}:
% https://q.uiver.app/?q=WzAsNCxbMCwxLCJBIl0sWzEsMCwiQiJdLFsxLDEsIkMiXSxbMCwwLCJBXFx0aW1lc19DIEIiXSxbMCwyLCJmIiwyXSxbMSwyLCJnIl0sWzMsMCwiXFxwaV8yIiwyLHsic3R5bGUiOnsiYm9keSI6eyJuYW1lIjoiZGFzaGVkIn19fV0sWzMsMSwiXFxwaV8xIiwwLHsic3R5bGUiOnsiYm9keSI6eyJuYW1lIjoiZGFzaGVkIn19fV0sWzMsMiwiIiwxLHsic3R5bGUiOnsibmFtZSI6ImNvcm5lciJ9fV1d
\[\begin{tikzcd}
	{A\times_C B} & B \\
	A & C
	\arrow["f"', from=2-1, to=2-2]
	\arrow["g", from=1-2, to=2-2]
	\arrow["{\pi_2}"', dashed, from=1-1, to=2-1]
	\arrow["{\pi_1}", dashed, from=1-1, to=1-2]
	\arrow["\lrcorner"{anchor=center, pos=0.125}, draw=none, from=1-1, to=2-2]
\end{tikzcd}\]
Note that we overload the notation of projection morphisms of products ($\pi_1, \pi_2$ in~\cref{def:prodobj})
to work with pullbacks as well.
\end{notation}

\begin{remark}\label{rem:pullback-intuition}
Similar to products (\cref{rem:prod-intuition}), in usually cases, a pullback object (\cref{def:pullback})
$A\times_C B$ is the \textit{simplest} object that \textit{contains} the information of both $A$ and $B$,
given that a shared subset of $A$ and $B$, called $C$, is also shared in $A\times_C B$.

There is another way to think about $C$: that we talk about $A$ and $B$ under the \textit{context} $C$,
and the \textit{contextual product} of $A$ and $B$, called $A\times_C B$, is the product of
``information about $A$ that is not in $C$'' and ``information about $B$ that is not in $C$'',
under the context $C$.
\end{remark}

\begin{lem}\label{lem:prod-from-pullback}
In a category $\CC$ with products (\cref{def:prodobj}) and a terminal object (\cref{def:termobj})
$\textit1\in\CC$, for every $A,B\in\CC$ we have $A\times_1 B\cong A\times B$ up to unique isomorphism.
\end{lem}
\begin{proof}
We prove by constructing both commutative squares:
% https://q.uiver.app/?q=WzAsNyxbMCwxLCJBIl0sWzEsMCwiQiJdLFsxLDEsIlxcdGV4dGl0MSJdLFswLDAsIkEgXFx0aW1lc197XFx0ZXh0aXQxfSBCIl0sWzIsMCwiQSBcXHRpbWVzIEIiXSxbMiwxLCJBIl0sWzMsMCwiQiJdLFswLDIsIlxcdGV4dGl0MSIsMl0sWzEsMiwiXFx0ZXh0aXQxIl0sWzMsMCwiIiwyLHsic3R5bGUiOnsiYm9keSI6eyJuYW1lIjoiZGFzaGVkIn19fV0sWzMsMSwiIiwwLHsic3R5bGUiOnsiYm9keSI6eyJuYW1lIjoiZGFzaGVkIn19fV0sWzMsMiwiIiwxLHsic3R5bGUiOnsibmFtZSI6ImNvcm5lciJ9fV0sWzQsNSwiIiwwLHsic3R5bGUiOnsiYm9keSI6eyJuYW1lIjoiZGFzaGVkIn19fV0sWzQsNiwiIiwyLHsic3R5bGUiOnsiYm9keSI6eyJuYW1lIjoiZGFzaGVkIn19fV1d
\[\begin{tikzcd}
	{A \times_{\textit1} B} & B & {A \times B} & B \\
	A & \textit1 & A
	\arrow["\textit1"', from=2-1, to=2-2]
	\arrow["\textit1", from=1-2, to=2-2]
	\arrow[dashed, from=1-1, to=2-1]
	\arrow[dashed, from=1-1, to=1-2]
	\arrow["\lrcorner"{anchor=center, pos=0.125}, draw=none, from=1-1, to=2-2]
	\arrow[dashed, from=1-3, to=2-3]
	\arrow[dashed, from=1-3, to=1-4]
\end{tikzcd}\]
The fact that the right-bottom vertex of the pullback square is \textit1 makes the square
always commutative, so being a pullback means being terminal in all such squares,
which coincides with the definition of product objects.
\end{proof}

\begin{lem}
Pullback objects are unique up to unique isomorphism.
\end{lem}
\begin{proof}
Exactly the same as~\cref{lem:prod-unique}, by replacing $\times$ with $\times_C$.
\end{proof}

\begin{lem}\label{lem:pullback-along-id}
Consider a pullback square where one of the given morphism is identity:
% https://q.uiver.app/?q=WzAsNCxbMCwxLCJBIl0sWzEsMCwiQiJdLFsxLDEsIkEiXSxbMCwwLCJBXFx0aW1lc19BIEIiXSxbMCwyLCIiLDIseyJsZXZlbCI6Miwic3R5bGUiOnsiaGVhZCI6eyJuYW1lIjoibm9uZSJ9fX1dLFsxLDJdLFszLDAsIiIsMix7InN0eWxlIjp7ImJvZHkiOnsibmFtZSI6ImRhc2hlZCJ9fX1dLFszLDEsIiIsMCx7InN0eWxlIjp7ImJvZHkiOnsibmFtZSI6ImRhc2hlZCJ9fX1dLFszLDIsIiIsMSx7InN0eWxlIjp7Im5hbWUiOiJjb3JuZXIifX1dXQ==
\[\begin{tikzcd}
	{A\times_A B} & B \\
	A & A
	\arrow[Rightarrow, no head, from=2-1, to=2-2]
	\arrow[from=1-2, to=2-2]
	\arrow[dashed, from=1-1, to=2-1]
	\arrow[dashed, from=1-1, to=1-2]
	\arrow["\lrcorner"{anchor=center, pos=0.125}, draw=none, from=1-1, to=2-2]
\end{tikzcd}\]
Then, $A\times_A B\cong B$ up to unique isomorphism.
\end{lem}
\begin{proof}
We start by applying the definition of pullbacks, where the morphisms $i, j$ in
the following diagram are both unique:
% https://q.uiver.app/?q=WzAsNSxbMSwxLCJBIl0sWzIsMCwiQiJdLFsyLDEsIkEiXSxbMSwwLCJBXFx0aW1lc19BIEIiXSxbMCwwLCJCIl0sWzAsMiwiIiwyLHsibGV2ZWwiOjIsInN0eWxlIjp7ImhlYWQiOnsibmFtZSI6Im5vbmUifX19XSxbMSwyLCJnIl0sWzMsMCwiIiwyLHsic3R5bGUiOnsiYm9keSI6eyJuYW1lIjoiZGFzaGVkIn19fV0sWzMsMSwiaiIsMCx7InN0eWxlIjp7ImJvZHkiOnsibmFtZSI6ImRhc2hlZCJ9fX1dLFszLDIsIiIsMSx7InN0eWxlIjp7Im5hbWUiOiJjb3JuZXIifX1dLFs0LDMsImkiLDAseyJzdHlsZSI6eyJib2R5Ijp7Im5hbWUiOiJkYXNoZWQifX19XSxbNCwwLCJnIiwxXV0=
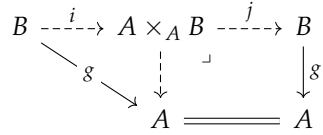
\[\begin{tikzcd}
	B & {A\times_A B} & B \\
	& A & A
	\arrow[Rightarrow, no head, from=2-2, to=2-3]
	\arrow["g", from=1-3, to=2-3]
	\arrow[dashed, from=1-2, to=2-2]
	\arrow["j", dashed, from=1-2, to=1-3]
	\arrow["\lrcorner"{anchor=center, pos=0.125}, draw=none, from=1-2, to=2-3]
	\arrow["i", dashed, from=1-1, to=1-2]
	\arrow["g"{description}, from=1-1, to=2-2]
\end{tikzcd}\]
It is left to show that $j\circ i=\id_B$. This is by the commutativity
of the above diagram, where $g\circ j\circ i=\id_A\circ g=g$.
\end{proof}

\subsection{Duality and coproducts}
\begin{defn}\label{def:oppocat}
For a category $\CC$, the \textit{opposite category} $\OpCat\CC$ of $\CC$
is $\CC$ but the domain and codomain of every morphism are swapped.
\end{defn}

\begin{defn}[Initial]\label{def:initialobj}
For a category $\CC$, an object $\textit0\in\CC$ is called \textit{initial} if it is the
terminal object (\cref{def:termobj}) in $\OpCat\CC$ (\cref{def:oppocat}), visualized below:
% https://q.uiver.app/?q=WzAsNixbMSwxLCJcXHRleHRpdCAwIl0sWzAsMCwiXFxidWxsZXQiXSxbMSwwLCJcXGJ1bGxldCJdLFsyLDAsIlxcYnVsbGV0Il0sWzAsMSwiXFxidWxsZXQiXSxbMiwxLCJcXGJ1bGxldCJdLFswLDEsIiIsMix7InN0eWxlIjp7ImJvZHkiOnsibmFtZSI6ImRhc2hlZCJ9fX1dLFswLDIsIiIsMCx7InN0eWxlIjp7ImJvZHkiOnsibmFtZSI6ImRhc2hlZCJ9fX1dLFswLDMsIiIsMCx7InN0eWxlIjp7ImJvZHkiOnsibmFtZSI6ImRhc2hlZCJ9fX1dLFswLDQsIiIsMCx7InN0eWxlIjp7ImJvZHkiOnsibmFtZSI6ImRhc2hlZCJ9fX1dLFswLDUsIiIsMCx7InN0eWxlIjp7ImJvZHkiOnsibmFtZSI6ImRhc2hlZCJ9fX1dXQ==
\[\begin{tikzcd}
	\bullet & \bullet & \bullet \\
	\bullet & {\textit 0} & \bullet
	\arrow[dashed, from=2-2, to=1-1]
	\arrow[dashed, from=2-2, to=1-2]
	\arrow[dashed, from=2-2, to=1-3]
	\arrow[dashed, from=2-2, to=2-1]
	\arrow[dashed, from=2-2, to=2-3]
\end{tikzcd}\]
\lessSpace{-1}
\end{defn}

\begin{example}[Parker]
The empty set $\emptyset$ is the initial object of $\SET$
since the maps from $\emptyset$ are all trivial.
\end{example}

\begin{defn}[Coproduct]\label{def:coprodobj}
For a category $\CC$ and $A,B\in\CC$ we say take the product object $(A\times B)\in\OpCat\CC$
to be the \textit{coproduct object} $(A \sqcup B)\in\CC$,
visualized along with the product object $(A\times B)\in\CC$ below:
% https://q.uiver.app/?q=WzAsOCxbMCwwLCJBIl0sWzIsMCwiQiJdLFsxLDAsIkFcXHNxY3VwIEIiXSxbMSwxLCJcXGJ1bGxldCJdLFszLDEsIkEiXSxbNSwxLCJCIl0sWzQsMSwiQVxcdGltZXMgQiJdLFs0LDAsIlxcYnVsbGV0Il0sWzAsMl0sWzEsMl0sWzAsM10sWzEsM10sWzIsMywiIiwxLHsic3R5bGUiOnsiYm9keSI6eyJuYW1lIjoiZGFzaGVkIn19fV0sWzYsNF0sWzYsNV0sWzcsNiwiIiwwLHsic3R5bGUiOnsiYm9keSI6eyJuYW1lIjoiZGFzaGVkIn19fV0sWzcsNF0sWzcsNV1d
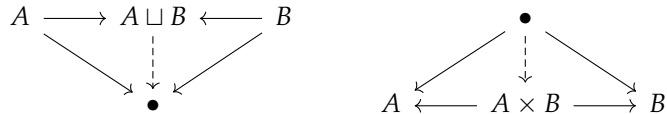
\[\begin{tikzcd}
	A & {A\sqcup B} & B && \bullet \\
	& \bullet && A & {A\times B} & B
	\arrow[from=1-1, to=1-2]
	\arrow[from=1-3, to=1-2]
	\arrow[from=1-1, to=2-2]
	\arrow[from=1-3, to=2-2]
	\arrow[dashed, from=1-2, to=2-2]
	\arrow[from=2-5, to=2-4]
	\arrow[from=2-5, to=2-6]
	\arrow[dashed, from=1-5, to=2-5]
	\arrow[from=1-5, to=2-4]
	\arrow[from=1-5, to=2-6]
\end{tikzcd}\]
Note that flipping the arrows in one of the diagrams results in the other.
\end{defn}

\begin{convention}[Duality]\label{conv:duality}
If one construction in a category $\CC$ is another construction in the category $\OpCat\CC$,
we say that these two constructions are \textit{dual} to each other.
For example, product objects (\cref{def:prodobj}) are dual to coproduct objects (\cref{def:coprodobj}),
terminal objects (\cref{def:termobj}) are dual to initial objects (\cref{def:initialobj}).
\end{convention}

\subsection{The category of contexts}
\begin{defn}\label{def:cat-ctx}
A \textit{category of contexts} $\CC$ is a category whose objects are (telescopic) contexts and morphisms are substitution objects
(also known as context morphisms).
For instance, if there are objects $\Gamma,\Delta\in\CC$ that correspond to some contexts in type theory,
the morphism $\sigma\in\CC(\Gamma,\Delta)$ corresponds to a substitution object that is typed in $\Gamma$
and instantiates (\cref{rem:subst-ctx}) $\Delta$.

For an object $\Gamma\in\CC$, we define $\id_\Gamma$ to be the identity substitution and
$\sigma\circ\gamma$ as ``applying the substitution $\gamma$ on the terms in $\sigma$''.

The category of contexts of a type theory $T$ is also known as the
\textit{syntactical category}~\cite[definition 2]{DTT-Models} of $T$.
\end{defn}

\begin{defn}\label{def:classifycat}
The \textit{classifying category} $Cl(T)$ of a type theory $T$ is the
\textit{most general} interpretation of $T$.
The name is due to Pitts~\cite[\S 4.2]{catl}.
In most cases, not only all constructions in $T$ are interpreted in $Cl(T)$,
but also all the constructions in $Cl(T)$ are interpreted in $T$.
This property is also known as \textit{completeness}.
\end{defn}

\begin{remark}
If we think of the interpretations of a type theory as the objects of a category and the
interpretation of interpretations as morphisms, then the classifying category is the initial object
of this category. Note that this is just an idea -- formal treatment of the definition requires a
formal definition of \textit{interpretation} (unrelated to the major interest of this document).

This is also related to the notion of \textit{functorial semantics},
because interpretations of a type theory are usually themselves categories,
so the morphisms between them are functors.
\end{remark}

\begin{history}
The syntactical category (\cref{def:cat-ctx}) of a type theory is supposed to be its
classifying category (\cref{def:classifycat}). This is known as the \textit{injtiality conjecture},
which proved for the \textit{calculus of constructions}~\cite{CoC} in~\cite{initialityCoC}.
Then, it is considered true for Martin-L\"of type theory, a (slightly) different type theory,
and some extensions for a long time without a sufficiently formal proof.
Some mathematicians (like Voevodsky~\cite{VV-IC}) particularly concerned about the lack of formalism here,
until the Agda formalization by Guillaume Brunerie and Peter Lumsdaine~\cite{initialityAgda}.

There are some other initiality work published between the two cited works,
but~\cite{initialityAgda} criticized (most of) them for being specific to a certain type theory,
handwaving details, and not being extensible.
\end{history}

\begin{notation}
We write $\lrbracket\Gamma$ for the semantical interpretation of a type theoretical construct $\Gamma$
in a category. This notation will be overloaded for contexts and substitution objects.
Later, in~\cref{sec:semantics}, the same notation will be used on types and terms, too.
\end{notation}

\begin{example}\label{ex:cat-ctx}
Recall the last deduction in~\cref{rem:subst-ctx}.
It corresponds to the composition of substitution objects in the category of contexts $\CC$:
for $\lrbracket\gamma\in\CC(\lrbracket{\Gamma'},\lrbracket{\Gamma''})$ and
$\lrbracket\sigma\in\CC(\lrbracket\Gamma,\lrbracket{\Gamma'})$, there is
$\lrbracket{\gamma\sigma}=\lrbracket\gamma \circ \lrbracket\sigma\in\CC(\lrbracket{\Gamma},\lrbracket{\Gamma''})$.
Visualization omitting the brackets:
% https://q.uiver.app/?q=WzAsMyxbMCwxLCJcXEdhbW1hIl0sWzIsMCwiXFxHYW1tYSciXSxbNCwxLCJcXEdhbW1hJyciXSxbMCwxLCJcXHNpZ21hIiwxXSxbMSwyLCJcXGdhbW1hIiwxXSxbMCwyLCJcXGdhbW1hXFxzaWdtYSIsMV1d
\[\begin{tikzcd}
	&& {\Gamma'} \\
	\Gamma &&&& {\Gamma''}
	\arrow["\sigma"{description}, from=2-1, to=1-3]
	\arrow["\gamma"{description}, from=1-3, to=2-5]
	\arrow["\gamma\sigma"{description}, from=2-1, to=2-5]
\end{tikzcd}\]
\lessSpace{-1}
\end{example}

\begin{remark}\label{rem:ty-subst}
Recall the second last deduction in \cref{rem:subst-ctx}.
Consider an operation $\pi_A$ for a dependent type $A$ (we did not say that $A$ belong to
any category, but we can still talk about functions acting on types)
that gives the context where $A$ is formed within,
the following (informal) diagram commutes in a syntactical category (all of the vertices are from the
second last judgment in \cref{rem:subst-ctx}):
% https://q.uiver.app/?q=WzAsNCxbMCwwLCJBXFxzaWdtYSJdLFsxLDAsIkEiXSxbMCwxLCJcXEdhbW1hIl0sWzEsMSwiXFxEZWx0YSJdLFswLDEsIlxcc2lnbWEiXSxbMCwyLCJcXHBpX3tBXFxzaWdtYX0iLDJdLFsyLDMsIlxcc2lnbWEiLDJdLFsxLDMsIlxccGlfQSJdLFswLDMsIiIsMSx7InN0eWxlIjp7Im5hbWUiOiJjb3JuZXIifX1dXQ==
\[\begin{tikzcd}
	A\sigma & A \\
	\Gamma & \Delta
	\arrow["\sigma", from=1-1, to=1-2]
	\arrow["{\pi_{A\sigma}}"', from=1-1, to=2-1]
	\arrow["\sigma"', from=2-1, to=2-2]
	\arrow["{\pi_A}", from=1-2, to=2-2]
	\arrow["\lrcorner"{anchor=center, pos=0.125}, draw=none, from=1-1, to=2-2]
\end{tikzcd}\]
The formal version is discussed later in~\cref{def:ctxext}, where all the technical details
(such as the representation of dependent types as contexts and the definition of $\pi_A$) are expanded.
\end{remark}

% \begin{exercise}
% Prove the universal property of the pullback (\cref{def:pullback}) square in~\cref{rem:ty-subst}.
% A referential proof is in~\cite[lemma 6]{DTT-Models}.
% \end{exercise}

% \begin{remark}[STLC]\label{rem:stlc}
% A category of contexts (\cref{def:cat-ctx}) can be used to interpret simple type theories.
% Since a type theoretical context is just an ordered list of types,
% a category theoretical type is just a context with only one element.
% Similarly, terms are just substitution objects whose codomain is a type.
% This is because in a simple type theory, all types are formed \textit{without} a context:
% \begin{mathpar}
% \inferrule{\isType A \\ \isType B}{\isType{(A\times B)}} \and
% \inferrule{\isType A \\ \isType B}{\isType{(A\to B)}} \and
% \inferrule{\isType A}{\isType{(\text{List}~A)}} \and
% \inferrule{\isType A}{\isType{(\text{Maybe}~A)}} \and
% \inferrule{}{\isType{\text{Int}}} \and
% \inferrule{}{\isType{\text{Bool}}} \and
% \inferrule{}{\isType\top} \and
% \inferrule{}{\isType\bot}
% \end{mathpar}
% The formation of types is interpreted as objects in a category of contexts,
% and the typing judgments of form $\Gvdash u:A$, $\Gvdash \sigma:\Delta$ are interpreted
% as morphisms in a category of contexts.
% \end{remark}

% \begin{remark}
% It is possible to describe a \textit{dependent type} in a category of contexts $\CC$
% by interpreting the judgment \fbox{$\Gvdash\isType A$} as \fbox{$\Gamma,A\vdash$}
% so that it corresponds to an object in $\CC$.
% \end{remark}

\begin{remark}
By developing a categorical model directly from the classifying category,
we will learn nothing about the relationship between category theory and type theory.

To find a better interpretation, we instead develop a category theoretical construction
directly in category theory and then interpret type theoretical constructions with it.
\end{remark}

\section{Basic category theory}\label{sec:intro}
\begin{defn}[Presheaf]\label{def:presh}
For a category $\CC$, functor $T:\OpCat\CC\to\SET$ is called a \textit{presheaf}.
Sometimes we also say that functor $T':\OpCat\CC\to\VV$ to be a $\VV$-valued presheaf.
\end{defn}

\begin{convention}[Presheaves]\label{conv:presh}
Not all categories $\VV$ are suitable as the codomain of presheaves (\cref{def:presh}).
We normally use those whose objects are some kind of ``containers'' (like sets, categories, etc.)
so that for each object and morphism in $\CC$, we obtain information about them from the presheaves over $\CC$.
\end{convention}

\begin{defn}[HomFunctor]\label{def:homfunc}
For a category $\CC$ and an object $B\in\CC$, the \textit{hom functor} $\CC(-,B):\OpCat\CC\to\SET$
sends each object $A\in\CC$ to the hom set $\CC(A,B)$. The morphism $h\in\CC(X,Y)$ is sent to the function
$(-\circ h):\CC(Y,B)\to\CC(X,B)$. A poor visualization:
% https://q.uiver.app/?q=WzAsMyxbMCwwLCJBXzAiXSxbMiwwLCJCIl0sWzQsMCwiQV8xIl0sWzAsMSwiaF8xIiwxXSxbMCwxLCJoXzAiLDEseyJjdXJ2ZSI6LTJ9XSxbMCwxLCJoXzIiLDEseyJjdXJ2ZSI6Mn1dLFsyLDEsImdfMCIsMSx7ImN1cnZlIjotMn1dLFsyLDEsImdfMSIsMSx7ImN1cnZlIjoyfV1d
\[\begin{tikzcd}
	{A_0} && B && {A_1}
	\arrow["{h_1}"{description}, from=1-1, to=1-3]
	\arrow["{h_0}"{description}, curve={height=-12pt}, from=1-1, to=1-3]
	\arrow["{h_2}"{description}, curve={height=12pt}, from=1-1, to=1-3]
	\arrow["{g_0}"{description}, curve={height=-12pt}, from=1-5, to=1-3]
	\arrow["{g_1}"{description}, curve={height=12pt}, from=1-5, to=1-3]
\end{tikzcd}\]
The hom functor sends $A_0$ to the set $\{h_0,h_1,h_2\}$ and $A_1$ to the set $\{g_0,g_1\}$.
Similarly, we also say $\CC(B,-):\CC\to\SET$ to be a hom functor.
\end{defn}

\begin{example}
Hom functors (\cref{def:homfunc}) are presheaves (\cref{def:presh}).
For a category $\CC$, a presheaf $T:\OpCat\CC\to\SET$ is called a
\textit{representable presheaf} if it is naturally isomorphic to $\CC(-,B)$.
\end{example}

\begin{defn}[Overcategory]\label{def:slice}
For a category $\CC$ and $B\in\CC$, we define a category $\CC_{/B}$, called the \textit{category over} $B$,
whose objects are morphisms in $\CC$ towards $B$.
Morphisms in $\CC_{/B}$ are inherited from $\CC$ in the following way: given morphisms
$f\in\CC(A_0,B)$ and $g\in\CC(A_1,B)$, a morphism in $\CC_{/B}$ between objects $f$ and $g$ is a morphism
$h\in\CC(A_0,A_1)$ (note~\cref{cor:dom-slice}) such that $g\circ h=f$. This process is visualized below:
% https://q.uiver.app/?q=WzAsMyxbMCwwLCJBXzAiXSxbMSwxLCJCIl0sWzIsMCwiQV8xIl0sWzAsMSwiZiIsMV0sWzAsMiwiaCIsMV0sWzIsMSwiZyIsMV1d
\[\begin{tikzcd}
	{A_0} && {A_1} \\
	& B
	\arrow["f"{description}, from=1-1, to=2-2]
	\arrow["h"{description}, from=1-1, to=1-3]
	\arrow["g"{description}, from=1-3, to=2-2]
\end{tikzcd}\]
Identities and compositions of $\CC_{/B}$ are also inherited from $\CC$.
We visualize $\CC_{/B}$ with the following diagram with $A_0,A_1,A_2,A_3\in\CC$:
% https://q.uiver.app/?q=WzAsOCxbNCwxLCJCIl0sWzMsMCwiQV8xIl0sWzUsMCwiQV8yIl0sWzEsMCwiQV8wIl0sWzcsMCwiQV8zIl0sWzcsMV0sWzAsMCwiXFx0ZXh0e2xpbmV9Il0sWzgsMF0sWzEsMCwiZl8xIiwxLHsiY3VydmUiOjF9XSxbMiwwLCJmXzIiLDEseyJjdXJ2ZSI6LTF9XSxbMywwLCJmXzAiLDEseyJjdXJ2ZSI6Mn1dLFs0LDAsImZfMyIsMSx7ImN1cnZlIjotMn1dLFsyLDQsImdfezIzfSIsMSx7InN0eWxlIjp7InRhaWwiOnsibmFtZSI6ImFycm93aGVhZCJ9fX1dLFsxLDIsImdfezEyfSIsMSx7InN0eWxlIjp7InRhaWwiOnsibmFtZSI6ImFycm93aGVhZCJ9fX1dLFsxLDQsImdfezEzfSIsMSx7ImN1cnZlIjotMiwic3R5bGUiOnsidGFpbCI6eyJuYW1lIjoiYXJyb3doZWFkIn19fV0sWzEsMywiZ197MDF9IiwxLHsic3R5bGUiOnsidGFpbCI6eyJuYW1lIjoiYXJyb3doZWFkIn19fV0sWzMsMiwiZ197MDJ9IiwxLHsiY3VydmUiOi0yLCJzdHlsZSI6eyJ0YWlsIjp7Im5hbWUiOiJhcnJvd2hlYWQifX19XSxbMyw0LCJnX3swM30iLDEseyJjdXJ2ZSI6LTQsInN0eWxlIjp7InRhaWwiOnsibmFtZSI6ImFycm93aGVhZCJ9fX1dLFs2LDcsIiIsMSx7Im9mZnNldCI6Mywic3R5bGUiOnsiYm9keSI6eyJuYW1lIjoiZGFzaGVkIn0sImhlYWQiOnsibmFtZSI6Im5vbmUifX19XV0=
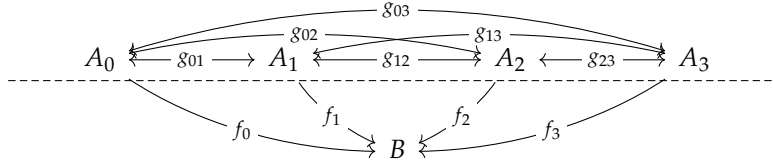
\[\begin{tikzcd}
	{} & {A_0} && {A_1} && {A_2} && {A_3} & {} \\
	&&&& B &&& {}
	\arrow["{f_1}"{description}, curve={height=6pt}, from=1-4, to=2-5]
	\arrow["{f_2}"{description}, curve={height=-6pt}, from=1-6, to=2-5]
	\arrow["{f_0}"{description}, curve={height=12pt}, from=1-2, to=2-5]
	\arrow["{f_3}"{description}, curve={height=-12pt}, from=1-8, to=2-5]
	\arrow["{g_{23}}"{description}, tail reversed, from=1-6, to=1-8]
	\arrow["{g_{12}}"{description}, tail reversed, from=1-4, to=1-6]
	\arrow["{g_{13}}"{description}, curve={height=-12pt}, tail reversed, from=1-4, to=1-8]
	\arrow["{g_{01}}"{description}, tail reversed, from=1-4, to=1-2]
	\arrow["{g_{02}}"{description}, curve={height=-12pt}, tail reversed, from=1-2, to=1-6]
	\arrow["{g_{03}}"{description}, curve={height=-24pt}, tail reversed, from=1-2, to=1-8]
	\arrow[shift right=3, dashed, no head, from=1-1, to=1-9]
\end{tikzcd}\]
The morphisms below the dashed line ($f_i$ for $i\in[0,3]$) are objects of $\CC_{/B}$,
while the others ($g_{ij}$ for $i,j\in[0,3]$) are morphisms of $\CC_{/B}$.
We use bidirectional arrows to mean ``morphisms of both directions''.
$\CC_{/B}$ is also known as an \textit{overcategory} or the \textit{slice category over $B$}.
\end{defn}

\begin{prop}\label{cor:dom-slice}
\lessSpace{-2}
\begin{align*}
g\in\CC_{/B}(f,f') &\implies g\in\CC(dom(f), dom(f')) \\
g\in\CC(dom(f), dom(f')) &\centernot\implies g\in\CC_{/B}(f,f')
\end{align*}
\lessSpace{-2}
\end{prop}

\begin{lem}\label{lem:overcat-equiv}
For a category $\CC$ and its terminal object (\cref{def:termobj}) 1, $\CC\simeq\CC_{/1}$.
\end{lem}
\begin{proof}
We take the objects in $\CC$ to its unique morphism towards the terminal object in the forward direction,
and take the domain of these morphisms in the backward direction.
\end{proof}

\begin{example}
For $\CC$ a category of contexts (\cref{def:cat-ctx}) and $\Gamma\in\CC$,
the category of ``substitutions typed in $\Gamma$'' is the overcategory (\cref{def:slice}) $\CC_{/\Gamma}$.
\end{example}

\begin{defn}[ArrowCat]\label{def:arrcat}
For an arbitrary category $\CC$, the \textit{arrow category} $\CC^\to$ (or $\text{Arr}(\CC)$) is the category
whose objects are morphisms in $\CC$. The morphisms in $\CC^\to$ are square-shaped commutative diagrams in $\CC$.
For example, for $f_0,f_1\in\CC^\to$, the hom set $\CC^\to(f_0,f_1)$ are pairs of morphisms $(g_0,h_0)$
in $\CC$ such that the following square commutes:
% https://q.uiver.app/?q=WzAsNCxbMCwwLCJcXGJ1bGxldCJdLFswLDEsIlxcYnVsbGV0Il0sWzEsMCwiXFxidWxsZXQiXSxbMSwxLCJcXGJ1bGxldCJdLFswLDEsImZfMCIsMl0sWzAsMiwiZ18wIl0sWzIsMywiZl8xIiwyXSxbMSwzLCJoXzAiXV0=
\[\begin{tikzcd}
	\bullet & \bullet \\
	\bullet & \bullet
	\arrow["{f_0}"', from=1-1, to=2-1]
	\arrow["{g_0}", from=1-1, to=1-2]
	\arrow["{f_1}"', from=1-2, to=2-2]
	\arrow["{h_0}", from=2-1, to=2-2]
\end{tikzcd}\]
Composition in $\CC$ is the composition of commutative diagrams. For example, the composition of
$(g_0,h_0)\in\CC^\to(f_0,f_1)$ and $(g_1,h_1)\in\CC^\to(f_1,f_2)$ can be visualized in this way:
% https://q.uiver.app/?q=WzAsNixbMCwwLCJcXGJ1bGxldCJdLFswLDEsIlxcYnVsbGV0Il0sWzEsMCwiXFxidWxsZXQiXSxbMSwxLCJcXGJ1bGxldCJdLFsyLDAsIlxcYnVsbGV0Il0sWzIsMSwiXFxidWxsZXQiXSxbMCwxLCJmXzAiLDJdLFswLDIsImdfMCJdLFsyLDMsImZfMSIsMl0sWzEsMywiaF8wIl0sWzIsNCwiZ18xIl0sWzMsNSwiaF8xIl0sWzQsNSwiZl8yIiwyXV0=
\[\begin{tikzcd}
	\bullet & \bullet & \bullet \\
	\bullet & \bullet & \bullet
	\arrow["{f_0}"', from=1-1, to=2-1]
	\arrow["{g_0}", from=1-1, to=1-2]
	\arrow["{f_1}"', from=1-2, to=2-2]
	\arrow["{h_0}", from=2-1, to=2-2]
	\arrow["{g_1}", from=1-2, to=1-3]
	\arrow["{h_1}", from=2-2, to=2-3]
	\arrow["{f_2}"', from=1-3, to=2-3]
\end{tikzcd}\]
\lessSpace{-1}
\end{defn}

\begin{defn}[CodProj]\label{def:mor-proj}
The codomain projection $cod:\CC^\to \to \CC$
and the domain projection $dom:\CC^\to \to \CC$ (\cref{not:ct})
are functors from the arrow category (\cref{def:arrcat}) of $\CC$ to $\CC$.
Objects in $\CC^\to$ are mapped to their (co)domain, and morphisms are mapped like this:
consider a morphism $s\in\CC^\to(a, b)$ with the following data:
% https://q.uiver.app/?q=WzAsNCxbMSwxLCJcXGJ1bGxldCIsWzI0MCw2MCw1MCwxXV0sWzEsMCwiXFxidWxsZXQiLFsxMjAsNjAsMzMsMV1dLFswLDEsIlxcYnVsbGV0Il0sWzAsMCwiXFxidWxsZXQiXSxbMSwwLCJmIl0sWzIsMCwiYSIsMix7ImNvbG91ciI6WzI0MCw2MCw1MF19LFsyNDAsNjAsNTAsMV1dLFszLDIsImciLDJdLFszLDEsImIiLDAseyJjb2xvdXIiOlsxMjAsNjAsMzNdfSxbMTIwLDYwLDMzLDFdXSxbMywwLCJzIiwxLHsic3R5bGUiOnsiYm9keSI6eyJuYW1lIjoibm9uZSJ9LCJoZWFkIjp7Im5hbWUiOiJub25lIn19fV1d
\[\begin{tikzcd}
	\bullet & \textcolor{rgb,255:red,34;green,135;blue,34}{\bullet} \\
	\bullet & \textcolor{rgb,255:red,51;green,51;blue,204}{\bullet}
	\arrow["f", from=1-2, to=2-2]
	\arrow["a"', color={rgb,255:red,51;green,51;blue,204}, from=2-1, to=2-2]
	\arrow["g"', from=1-1, to=2-1]
	\arrow["b", color={rgb,255:red,34;green,135;blue,34}, from=1-1, to=1-2]
	\arrow["s"{description}, draw=none, from=1-1, to=2-2]
\end{tikzcd}\]
We have $cod(s)=f$ and $dom(s)=g$. This completes the definition of the two functors.
The functoriality follows directly from morphism composition in $\CC$.
\end{defn}

\subsection{Grothendieck fibrations}
\begin{defn}[Cartesianness]\label{def:cart-mor}
For arbitrary categories $\EE,\CC$, a functor $p:\EE\to\CC$, and objects $\Gamma\in\CC$ and $D,E\in\EE$,
a morphism $f\in\EE(D,E)$ is called \textit{($p$-)Grothendieck cartesian} if for all
$f'\in\EE(D',E)$ (where $D'\in\EE$) such that $p(f)=p(f')\in\CC(p(D),p(E))$
(this implies $p(D)=p(D')\in\CC$), there is a unique $g\in\EE(D',D)$ such that $p(g)=\id_{p(D)}$ and $f'=f\circ g$.
We visualize the ($p$-)Grothendieck cartesianness of $f$ as:
% https://q.uiver.app/?q=WzAsNSxbMiwwLCJwKEUpIixbMjQwLDYwLDUwLDFdXSxbMiwxLCJwKEQpIixbMzAsNjAsNDEsMV1dLFsxLDEsIkQiLFszMCw2MCw0MSwxXV0sWzAsMSwiRCciLFszMCw2MCw0MSwxXV0sWzAsMCwiRSIsWzI0MCw2MCw1MCwxXV0sWzEsMCwicChmKSIsMix7ImNvbG91ciI6WzMwMCw2MCwzOV19LFszMDAsNjAsMzksMV1dLFszLDIsImciLDIseyJjb2xvdXIiOlszMCw2MCw0MV0sInN0eWxlIjp7ImJvZHkiOnsibmFtZSI6ImRhc2hlZCJ9fX0sWzMwLDYwLDQxLDFdXSxbMyw0LCJmJyIsMCx7ImNvbG91ciI6WzMwMCw2MCwzOV19LFszMDAsNjAsMzksMV1dLFsyLDQsImYiLDEseyJjb2xvdXIiOlszMDAsNjAsMzldfSxbMzAwLDYwLDM5LDFdXV0=
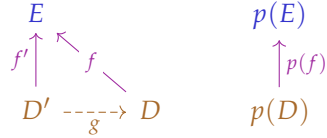
\[\begin{tikzcd}
	\textcolor{rgb,255:red,51;green,51;blue,204}{E} && \textcolor{rgb,255:red,51;green,51;blue,204}{p(E)} \\
	\textcolor{rgb,255:red,167;green,105;blue,42}{D'} & \textcolor{rgb,255:red,167;green,105;blue,42}{D} & \textcolor{rgb,255:red,167;green,105;blue,42}{p(D)}
	\arrow["{p(f)}"', color={rgb,255:red,159;green,40;blue,159}, from=2-3, to=1-3]
	\arrow["g"', color={rgb,255:red,167;green,105;blue,42}, dashed, from=2-1, to=2-2]
	\arrow["{f'}", color={rgb,255:red,159;green,40;blue,159}, from=2-1, to=1-1]
	\arrow["f"{description}, color={rgb,255:red,159;green,40;blue,159}, from=2-2, to=1-1]
\end{tikzcd}\]
The diagram shows why $f$ is sometimes known as a \textit{terminal lifting}.
This definition is due to Jacobs~\cite[definition 2.1 (i)]{CompCat} and Grothendieck~\cite{Fibration}.
\end{defn}

\begin{history}\label{his:cart}
The~\cref{def:cart-mor} is called \textit{cartesianness} in~\cite{CompCat}, but we prefix it with ``Grothendieck''
because there is another (newer) definition of cartesianness of morphisms in category theory,
which is discussed in~\cref{def:gen-cart}.
Grothendieck cartesianness is also known as \textit{weak cartesianness}, where~\cref{def:gen-cart}
is also known as \textit{strong cartesianness}.
\end{history}

\begin{remark}
Recall the intuition in~\cref{rem:pullback-intuition} and consider the functor $p:\EE\to\CC$.
A (p-)Grothendieck cartesian morphism (\cref{def:cart-mor}) $f$ is nice because among all morphisms $f'$
such that $p(f')=p(f)$, we can turn $f'$ into $f$ by removing information,
so $f$ is the simplest such morphism and therefore best for \textit{representing} $p(f)$ in $\CC$.
\end{remark}

\begin{example}\label{ex:codproj-cart}
Consider the codomain projection functor $cod:\CC^\to\to\CC$ (\cref{def:mor-proj})
and recall that morphisms in $\CC^\to$ are commutative squares (\cref{def:arrcat}).
Pullback squares (\cref{def:pullback}) in $\CC$ are cartesian morphisms (\cref{def:cart-mor}).
For $a,b\in\CC^\to$ such that $cod(a)=A\in\CC$ and $cod(b)=B\in\CC$, the cartesianness of the
morphism $s\in\CC^\to(b,a)$ (where $cod(s)=f$) in the following diagram is given by the pullback:
% https://q.uiver.app/?q=WzAsNixbMiwxLCJBIixbMjQwLDYwLDUwLDFdXSxbMiwwLCJCIixbMTIwLDYwLDMzLDFdXSxbMSwxLCJBIixbMjQwLDYwLDUwLDFdXSxbMSwwLCJCIixbMTIwLDYwLDMzLDFdXSxbMCwxLCJcXGJ1bGxldCJdLFswLDAsIlxcYnVsbGV0Il0sWzEsMCwiZiJdLFszLDIsImYiXSxbNCwyLCJhIiwyLHsiY29sb3VyIjpbMjQwLDYwLDUwXX0sWzI0MCw2MCw1MCwxXV0sWzUsNCwiIiwyLHsic3R5bGUiOnsiYm9keSI6eyJuYW1lIjoiZGFzaGVkIn19fV0sWzUsMywiYiIsMCx7ImNvbG91ciI6WzEyMCw2MCwzM119LFsxMjAsNjAsMzMsMV1dLFs1LDIsInMiLDEseyJzdHlsZSI6eyJuYW1lIjoiY29ybmVyIn19XV0=
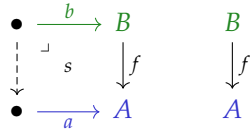
\[\begin{tikzcd}
	\bullet & \textcolor{rgb,255:red,34;green,135;blue,34}{B} & \textcolor{rgb,255:red,34;green,135;blue,34}{B} \\
	\bullet & \textcolor{rgb,255:red,51;green,51;blue,204}{A} & \textcolor{rgb,255:red,51;green,51;blue,204}{A}
	\arrow["f", from=1-3, to=2-3]
	\arrow["f", from=1-2, to=2-2]
	\arrow["a"', color={rgb,255:red,51;green,51;blue,204}, from=2-1, to=2-2]
	\arrow[dashed, from=1-1, to=2-1]
	\arrow["b", color={rgb,255:red,34;green,135;blue,34}, from=1-1, to=1-2]
	\arrow["s"{description}, "\lrcorner"{anchor=center, pos=0.125}, draw=none, from=1-1, to=2-2]
\end{tikzcd}\]
We now verify the cartesianness. Since $cod(s)=f$, for every diagram $s'\in\CC^\to(b', a)$
(where $cod(s')=f$), we need to construct a unique diagram $o\in\CC^\to(b', b)$
such that $s'=s\circ o$. We can put the diagrams $s, s', o$ together like this:
% https://q.uiver.app/?q=WzAsMTIsWzUsMSwiQSIsWzI0MCw2MCw1MCwxXV0sWzUsMCwiQiIsWzEyMCw2MCwzMywxXV0sWzQsMSwiXFxidWxsZXQiXSxbNCwwLCJcXGJ1bGxldCJdLFsyLDAsIlxcYnVsbGV0Il0sWzMsMCwiQiIsWzEyMCw2MCwzMywxXV0sWzIsMSwiXFxidWxsZXQiXSxbMywxLCJBIixbMjQwLDYwLDUwLDFdXSxbMSwwLCJCIixbMTIwLDYwLDMzLDFdXSxbMCwwLCJcXGJ1bGxldCJdLFsxLDEsIkIiLFsxMjAsNjAsMzMsMV1dLFswLDEsIlxcYnVsbGV0Il0sWzEsMCwiZiIsMCx7ImNvbG91ciI6WzMzNyw2MCw0NF19LFszMzcsNjAsNDQsMV1dLFsyLDAsImEiLDIseyJjb2xvdXIiOlsyNDAsNjAsNTBdfSxbMjQwLDYwLDUwLDFdXSxbMywyLCIiLDIseyJzdHlsZSI6eyJib2R5Ijp7Im5hbWUiOiJkYXNoZWQifX19XSxbMywxLCJiIiwwLHsiY29sb3VyIjpbMTIwLDYwLDMzXX0sWzEyMCw2MCwzMywxXV0sWzMsMCwicyIsMSx7InN0eWxlIjp7Im5hbWUiOiJjb3JuZXIifX1dLFs0LDUsImInIl0sWzYsNywiYSIsMix7ImNvbG91ciI6WzI0MCw2MCw1MF19LFsyNDAsNjAsNTAsMV1dLFs1LDcsImYiLDAseyJjb2xvdXIiOlszMzcsNjAsNDRdfSxbMzM3LDYwLDQ0LDFdXSxbNCw2XSxbOSw4LCJiJyJdLFs4LDEwLCIiLDAseyJsZXZlbCI6MiwiY29sb3VyIjpbMTIwLDYwLDMzXSwic3R5bGUiOnsiaGVhZCI6eyJuYW1lIjoibm9uZSJ9fX1dLFs5LDExLCIiLDIseyJzdHlsZSI6eyJib2R5Ijp7Im5hbWUiOiJkYXNoZWQifX19XSxbMTEsMTAsImIiLDIseyJjb2xvdXIiOlsxMjAsNjAsMzNdfSxbMTIwLDYwLDMzLDFdXSxbOSwxMCwibyIsMSx7InN0eWxlIjp7ImJvZHkiOnsibmFtZSI6Im5vbmUifSwiaGVhZCI6eyJuYW1lIjoibm9uZSJ9fX1dLFs0LDcsInMnIiwxLHsic3R5bGUiOnsiYm9keSI6eyJuYW1lIjoibm9uZSJ9LCJoZWFkIjp7Im5hbWUiOiJub25lIn19fV1d
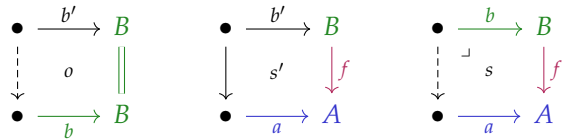
\[\begin{tikzcd}
	\bullet & \textcolor{rgb,255:red,34;green,135;blue,34}{B} & \bullet & \textcolor{rgb,255:red,34;green,135;blue,34}{B} & \bullet & \textcolor{rgb,255:red,34;green,135;blue,34}{B} \\
	\bullet & \textcolor{rgb,255:red,34;green,135;blue,34}{B} & \bullet & \textcolor{rgb,255:red,51;green,51;blue,204}{A} & \bullet & \textcolor{rgb,255:red,51;green,51;blue,204}{A}
	\arrow["f", color={rgb,255:red,180;green,45;blue,96}, from=1-6, to=2-6]
	\arrow["a"', color={rgb,255:red,51;green,51;blue,204}, from=2-5, to=2-6]
	\arrow[dashed, from=1-5, to=2-5]
	\arrow["b", color={rgb,255:red,34;green,135;blue,34}, from=1-5, to=1-6]
	\arrow["s"{description}, "\lrcorner"{anchor=center, pos=0.125}, draw=none, from=1-5, to=2-6]
	\arrow["{b'}", from=1-3, to=1-4]
	\arrow["a"', color={rgb,255:red,51;green,51;blue,204}, from=2-3, to=2-4]
	\arrow["f", color={rgb,255:red,180;green,45;blue,96}, from=1-4, to=2-4]
	\arrow[from=1-3, to=2-3]
	\arrow["{b'}", from=1-1, to=1-2]
	\arrow[color={rgb,255:red,34;green,135;blue,34}, Rightarrow, no head, from=1-2, to=2-2]
	\arrow[dashed, from=1-1, to=2-1]
	\arrow["b"', color={rgb,255:red,34;green,135;blue,34}, from=2-1, to=2-2]
	\arrow["o"{description}, draw=none, from=1-1, to=2-2]
	\arrow["{s'}"{description}, draw=none, from=1-3, to=2-4]
\end{tikzcd}\]
From the visualization we can tell that the goal is to find a unique morphism in $\CC(dom(b'), dom(b))$.
Now we rearrange the diagrams:
% https://q.uiver.app/?q=WzAsNSxbMiwxLCJBIixbMjQwLDYwLDUwLDFdXSxbMiwwLCJCIixbMTIwLDYwLDMzLDFdXSxbMSwxLCJcXGJ1bGxldCJdLFsxLDAsIlxcYnVsbGV0Il0sWzAsMCwiXFxidWxsZXQiXSxbMSwwLCJmIiwwLHsiY29sb3VyIjpbMzM3LDYwLDQ0XX0sWzMzNyw2MCw0NCwxXV0sWzIsMCwiYSIsMix7ImNvbG91ciI6WzI0MCw2MCw1MF19LFsyNDAsNjAsNTAsMV1dLFszLDIsIiIsMix7InN0eWxlIjp7ImJvZHkiOnsibmFtZSI6ImRhc2hlZCJ9fX1dLFszLDEsImIiLDIseyJjb2xvdXIiOlsxMjAsNjAsMzNdfSxbMTIwLDYwLDMzLDFdXSxbMywwLCJzIiwxLHsic3R5bGUiOnsibmFtZSI6ImNvcm5lciJ9fV0sWzQsMl0sWzQsMSwiYiciLDAseyJjdXJ2ZSI6LTF9XSxbNCwzLCIiLDEseyJzdHlsZSI6eyJib2R5Ijp7Im5hbWUiOiJkYXNoZWQifX19XV0=
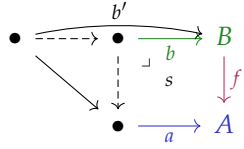
\[\begin{tikzcd}
	\bullet & \bullet & \textcolor{rgb,255:red,34;green,135;blue,34}{B} \\
	& \bullet & \textcolor{rgb,255:red,51;green,51;blue,204}{A}
	\arrow["f", color={rgb,255:red,180;green,45;blue,96}, from=1-3, to=2-3]
	\arrow["a"', color={rgb,255:red,51;green,51;blue,204}, from=2-2, to=2-3]
	\arrow[dashed, from=1-2, to=2-2]
	\arrow["b"', color={rgb,255:red,34;green,135;blue,34}, from=1-2, to=1-3]
	\arrow["s"{description}, "\lrcorner"{anchor=center, pos=0.125}, draw=none, from=1-2, to=2-3]
	\arrow[from=1-1, to=2-2]
	\arrow["{b'}", curve={height=-6pt}, from=1-1, to=1-3]
	\arrow[dashed, from=1-1, to=1-2]
\end{tikzcd}\]
This is precisely the characterization diagram of pullbacks (\cref{def:pullback}),
which gives us the desired unique morphism.
\end{example}

\begin{defn}[Fibration]\label{def:fibr}
For arbitrary categories $\EE,\CC$, objects $\Gamma\in\CC$ and $D\in\EE$, a functor $p:\EE\to\CC$ is called a
\textit{Grothendieck fibration} or a \textit{fibered category} if the following holds:
\begin{itemize}
\item Composition of morphisms in $\EE$ preserves Grothendieck cartesianness.
\item For every $E\in\EE$ and $\sigma\in\CC(\Gamma,p(E))$, there exists a Grothendieck cartesian morphism
(\cref{def:cart-mor}) $f\in\EE(D, E)$ such that $p(f)=\sigma$.
Such $f$ is called the \textit{cartesian lifting} of $\sigma$, denoted $\overline \sigma(E)$.
The domain $D$ of $f$ is also uniquely determined, denoted $\sigma^*(E)$.
\end{itemize}
We visualize latter condition as:
\[\begin{tikzcd}
	E & {p(E)} \\
	{\sigma^*(E)} & \Gamma
	\arrow[""{name=0, anchor=center, inner sep=0}, "\sigma"', from=2-2, to=1-2]
	\arrow[""{name=1, anchor=center, inner sep=0}, "{\overline \sigma(E)}", from=2-1, to=1-1]
	\arrow["p"', shorten <=7pt, shorten >=7pt, Rightarrow, from=1, to=0]
\end{tikzcd}\]
The entire diagram is uniquely determined by only two variables: $\sigma$ and $E$.
This definition is due to Jacobs~\cite[definition 2.1 (iii)]{CompCat} and Grothendieck~\cite{Fibration}.
\end{defn}

\begin{convention}
We will refer to ``Grothendieck fibrations'' as ``fibrations'' in the rest of this document
since we do not discuss other kinds of fibrations.
\end{convention}

\begin{example}\label{ex:codproj}
The codomain projection functor $cod:\CC^\to\to\CC$ (\cref{def:mor-proj}) is a fibration (\cref{def:fibr})
if $\CC$ has all pullbacks (\cref{def:pullback}). This is because pullbacks are cartesian (\cref{ex:codproj-cart}),
and ``have all pullback'' means for every morphism there exists a cartesian lifting (\cref{def:fibr}).
\end{example}

\begin{example}\label{ex:domproj}
The domain projection functor $dom:\CC^\to\to\CC$ (\cref{def:mor-proj}) is a fibration (\cref{def:fibr}).
For $a,b\in\CC^\to$ such that $dom(a)=A\in\CC$ and $dom(b)=B\in\CC$,
we visualize the cartesianness of the commutative diagram, which is a morphism in $\CC^\to(b,a)$, as follows:
% https://q.uiver.app/?q=WzAsNixbMCwxLCJBIixbMjQwLDYwLDUwLDFdXSxbMCwwLCJCIixbMTIwLDYwLDMzLDFdXSxbMSwxLCJBIixbMjQwLDYwLDUwLDFdXSxbMSwwLCJCIixbMTIwLDYwLDMzLDFdXSxbMiwxLCJcXGJ1bGxldCJdLFsyLDAsIlxcYnVsbGV0Il0sWzEsMCwiZiIsMl0sWzIsNCwiYSIsMix7ImNvbG91ciI6WzI0MCw2MCw1MF19LFsyNDAsNjAsNTAsMV1dLFs1LDQsImciXSxbMyw1LCJiIiwwLHsiY29sb3VyIjpbMTIwLDYwLDMzXX0sWzEyMCw2MCwzMywxXV0sWzMsMiwiZiIsMl1d
\[\begin{tikzcd}
	\textcolor{rgb,255:red,34;green,135;blue,34}{B} & \textcolor{rgb,255:red,34;green,135;blue,34}{B} & \bullet \\
	\textcolor{rgb,255:red,51;green,51;blue,204}{A} & \textcolor{rgb,255:red,51;green,51;blue,204}{A} & \bullet
	\arrow["f"', from=1-1, to=2-1]
	\arrow["a"', color={rgb,255:red,51;green,51;blue,204}, from=2-2, to=2-3]
	\arrow["g", from=1-3, to=2-3]
	\arrow["b", color={rgb,255:red,34;green,135;blue,34}, from=1-2, to=1-3]
	\arrow["f"', from=1-2, to=2-2]
\end{tikzcd}\]
Unlike in \cref{ex:codproj}, $\CC$ does not need to have pullbacks because the cartesianness of $f$
only requires the commutativity of the square.
\end{example}

\begin{exercise}
Verify the cartesianness of the commutative diagram in~\cref{ex:domproj}.
\end{exercise}

\subsection{Towards a comprehension category}
\begin{glorious-defn}[Jacobs' CompCat]\label{def:compcat}
A \textit{comprehension category} is a structure consisting of a category $\CC$,
a functor $\FF:\EE\to\CC^\to$ preserving Grothendieck cartesianness of morphisms, and $p:=cod\circ\FF:\EE\to\CC$
a fibration (\cref{def:fibr}). We denote $\FF_0=dom\circ\FF$ due to Jacobs~\cite[notation 4.2]{CompCat}.
The definition of a comprehension category commutes the following diagram:
% https://q.uiver.app/?q=WzAsOSxbNiwxLCJcXERlbHRhIixbMjQwLDYwLDUwLDFdXSxbNiwwLCJcXEdhbW1hIixbMTIwLDYwLDMzLDFdXSxbNCwxLCJcXERlbHRhIixbMjQwLDYwLDUwLDFdXSxbNCwwLCJcXEdhbW1hIixbMTIwLDYwLDMzLDFdXSxbMywxLCJcXGJ1bGxldCIsWzI0MCw2MCw1MCwxXV0sWzMsMCwiXFxidWxsZXQiLFsxMjAsNjAsMzMsMV1dLFsxLDAsIlxcYnVsbGV0IixbMTIwLDYwLDMzLDFdXSxbMSwxLCJcXGJ1bGxldCIsWzI0MCw2MCw1MCwxXV0sWzAsMSwiXFxidWxsZXQiXSxbMSwwLCIiLDAseyJjb2xvdXIiOlsyNzAsNjAsNDZdfV0sWzMsMiwiIiwwLHsiY29sb3VyIjpbMjcwLDYwLDQ2XX1dLFs0LDIsIiIsMix7ImNvbG91ciI6WzI0MCw2MCw1MF19XSxbNSw0LCIiLDIseyJjb2xvdXIiOlsyNzAsNjAsNDZdLCJzdHlsZSI6eyJib2R5Ijp7Im5hbWUiOiJkYXNoZWQifX19XSxbNSwyLCJcXENDXlxcdG8iLDEseyJzdHlsZSI6eyJuYW1lIjoiY29ybmVyIn19XSxbNiw3LCIiLDAseyJjb2xvdXIiOlsyNzAsNjAsNDZdfV0sWzgsN10sWzgsNiwiIiwxLHsic3R5bGUiOnsiYm9keSI6eyJuYW1lIjoiZGFzaGVkIn19fV0sWzUsMywiIiwwLHsiY29sb3VyIjpbMTIwLDYwLDQ2XX1dLFsxMCw5LCIiLDEseyJzaG9ydGVuIjp7InNvdXJjZSI6MTAsInRhcmdldCI6MTB9LCJzdHlsZSI6eyJoZWFkIjp7Im5hbWUiOiJub25lIn19fV0sWzE0LDEyLCJcXEZGXzAiLDEseyJzaG9ydGVuIjp7InNvdXJjZSI6MTAsInRhcmdldCI6MTB9LCJsZXZlbCI6MX1dLFsxNCw5LCJwIiwxLHsiY3VydmUiOjUsInNob3J0ZW4iOnsic291cmNlIjoxMCwidGFyZ2V0IjoxMH0sImxldmVsIjoxfV0sWzE2LDcsIlxcRUUiLDEseyJzaG9ydGVuIjp7InRhcmdldCI6MjB9LCJzdHlsZSI6eyJib2R5Ijp7Im5hbWUiOiJub25lIn0sImhlYWQiOnsibmFtZSI6Im5vbmUifX19XSxbNiwxNywiXFxGRiIsMSx7Im9mZnNldCI6LTIsImN1cnZlIjotMiwibGV2ZWwiOjF9XV0=&macro_url=https%3A%2F%2Fgist.githubusercontent.com%2Fice1000%2F47b7ea52f8c351607831f7f4afa9a79b%2Fraw%2Ff57182e539831aa2c41b1071cb001d2c4a725018%2Fquiver-macros.tex
\[\begin{tikzcd}
	& \textcolor{rgb,255:red,34;green,135;blue,34}{\bullet} && \textcolor{rgb,255:red,34;green,135;blue,34}{\bullet} & \textcolor{rgb,255:red,34;green,135;blue,34}{\Gamma} && \textcolor{rgb,255:red,34;green,135;blue,34}{\Gamma} \\
	\bullet & \textcolor{rgb,255:red,51;green,51;blue,204}{\bullet} && \textcolor{rgb,255:red,51;green,51;blue,204}{\bullet} & \textcolor{rgb,255:red,51;green,51;blue,204}{\Delta} && \textcolor{rgb,255:red,51;green,51;blue,204}{\Delta}
	\arrow[""{name=0, anchor=center, inner sep=0}, draw={rgb,255:red,117;green,47;blue,188}, from=1-7, to=2-7]
	\arrow[""{name=1, anchor=center, inner sep=0}, draw={rgb,255:red,117;green,47;blue,188}, from=1-5, to=2-5]
	\arrow[draw={rgb,255:red,51;green,51;blue,204}, from=2-4, to=2-5]
	\arrow[""{name=2, anchor=center, inner sep=0}, draw={rgb,255:red,117;green,47;blue,188}, dashed, from=1-4, to=2-4]
	\arrow["{\CC^\to}"{description}, "\lrcorner"{anchor=center, pos=0.125}, draw=none, from=1-4, to=2-5]
	\arrow[""{name=3, anchor=center, inner sep=0}, draw={rgb,255:red,117;green,47;blue,188}, from=1-2, to=2-2]
	\arrow[from=2-1, to=2-2]
	\arrow[""{name=4, anchor=center, inner sep=0}, dashed, from=2-1, to=1-2]
	\arrow[""{name=5, anchor=center, inner sep=0}, draw={rgb,255:red,47;green,188;blue,47}, from=1-4, to=1-5]
	\arrow[shorten <=6pt, shorten >=6pt, Rightarrow, no head, from=1, to=0]
	\arrow["{\FF_0}"{description}, shorten <=6pt, shorten >=6pt, from=3, to=2]
	\arrow["p"{description}, curve={height=30pt}, shorten <=17pt, shorten >=17pt, from=3, to=0]
	\arrow["\EE"{description}, Rightarrow, draw=none, from=4, to=2-2]
	\arrow["\FF"{description}, shift left=2, curve={height=-12pt}, from=1-2, to=5]
\end{tikzcd}\]
We refer to the functor $\FF$ as the \textit{(context) comprehension}.
The intuition of this name is elaborated in~\cref{def:ctxext}.
\end{glorious-defn}

\begin{remark}
Although $p:=cod\circ\FF$ is a fibration in~\cref{def:compcat}, $cod$ is not necessarily a
fibration because $\CC$ may not have all pullbacks (see~\cref{ex:codproj-cart,ex:codproj}).
We require the functor $\FF$ to preserve Grothendieck cartesianness so that only the image of $\FF$ are Grothendieck cartesian morphisms.
\end{remark}

\begin{glorious-defn}[DisplayMap]\label{def:displaymap}
Consider a comprehension category (\cref{def:compcat}) $\FF:\EE\to\CC^\to$ (with $p=cod\circ \FF$),
objects $\Gamma\in\CC$ and $A\in\EE$ such that $p(A)=\Gamma$.
We denote $\FF(A)$ as $\pi_A$ so that $\pi_A\in\CC_{/\Gamma}$ and refer to it as the \textit{display map} of $A$.
% https://q.uiver.app/?q=WzAsMyxbMywwLCJcXEdhbW1hIixbMTIwLDYwLDMzLDFdXSxbMSwwLCJcXEZGXzAoQSkiLFsxMjAsNjAsMzMsMV1dLFswLDAsIkEiLFsxMjAsNjAsMzMsMV1dLFsxLDAsIlxccGlfQSIsMSx7ImNvbG91ciI6WzEyMCw2MCwzM119LFsxMjAsNjAsMzMsMV1dXQ==&macro_url=https%3A%2F%2Fgist.githubusercontent.com%2Fice1000%2F47b7ea52f8c351607831f7f4afa9a79b%2Fraw%2Fbed9ddea8cd95fb03de082dc6b3fc55b9d754c90%2Fquiver-macros.tex
\[\begin{tikzcd}
	\textcolor{rgb,255:red,34;green,135;blue,34}{A} & \textcolor{rgb,255:red,34;green,135;blue,34}{\FF_0(A)} && \textcolor{rgb,255:red,34;green,135;blue,34}{\Gamma}
	\arrow["{\pi_A}"{description}, color={rgb,255:red,34;green,135;blue,34}, from=1-2, to=1-4]
\end{tikzcd}\]
Suppose a morphism $\sigma\in\CC(\Gamma,\Delta)$ is uniquely lifted (\cref{def:cart-mor}) to $\overline\sigma(B)\in\EE(A, B)$.
Then, since $\FF$ preserves Grothendieck cartesianness, $\FF(\overline\sigma(B))\in\CC^\to(\pi_A, \pi_B)$ is a cartesian morphism in $\CC^\to$,
which is a pullback square in $\CC$.
A visualization of $(\Gamma,A)$ in the (sub)diagram in \cref{def:compcat}:
% https://q.uiver.app/?q=WzAsNyxbMywxLCJcXERlbHRhIixbMjQwLDYwLDUwLDFdXSxbMywwLCJcXEdhbW1hIixbMTIwLDYwLDMzLDFdXSxbMSwwLCJcXEZGXzAoQSkiLFsxMjAsNjAsMzMsMV1dLFswLDAsIkEiLFsxMjAsNjAsMzMsMV1dLFswLDEsIkIiLFsyNDAsNjAsNTAsMV1dLFsyLDFdLFsxLDEsIlxcRkZfMChCKSIsWzI0MCw2MCw1MCwxXV0sWzEsMCwiXFxzaWdtYSJdLFsyLDEsIlxccGlfQSIsMSx7ImNvbG91ciI6WzEyMCw2MCwzM119LFsxMjAsNjAsMzMsMV1dLFsyLDAsIlxcQ0NeXFx0byIsMSx7InN0eWxlIjp7ImJvZHkiOnsibmFtZSI6Im5vbmUifSwiaGVhZCI6eyJuYW1lIjoibm9uZSJ9fX1dLFszLDQsIlxcb3ZlcmxpbmVcXHNpZ21hKEIpIl0sWzIsNSwiIiwxLHsic3R5bGUiOnsibmFtZSI6ImNvcm5lciJ9fV0sWzYsMCwiXFxwaV9CIiwxLHsiY29sb3VyIjpbMjQwLDYwLDUwXX0sWzI0MCw2MCw1MCwxXV0sWzIsNiwiIiwyLHsic3R5bGUiOnsiYm9keSI6eyJuYW1lIjoiZGFzaGVkIn19fV1d&macro_url=https%3A%2F%2Fgist.githubusercontent.com%2Fice1000%2F47b7ea52f8c351607831f7f4afa9a79b%2Fraw%2Fbed9ddea8cd95fb03de082dc6b3fc55b9d754c90%2Fquiver-macros.tex
\[\begin{tikzcd}
	\textcolor{rgb,255:red,34;green,135;blue,34}{A} & \textcolor{rgb,255:red,34;green,135;blue,34}{\FF_0(A)} && \textcolor{rgb,255:red,34;green,135;blue,34}{\Gamma} \\
	\textcolor{rgb,255:red,51;green,51;blue,204}{B} & \textcolor{rgb,255:red,51;green,51;blue,204}{\FF_0(B)} & {} & \textcolor{rgb,255:red,51;green,51;blue,204}{\Delta}
	\arrow["\sigma", from=1-4, to=2-4]
	\arrow["{\pi_A}"{description}, color={rgb,255:red,34;green,135;blue,34}, from=1-2, to=1-4]
	\arrow["{\CC^\to}"{description}, draw=none, from=1-2, to=2-4]
	\arrow["{\overline\sigma(B)}", from=1-1, to=2-1]
	\arrow["\lrcorner"{anchor=center, pos=0.125}, draw=none, from=1-2, to=2-3]
	\arrow["{\pi_B}"{description}, color={rgb,255:red,51;green,51;blue,204}, from=2-2, to=2-4]
	\arrow[dashed, from=1-2, to=2-2]
\end{tikzcd}\]
Note that the right-half of the diagram is similar to a rotated version of the diagram in~\cref{rem:ty-subst}.
\end{glorious-defn}

\begin{defn}[CtxExt]\label{def:ctxext}
Consider a comprehension category (\cref{def:compcat}) $\FF:\EE\to\CC^\to$ (with $\FF_0=dom\circ \FF$),
for every object $A\in\EE$, we can uniquely determine a display map (\cref{def:displaymap}) $\pi_A\in\CC_{/\Gamma}$ with $\FF$,
whose domain $\FF_0(A)$ is another uniquely determined object in $\CC$. We refer to this object as the
\textit{context extended from $\Gamma$ with $A$}, denoted $(\Gamma,A)\in\CC$.
% https://q.uiver.app/?q=WzAsMyxbMywwLCJcXEdhbW1hIixbMTIwLDYwLDMzLDFdXSxbMSwwLCJcXEdhbW1hLEEiLFsxMjAsNjAsMzMsMV1dLFswLDAsIkEiLFsxMjAsNjAsMzMsMV1dLFsxLDAsIlxccGlfQSIsMSx7ImNvbG91ciI6WzEyMCw2MCwzM119LFsxMjAsNjAsMzMsMV1dXQ==&macro_url=https%3A%2F%2Fgist.githubusercontent.com%2Fice1000%2F47b7ea52f8c351607831f7f4afa9a79b%2Fraw%2Fbed9ddea8cd95fb03de082dc6b3fc55b9d754c90%2Fquiver-macros.tex
\[\begin{tikzcd}
	\textcolor{rgb,255:red,34;green,135;blue,34}{A} & \textcolor{rgb,255:red,34;green,135;blue,34}{\Gamma,A} && \textcolor{rgb,255:red,34;green,135;blue,34}{\Gamma}
	\arrow["{\pi_A}"{description}, color={rgb,255:red,34;green,135;blue,34}, from=1-2, to=1-4]
\end{tikzcd}\]
\lessSpace{-1}
\end{defn}

\begin{remark}[Spoiler I]\label{rem:spoiler1}
The~\cref{def:compcat} of a comprehension category gives us a basic sketch of a (dependent) type theory:
\begin{enumerate}
\item Contexts are interpreted as objects in $\CC$.
\item Substitution objects are interpreted as morphisms in $\CC$.
\item Objects in $\EE$ should be considered ``dependent types''.
It is in $\EE$ that type formation happens.
\item Introduction and elimination happen in $\CC$.
\item $p(A)\in\CC$ is the context where the type $A$ is formed within.
So, $A$ is a type formed in the context $p(A)$.
\item $\FF$ allows us to add new types to a context, see~\cref{def:ctxext}.
\end{enumerate}
We complete the interpretation formally in~\cref{sec:semantics}.
\end{remark}

\subsection{Slightly higher categories}
\begin{defn}[2-Cat]\label{def:bicat}
A \textit{strict 2-category} $\CC$ is a category such that for objects $A,B\in\CC$, the hom sets $f,g\in\CC(A,B)$
are also categories (called \textit{hom-categories}), and the morphisms in all these categories compose if
their domain and codomain match (this includes a lot of cases).

The objects of $\CC$ are called \textit{0-cells}, the morphisms of $\CC$ are called \textit{1-cells}
or \textit{1-morphisms}, and the morphisms in $\CC(A,B)$ are called \textit{2-cells} or \textit{2-morphisms}.
To visualize a strict 2-category, we draw a random commutative square in a strict 2-category,
where for $i\in[0,3]$, $F_i$ is a 1-cell and $t_i$ is a 2-cell:
% https://q.uiver.app/?q=WzAsOCxbMCwxLCJcXGJ1bGxldCJdLFsyLDEsIlxcYnVsbGV0Il0sWzYsMSwiXFxidWxsZXQiXSxbNCwxLCJcXGJ1bGxldCJdLFswLDAsIlxcYnVsbGV0Il0sWzIsMCwiXFxidWxsZXQiXSxbNiwwLCJcXGJ1bGxldCJdLFs0LDAsIlxcYnVsbGV0Il0sWzAsMSwiRl8wIiwxXSxbMiwzLCJGXzEiLDFdLFs0LDUsIkZfMyIsMV0sWzYsNywiRl8yIiwxXSxbMTAsOCwidF8zIiwyLHsib2Zmc2V0Ijo1LCJzaG9ydGVuIjp7InNvdXJjZSI6MjAsInRhcmdldCI6MjB9fV0sWzExLDksInRfMSIsMCx7Im9mZnNldCI6LTUsInNob3J0ZW4iOnsic291cmNlIjoyMCwidGFyZ2V0IjoyMH19XSxbMTAsMTEsInRfMiIsMSx7ImN1cnZlIjotMiwic2hvcnRlbiI6eyJzb3VyY2UiOjEwLCJ0YXJnZXQiOjEwfX1dLFs4LDksInRfMCIsMSx7ImN1cnZlIjotMywic2hvcnRlbiI6eyJzb3VyY2UiOjEwLCJ0YXJnZXQiOjEwfX1dXQ==
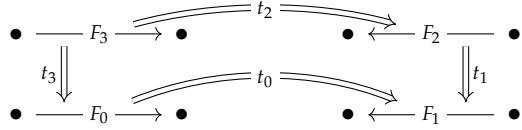
\[\begin{tikzcd}
	\bullet && \bullet && \bullet && \bullet \\
	\bullet && \bullet && \bullet && \bullet
	\arrow[""{name=0, anchor=center, inner sep=0}, "{F_0}"{description}, from=2-1, to=2-3]
	\arrow[""{name=1, anchor=center, inner sep=0}, "{F_1}"{description}, from=2-7, to=2-5]
	\arrow[""{name=2, anchor=center, inner sep=0}, "{F_3}"{description}, from=1-1, to=1-3]
	\arrow[""{name=3, anchor=center, inner sep=0}, "{F_2}"{description}, from=1-7, to=1-5]
	\arrow["{t_3}"', shift right=5, shorten <=4pt, shorten >=4pt, Rightarrow, from=2, to=0]
	\arrow["{t_1}", shift left=5, shorten <=4pt, shorten >=4pt, Rightarrow, from=3, to=1]
	\arrow["{t_2}"{description}, curve={height=-12pt}, shorten <=13pt, shorten >=13pt, Rightarrow, from=2, to=3]
	\arrow["{t_0}"{description}, curve={height=-18pt}, shorten <=13pt, shorten >=13pt, Rightarrow, from=0, to=1]
\end{tikzcd}\]
\lessSpace{-1}
\end{defn}

\begin{remark}
A formal definition of strict 2-categories is actually much more complex than~\cref{def:bicat},
where we omit the preservation of identities and the detailed definition of compositions.
By that, we capture the essential features of strict 2-categories and get rid of the bureaucracy of dealing with associators.
\end{remark}

\begin{remark}[Enrichment]\label{rem:bicat-enrich}
A strict 2-category (\cref{def:bicat}) is a special case of an \textit{enriched category}.
We will define the notion of \textit{enrichment} in~\cref{sec:more-cat}, \cref{ex:enrichcat}.
\end{remark}

\begin{example}[Cat]\label{ex:bigcat}
The category of small categories\footnote{If you do not know what are \textit{small} categories,
think of them as categories.}, $\CAT$, is a strict 2-category (\cref{def:bicat}), whose objects are small categories,
1-cells are functors, and 2-cells are natural transformations.
\end{example}

\begin{defn}[Pseudofunctor]\label{def:pseudof}
For strict 2-categories (\cref{def:bicat}) $C, D$ and $x,y\in C$, a \textit{pseudofunctor} $\FF:C\to D$
maps $\text{Ob}(C)$ to $\text{Ob}(D)$ and $C(x,y)$ to $D(\FF(x), \FF(y))$ such that the functoriality holds
only up to isomorphisms, not strict identities. In particular:
\begin{itemize}
\item For every $x\in C$, the identity 1-cell (\cref{def:bicat}) $\id_x\in C(x, x)$ is mapped to
another 1-cell $\FF(\id_x)\in D(x, x)$ such that in the category $D(x, x)$,
there exists a 2-cell $o\in \text{Hom}_{D(x, x)}(\FF(\id_x),\id_{\FF(x)})$ that is an isomorphism.
\item For every $x, y, z\in C$ and $f\in C(x, y), g\in C(y, z)$, there is a natural isomorphism
$\FF(g) \circ \FF(f)\simeq \FF(g\circ f)$ between functors in $D(\FF(x), \FF(z))$.
\end{itemize}
The latter condition is visualized below:
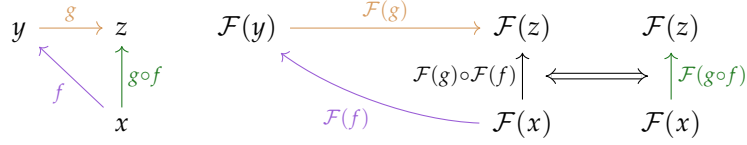
\[\begin{tikzcd}
	y & z & {\FF(y)} &&& {\FF(z)} & {\FF(z)} \\
	& x &&&& {\FF(x)} & {\FF(x)}
	\arrow["g", color={rgb,255:red,214;green,153;blue,92}, from=1-1, to=1-2]
	\arrow["f", color={rgb,255:red,153;green,92;blue,214}, from=2-2, to=1-1]
	\arrow["{g\circ f}"', color={rgb,255:red,31;green,122;blue,31}, from=2-2, to=1-2]
	\arrow["{\FF(g)}", color={rgb,255:red,214;green,153;blue,92}, from=1-3, to=1-6]
	\arrow["{\FF(f)}", color={rgb,255:red,153;green,92;blue,214}, curve={height=-12pt}, from=2-6, to=1-3]
	\arrow[""{name=0, anchor=center, inner sep=0}, "{\FF(g) \circ \FF(f)}", from=2-6, to=1-6]
	\arrow[""{name=1, anchor=center, inner sep=0}, "{\FF(g\circ f)}"', color={rgb,255:red,31;green,122;blue,31}, from=2-7, to=1-7]
	\arrow[shorten <=8pt, shorten >=8pt, Rightarrow, 2tail reversed, from=0, to=1]
\end{tikzcd}\]
We refer to these properties as \textit{pseudofunctoriality}.
\end{defn}

\begin{warning}\label{ex:catval-presh}
A $\CAT$-valued (\cref{ex:bigcat}) pseudofunctor may not be a presheaf (\cref{def:presh})
since presheaves must be functors (\cref{def:pseudof}).
Maybe we can call it a \textit{pseudopresheaf}, but this terminology is unpopular (yet).
\end{warning}

\subsection{Preparation for a category with attributes}
\begin{defn}[Subcategory]\label{def:subcat}
For a category $C$, a \textit{subcategory} $H$ of $C$ is a category such that $\text{Ob}(H) \subset \text{Ob}(C)$
and for $A,B\in H$, $H(A,B)\subset C(A,B)$. The identities must exist in the hom sets of $H$,
and the hom sets of $H$ must close under composition.
\end{defn}

\begin{defn}\label{def:fullsub}
For a category $C$, a full subcategory $H$ of $C$ is a subcategory (\cref{def:subcat}) of $C$ such that
the natural inclusion (\cref{not:set}) functor $\iota : H\to C$ is fully faithful.
\end{defn}

\begin{defn}\label{def:fullsub2}
This is an alternative to~\cref{def:fullsub}.
For a category $C$, a full subcategory $H$ of $C$ is a subcategory of $C$ such that
for every $A,B\in H$ and every $f\in C(A,B)$, $f\in H(A,B)$.
\end{defn}

\begin{defn}[Fiber]\label{def:fiber}
For categories $\EE,\CC$ and a functor $\FF:\EE\to\CC$, the \textit{fiber} of $\FF$ over an object
$\Gamma\in\CC$ (denoted as $\EE_\Gamma$) is a full subcategory (\cref{def:fullsub,def:fullsub2}) of $\EE$ whose objects
are mapped to $\Gamma$ by $\FF$ and morphisms are mapped to $\id_\Gamma$ by $\FF$.
\end{defn}

\begin{terminology}
In~\cite[definition 2.2]{CompCat}, Jacobs said that $\EE_\Gamma$ (\cref{def:fiber}) has objects \textit{above} $\CC$,
and it has \textit{vertical} morphisms. The intuition of these terminologies is visualized
in a rotated version of the diagram in~\cref{def:cart-mor}, where the ($E,D,D'$)-triangle
is ``above'' the $(\Gamma,p(E))$-line.
\end{terminology}

\begin{defn}[Reindex]\label{def:reindex}
For a fibration $p:\EE\to\CC$, morphisms $\sigma\in\CC(\Gamma,\Delta)$ and $f\in\EE_\Delta(D,E)$
(\cref{def:fiber}) (so that $p(D)=p(E)=\Delta \in\CC$ and $p(f)=\id_\Delta$),
we can take the cartesian lifting (\cref{def:fibr}) of $\sigma$ to both $D$ and $E$ --
by that we obtain objects $\sigma^*(D), \sigma^*(E)\in\EE_\Gamma$ and (cartesian) morphisms
$\overline\sigma(D)\in\EE(\sigma^*(D), D), \overline\sigma(E)\in\EE(\sigma^*(E), E)$.
This means that $p(\overline\sigma(D))=p(\overline\sigma(E))=\sigma$.
% https://q.uiver.app/?q=WzAsNixbMywwLCJcXEdhbW1hIixbMTIwLDYwLDMzLDFdXSxbMywxLCJcXERlbHRhIixbMjcwLDYwLDUwLDFdXSxbMCwxLCJEIixbMjcwLDYwLDUwLDFdXSxbMiwxLCJFIixbMjcwLDYwLDUwLDFdXSxbMCwwLCJcXHNpZ21hXiooRCkiLFsxMjAsNjAsMzMsMV1dLFsyLDAsIlxcc2lnbWFeKihFKSIsWzEyMCw2MCwzMywxXV0sWzAsMSwiXFxzaWdtYSIsMCx7ImNvbG91ciI6WzAsNjAsNjBdfSxbMCw2MCw2MCwxXV0sWzQsMiwiXFxvdmVybGluZSBcXHNpZ21hKEQpIiwyLHsiY29sb3VyIjpbMCw2MCw2MF19LFswLDYwLDYwLDFdXSxbNSwzLCJcXG92ZXJsaW5lIFxcc2lnbWEoRSkiLDAseyJjb2xvdXIiOlswLDYwLDYwXX0sWzAsNjAsNjAsMV1dLFsyLDMsImYiLDEseyJjb2xvdXIiOlsyNzAsNjAsNTBdfSxbMjcwLDYwLDUwLDFdXV0=
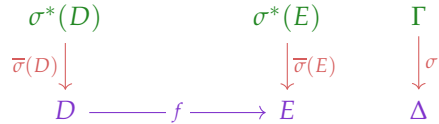
\[\begin{tikzcd}
	\textcolor{rgb,255:red,34;green,135;blue,34}{\sigma^*(D)} && \textcolor{rgb,255:red,34;green,135;blue,34}{\sigma^*(E)} & \textcolor{rgb,255:red,34;green,135;blue,34}{\Gamma} \\
	\textcolor{rgb,255:red,128;green,51;blue,204}{D} && \textcolor{rgb,255:red,128;green,51;blue,204}{E} & \textcolor{rgb,255:red,128;green,51;blue,204}{\Delta}
	\arrow["\sigma", color={rgb,255:red,214;green,92;blue,92}, from=1-4, to=2-4]
	\arrow["{\overline \sigma(D)}"', color={rgb,255:red,214;green,92;blue,92}, from=1-1, to=2-1]
	\arrow["{\overline \sigma(E)}", color={rgb,255:red,214;green,92;blue,92}, from=1-3, to=2-3]
	\arrow["f"{description}, color={rgb,255:red,128;green,51;blue,204}, from=2-1, to=2-3]
\end{tikzcd}\]
Now, consider the composition $(f\circ\overline\sigma(D))\in\EE(\sigma^*(D), E)$.
Note that $p(f\circ\overline\sigma(D))=p(f)\circ p(\overline\sigma(D))=\id_\Delta\circ\sigma=\sigma$.
This means we can apply the fact that $\overline\sigma(E)$ is cartesian to
uniquely characterize the morphism $\sigma^*(f)$ so that $p(\sigma^*(f))=\id_\Gamma$
and the following diagram commutes:
% https://q.uiver.app/?q=WzAsNixbMywwLCJcXEdhbW1hIixbMTIwLDYwLDMzLDFdXSxbMywxLCJcXERlbHRhIixbMjcwLDYwLDUwLDFdXSxbMCwxLCJEIixbMjcwLDYwLDUwLDFdXSxbMiwxLCJFIixbMjcwLDYwLDUwLDFdXSxbMCwwLCJcXHNpZ21hXiooRCkiLFsxMjAsNjAsMzMsMV1dLFsyLDAsIlxcc2lnbWFeKihFKSIsWzEyMCw2MCwzMywxXV0sWzAsMSwiXFxzaWdtYSIsMCx7ImNvbG91ciI6WzAsNjAsNjBdfSxbMCw2MCw2MCwxXV0sWzIsMywiZiIsMSx7ImNvbG91ciI6WzI3MCw2MCw1MF19LFsyNzAsNjAsNTAsMV1dLFs0LDIsIlxcb3ZlcmxpbmUgXFxzaWdtYShEKSIsMix7ImNvbG91ciI6WzAsNjAsNjBdfSxbMCw2MCw2MCwxXV0sWzUsMywiXFxvdmVybGluZSBcXHNpZ21hKEUpIiwwLHsiY29sb3VyIjpbMCw2MCw2MF19LFswLDYwLDYwLDFdXSxbNCw1LCJcXHNpZ21hXiooZikiLDEseyJjb2xvdXIiOlsxMjAsNjAsMzNdLCJzdHlsZSI6eyJib2R5Ijp7Im5hbWUiOiJkYXNoZWQifX19LFsxMjAsNjAsMzMsMV1dXQ==
\[\begin{tikzcd}
	\textcolor{rgb,255:red,34;green,135;blue,34}{\sigma^*(D)} && \textcolor{rgb,255:red,34;green,135;blue,34}{\sigma^*(E)} & \textcolor{rgb,255:red,34;green,135;blue,34}{\Gamma} \\
	\textcolor{rgb,255:red,128;green,51;blue,204}{D} && \textcolor{rgb,255:red,128;green,51;blue,204}{E} & \textcolor{rgb,255:red,128;green,51;blue,204}{\Delta}
	\arrow["\sigma", color={rgb,255:red,214;green,92;blue,92}, from=1-4, to=2-4]
	\arrow["f"{description}, color={rgb,255:red,128;green,51;blue,204}, from=2-1, to=2-3]
	\arrow["{\overline \sigma(D)}"', color={rgb,255:red,214;green,92;blue,92}, from=1-1, to=2-1]
	\arrow["{\overline \sigma(E)}", color={rgb,255:red,214;green,92;blue,92}, from=1-3, to=2-3]
	\arrow["{\sigma^*(f)}"{description}, color={rgb,255:red,34;green,135;blue,34}, dashed, from=1-1, to=1-3]
\end{tikzcd}\]
We can observe that the cartesian lifting of $\sigma\in\CC(\Gamma,\Delta)$ to objects in $\EE_\Delta$ are objects in $\EE_\Gamma$,
and the morphisms in $\EE_\Delta$ can also be lifted to $\EE_\Gamma$.
Therefore we obtain a functor $\sigma^*:\EE_\Delta \to \EE_\Gamma$
for every $\sigma\in\CC(\Gamma,\Delta)$, called the \textit{reindexing} functor.
\end{defn}

\begin{exercise}
State and verify the functoriality of the reindexing functor (\cref{def:reindex}).
\end{exercise}

\begin{remark}[Spoiler II]\label{rem:spoiler2}
In~\cref{def:reindex}, the morphism $\sigma\in\CC(\Gamma,\Delta)$ induces the functor $\sigma^*:\EE_\Delta\to\EE_\Gamma$.
The second last deduction in~\cref{rem:subst-ctx} and the diagram
in~\cref{rem:ty-subst} show that given a substitution $\Gvdash\sigma:\Delta$,
it can bring the type $A$ in $\Delta$ to $\Gamma$. Recall the previous spoiler~\cref{rem:spoiler1},
We can think of $\sigma^*$ as the ``action to apply the substitution $\sigma$ on a type''.
\end{remark}

\begin{example}[Weakening]\label{ex:weakening}
For a fibration (\cref{def:fibr}) $p:\EE\to\CC$, an object $\Gamma\in\CC$,
a display map (\cref{def:displaymap}) $\pi_A\in\CC_{/\Gamma}$ of the object $(\Gamma,A)\in\CC$,
the reindexing functor $\pi_A^*:\EE_{\Gamma}\to\EE_{\Gamma,A}$ (\cref{def:reindex}) of the display map $\pi_A$ corresponds to
the type theoretical weakening operator.
% https://q.uiver.app/?q=WzAsNixbMywwLCJcXEdhbW1hLEEiLFsxMjAsNjAsMzMsMV1dLFszLDEsIlxcR2FtbWEiLFsyNzAsNjAsNTAsMV1dLFswLDEsIkQiLFsyNzAsNjAsNTAsMV1dLFsyLDEsIkUiLFsyNzAsNjAsNTAsMV1dLFswLDAsIlxccGlfQV4qKEQpIixbMTIwLDYwLDMzLDFdXSxbMiwwLCJcXHBpX0FeKihFKSIsWzEyMCw2MCwzMywxXV0sWzAsMSwiXFxwaV9BIiwwLHsiY29sb3VyIjpbMCw2MCw2MF19LFswLDYwLDYwLDFdXSxbMiwzLCJmIiwxLHsiY29sb3VyIjpbMjcwLDYwLDUwXX0sWzI3MCw2MCw1MCwxXV0sWzQsMiwiXFxvdmVybGluZXtcXHBpX0F9KEQpIiwyLHsiY29sb3VyIjpbMCw2MCw2MF19LFswLDYwLDYwLDFdXSxbNSwzLCJcXG92ZXJsaW5le1xccGlfQX0oRSkiLDAseyJjb2xvdXIiOlswLDYwLDYwXX0sWzAsNjAsNjAsMV1dLFs0LDUsIlxccGlfQV4qKGYpIiwxLHsiY29sb3VyIjpbMTIwLDYwLDMzXSwic3R5bGUiOnsiYm9keSI6eyJuYW1lIjoiZGFzaGVkIn19fSxbMTIwLDYwLDMzLDFdXV0=
\[\begin{tikzcd}
	\textcolor{rgb,255:red,34;green,135;blue,34}{\pi_A^*(D)} && \textcolor{rgb,255:red,34;green,135;blue,34}{\pi_A^*(E)} & \textcolor{rgb,255:red,34;green,135;blue,34}{\Gamma,A} \\
	\textcolor{rgb,255:red,128;green,51;blue,204}{D} && \textcolor{rgb,255:red,128;green,51;blue,204}{E} & \textcolor{rgb,255:red,128;green,51;blue,204}{\Gamma}
	\arrow["{\pi_A}", color={rgb,255:red,214;green,92;blue,92}, from=1-4, to=2-4]
	\arrow["f"{description}, color={rgb,255:red,128;green,51;blue,204}, from=2-1, to=2-3]
	\arrow["{\overline{\pi_A}(D)}"', color={rgb,255:red,214;green,92;blue,92}, from=1-1, to=2-1]
	\arrow["{\overline{\pi_A}(E)}", color={rgb,255:red,214;green,92;blue,92}, from=1-3, to=2-3]
	\arrow["{\pi_A^*(f)}"{description}, color={rgb,255:red,34;green,135;blue,34}, dashed, from=1-1, to=1-3]
\end{tikzcd}\]
The above diagram demonstrates that for every type $E$ in context $\Gamma$, $\pi_A^*$ reindexes (\textit{weakens}) them to the extended context $(\Gamma,A)$.
\end{example}

\begin{defn}[Cleavage]\label{def:cleav}
For a fibration $p:\EE\to\CC$, a particular choice of $\sigma^*$ and $\overline \sigma$
(\cref{def:fibr}) for every $\sigma\in\CC(\Gamma,\Delta)$ is called a \textit{cleavage} of $p$.
In other words, a cleavage of $p$ is a mapping from $\sigma$ to the pair $(\sigma^*,\overline \sigma)$.

A cleavage of $p$ is equivalent to a subcategory (\cref{def:subcat}) of $\EE$.
\end{defn}

% \begin{prop}
% $p$ has a cleavage $\iff$ $p$ is a fibration.
% \end{prop}

\begin{remark}\label{rem:motivate-fiber}
In~\cref{def:fibr}, a fibration $p:\EE\to\CC$ is equipped with an operation that takes a morphism $\sigma$
and returns $p^{-1}(\sigma)$ a Grothendieck cartesian morphism. To some extent, $p$ is an ``invertible'' functor: for a morphism
$f'$ in $\EE$, we can send it to $\CC$ by $p$ and take it back to a cleavage (\cref{def:cleav}) $f$ in $\EE$.
The relationship between $f'$ and $f$ is similar to the morphisms of the same name in~\cref{def:cart-mor}.
\end{remark}

\begin{lem}[Faker]\label{lem:fake-funct}
For a fibration $p:\EE\to\CC$, we have a pseudofunctor (\cref{def:pseudof}) $\Psi:\OpCat\CC\to\CAT$ by
$\Psi(\Gamma) := \EE_\Gamma$ and for $\sigma\in\CC(\Gamma,\Delta)$, $\Psi(\sigma):=\sigma^*$.
\end{lem}

\begin{proof}
We verify the pseudofunctoriality of $\Psi$:
\begin{itemize}
\item For $\Gamma\in\CC$, $\Psi(\id_\Gamma)=\id_\Gamma^*\simeq\id_{\EE_\Gamma}$ a natural isomorphism.
\item For $\sigma\in\CC(\Gamma',\Gamma'')$ and $\gamma\in\CC(\Gamma,\Gamma')$
(so that $\sigma\circ \gamma\in\CC(\Gamma,\Gamma'')$), we know $\Psi(\sigma\circ \gamma)=(\sigma\circ \gamma)^*$
and $\sigma^* \circ \gamma^* =\Psi(\sigma)\circ\Psi(\gamma)$, and there is $(\sigma\circ \gamma)^* \simeq \sigma^*\circ\gamma^*$ a natural isomorphism.
\end{itemize}
\lessSpace{-1.5}
\end{proof}

\begin{defn}[Split]\label{def:split}
We say a fibration $p:\EE\to\CC$ is \textit{split} or \textit{cloven} if the pseudofunctor
$\Psi$ (\cref{lem:fake-funct}) defined on it satisfies functoriality instead of pseudofunctoriality.
This requires the following strict equalities:
\begin{itemize}
\item For $\Gamma\in\CC$, $\id_\Gamma^*=\id_{\EE_\Gamma}$.
\item For $\sigma\in\CC(\Gamma',\Gamma'')$ and $\gamma\in\CC(\Gamma,\Gamma')$, $(\sigma\circ\gamma)^* = \sigma^* \circ \gamma^*$.
\end{itemize}
\lessSpace{-1}
\end{defn}

\begin{example}
$\Psi$ in a split fibration (\cref{def:split}) is a $\CAT$-valued presheaf (\cref{def:presh}).
\end{example}

\begin{remark}[Coherence]\label{rem:coherence}
If our comprehension category is not split, we will have substitutions up to isomorphism
instead of strict equality as discussed by Curien~\cite{Coherence}.
\end{remark}

\begin{terminology}\label{term:fullsplit}
We say a comprehension category (\cref{def:compcat}) to be \textit{full} if the comprehension $\FF$
is fully faithful, and \textit{split} if the underlying fibration $p$ is split (\cref{def:split}).
\end{terminology}

\subsection{Cartmell's artifacts}
\begin{glorious-defn}[Cartmell's CwA]\label{def:cwa}
A \textit{category with attributes} (CwA) is a full split (\cref{term:fullsplit}) comprehension category.
Note that the codomain $\CC$ of the comprehension has a terminal object.

We refer to the domain $\EE$ of the comprehension as the \textit{attributes}.
\end{glorious-defn}

\begin{history}\label{his:diff-cwa-compcat}
CwA was already well-established when comprehension category was discussed.
The original definition of CwA is more similar to type categories (\cref{def:tycat}) where the
dependent types indexed by an object in $\CC$ forms a set instead of a fiber in another category $\EE$.
In this case, it is \textit{not} (but similar to) a full split comprehension category.
We discuss the difference in~\cref{rem:diff-cwa-compcat}.

Treating a CwA as a full split comprehension category is accepted in recent years, though.
\end{history}

\begin{defn}[Contextuality]\label{def:ctxly}
A CwA (\cref{def:cwa}) $\FF:\EE\to\CC^\to$ is \textit{contextual} if there exists a
\textit{length} function $\ell:\CC\to\mathbb N$ such that:
\begin{enumerate}
\item For every $\Gamma\in\CC$, $\ell(\Gamma)=0 \iff \Gamma=\textit1$ holds.
The terminal object $\textit1$ must be unique.
\item For every $\Gamma\in\CC$ and $A\in\EE_\Gamma$, $\ell(\Gamma,A)=\ell(\Gamma)+1$ holds.
\item For every $\Delta\in\CC$ where $\ell(\Delta)>0$, there exists uniquely
  $\Gamma\in\CC$ and $A\in\EE_\Gamma$ such that $\Delta=(\Gamma,A)$.
\end{enumerate}
In other words, the objects in $\CC$ are \textit{inductively generated} from the terminal object
and context extension (\cref{def:ctxext}).

A similar definition can be found in~\cite[definition 2]{CwF2}.
\end{defn}

\begin{convention}\label{conv:dontcare}
In a contextual CwA (\cref{def:ctxly}) $\FF:\EE\to\CC^\to$,
we use $\CC_{/\Gamma}$ (\cref{def:slice}) to denote the slice whose objects
\textit{only consists} of display maps (\cref{def:displaymap}),
not all morphisms of codomain $\Gamma$.
\end{convention}

\begin{glorious-defn}[Cartmell's CxlCat]\label{def:cxlcat}
A CwA $\FF:\EE\to\CC^\to$ satisfying contextuality (\cref{def:ctxly}) is called a \textit{contextual category}.
This is essentially the definition in~\cite[\S 14]{CxlCat}, where objects are described as a tree.

We will use the base category $\CC$ to refer to a contextual category and avoid talking about $\EE$,
because all the information we care about $\EE$ is already in $\CC$ by~\cref{conv:dontcare}.
\end{glorious-defn}

\begin{remark}
In~\cite[definition 7]{DTT-Models}, the length function (\cref{def:ctxly}) is described as a
\textit{grading} of objects in $\CC$, where the objects are indexed by a natural number.
It is denoted $\text{Ob}(\CC):=\coprod_{n:\mathbb N}\text{Ob}_n(\CC)$.
\end{remark}

\begin{lem}\label{lem:reindex-pullback}
In a contextual category $\CC$, the commutative square generated by
reindexing (\cref{def:reindex}) with a display map (\cref{def:displaymap})
is actually a pullback square:
% https://q.uiver.app/?q=WzAsNCxbMiwwLCJcXEdhbW1hIl0sWzIsMSwiXFxEZWx0YSJdLFswLDEsIlxcRGVsdGEsQSJdLFswLDAsIlxcR2FtbWEsQVxcc2lnbWEiXSxbMCwxLCJcXHNpZ21hIl0sWzIsMSwiXFxwaV9BIiwxXSxbMywyLCJcXHBpX0FeKihcXHNpZ21hKSIsMix7InN0eWxlIjp7ImJvZHkiOnsibmFtZSI6ImRhc2hlZCJ9fX1dLFszLDAsIiIsMix7InN0eWxlIjp7ImJvZHkiOnsibmFtZSI6ImRhc2hlZCJ9fX1dLFszLDEsIiIsMSx7InN0eWxlIjp7Im5hbWUiOiJjb3JuZXIifX1dXQ==&macro_url=https%3A%2F%2Fgist.githubusercontent.com%2Fice1000%2F47b7ea52f8c351607831f7f4afa9a79b%2Fraw%2Ff5bd21d4bd401a9f3672269a2816c728e53bd9c0%2Fquiver-macros.tex
\[\begin{tikzcd}
	{\Gamma,A\sigma} && \Gamma \\
	{\Delta,A} && \Delta
	\arrow["\sigma", from=1-3, to=2-3]
	\arrow["{\pi_A}"{description}, from=2-1, to=2-3]
	\arrow["{\pi_A^*(\sigma)}"', dashed, from=1-1, to=2-1]
	\arrow[dashed, from=1-1, to=1-3]
	\arrow["\lrcorner"{anchor=center, pos=0.125}, draw=none, from=1-1, to=2-3]
\end{tikzcd}\]
\end{lem}
\begin{proof}
Use exactly the same construction as in~\cref{def:displaymap},
with $\FF_0(A)$ replaced with $(\Gamma,Au)$, $\FF_0(B)$ replaced with $(\Delta,A)$.
\end{proof}

\begin{warning}
It is not the case that reindexing functors always generate pullback squares --
one of the morphisms must be a display map, as shown in~\cref{lem:reindex-pullback}.
\end{warning}

\begin{defn}[Democracy]\label{def:democ}
A CwA (\cref{def:cwa}) $\FF:\EE\to\CC^\to$ is \textit{democratic} if for every $\Gamma\in\CC$
there is an object $A\in\EE_{\textit1}$ such that $\Gamma\simeq(\textit1,A)$.
This definition is due to~\cite[definition 3]{CwF2}.
\end{defn}

\begin{remark}
Voevodsky proposed (\cite{C-Sys}) another name \textit{C-systems} for contextual categories (\cref{def:cxlcat})
because in the context of univalent foundations, there is the distinction between categories and precategories.
We will stick to the name contextual categories because we are not working in that context.
\end{remark}

\section{Categorical semantics}\label{sec:semantics}
\begin{defn}\label{def:tt}
A type theory consists of the following constructions, defined by \textit{typing judgments}:
\begin{enumerate}
\item Contexts, whose well-formedness is denoted $\Gvdash$. There is a well-formed \textit{empty context} \textit1.
\item Types formed in contexts, denoted \fbox{$\Gvdash\isType A$}, and type equality, denoted \fbox{$\Gvdash\isType{A=B}$}.
\item Context extension. Given \fbox{$\Gvdash\isType A$}, we can add $A$ to the end of $\Gamma$ to form a new context.
We denote this process using the following rule: \ttfragment{ctx-ext}
\item Substitution objects, whose well-formedness is denoted $\Gvdash \sigma:\Delta$.
There is a unique well-formed \textit{empty substitution} $\Gvdash []:\textit1$ for any context $\Gamma$.
\item Typing of terms. Given $A$ formed in $\Gamma$, ``term $u$ is of type $A$'' and
``terms $u$ and $v$ are equal instances of type $A$'' are described as: \ttfragment{term}
\item Substitution extension. This is similar to context extension: \ttfragment{subst-ext}
\end{enumerate}
These constitute an infrastructure for a type theory that can be extended in many ways.
\end{defn}

\begin{warning}[Universe]\label{warn:univ}
We do \textit{not} have any type that represents the \textit{universe} type
that usually appear in a type theory, such as the \textit{intuitionistic type theory}
due to Martin-L\"of~\cite{MLTT}, the \textit{calculus of constructions} due to
Coquand~\cite{CoC}, and the \textit{extended calculus of constructions} due to
Luo~\cite{LUTT,LUTT2}.

We \textit{cannot} have a type \fbox{$\vdash\isType\UU$} for all types
(including $\vdash\UU:\UU$) due to the paradox construction by Girard~\cite{Girard}
(and later simplified by Hurkens~\cite{Hurkens}).
\end{warning}

\begin{defn}\label{def:ctx-subst-sem}
We interpret contexts and substitutions in a CwA (\cref{def:cwa}) $\FF:\EE\to\CC^\to$.
\begin{enumerate}
\item A context $\Gamma$ is interpreted as an object in $\CC$, denoted $\lrbracket\Gamma\in\CC$.
In particular, $\lrbracket{\textit1}:=\textit1$, the terminal object in $\CC$
which is said to exist in~\cref{def:cat-ctx}.
\item A substitution $\Gvdash\sigma:\Delta$ is interpreted as a morphism in $\CC$, denoted
$\lrbracket\sigma:\CC(\lrbracket\Gamma,\lrbracket\Delta)$.
The empty substitution is the morphism from an object to the terminal object \textit1,
which always exists by the universal property of terminal objects.
\end{enumerate}
\lessSpace{-1.5}
\end{defn}

\begin{defn}\label{def:section}
For a morphism $a\in\CC(A, B)$, another morphism $b\in\CC(B, A)$ is called a
\textit{section} of $a$ if $ab=\id_B$.
\end{defn}

\begin{defn}\label{def:ty-term-sem}
We interpret types and terms with their equalities in a CwA (\cref{def:cwa}) $\FF:\EE\to\CC^\to$.
For a context $\Gamma$ in type theory and its interpretation $\lrbracket\Gamma\in\CC$,
\begin{enumerate}
\item The judgment \fbox{$\Gvdash\isType A$} is interpreted as an object
$\lrbracket A\in\EE_{\lrbracket\Gamma}$ (see~\cref{def:fiber}).
\item Two types $A,B$ formed in context $\Gamma$ are said to be equal (the judgment \fbox{$\Gvdash\isType{A=B}$})
if and only if their interpretations are uniquely isomorphic objects in $\EE_{\lrbracket\Gamma}$.
\item Context extension $\lrbracket\Gamma, \lrbracket A$ for $\lrbracket A\in\EE_{\lrbracket \Gamma}$ is defined
using the construction in~\cref{def:ctxext}, denoted $\lrbracket{\Gamma,A}\in\CC$.
\item The judgment $\Gvdash a:A$ is translated to $\Gvdash \id_\Gamma,a:(\Gamma,A)$
(appending $a$ to the identity substitution object of $\Gamma$),
whose interpretation is denoted $\lrbracket a\in\CC(\lrbracket\Gamma,\lrbracket{\Gamma,A})$.
Note that $\lrbracket a$ must be a section (\cref{def:section}) of $\pi_{\lrbracket A}$ to reflect the fact that
everything in $\Gamma$ is mapped identically.
\item Two terms $u,v$ of the same type $A$ are said to be equal (the judgment $\Gvdash u=v:A$) if
their interpretations are uniquely isomorphic objects in $\CC(\lrbracket\Gamma, \lrbracket{\Gamma,A})$.
\end{enumerate}
From now on, we will work with contextual categories (\cref{def:cxlcat}) and focus on
extended contexts instead of types in the fiber due to~\cref{conv:dontcare}.
\end{defn}

\begin{remark}
The interpretation of term/type equality in~\cref{def:ty-term-sem} is slightly different from the
one in~\cite[\S 6.4]{catl}, where they require \textit{sameness} instead of unique isomorphism.
% Our choice can be equalized to Pitts' by factoring $\CC$ with unique isomorphism.
\end{remark}

\begin{exercise}\label{exer:subst-ext}
Interpret ``substitution extension'' (the operation that takes a substitution
$\Gamma\vdash\sigma:\Delta$ along with a term $\Delta\vdash u:A$,
and returns an ``extended'' substitution $\Gamma\vdash(\sigma,u):(\Delta,A)$) in a CwA.
\end{exercise}

\begin{exercise}[Weakening]\label{exer:weakening}
Interpret in a CwA the context weakening rule: \ttfragment{weakening}
Hint: \cref{ex:weakening} already shows how to take the type $A$ to the weakened context.
\end{exercise}

\begin{remark}[Strictness]
The substitution operation is interpreted as morphisms in $\CC$. So, if two contexts
are equivalent up to substitution, their cartesian lifting (\cref{def:fibr}) will be the same
(instead of being naturally isomorphic categories) thanks to the strictness condition in~\cref{def:split}.
\end{remark}

\subsection{Star at the top: the unit type}
\begin{defn}[Unit]\label{def:top}
We extend~\cref{def:tt} with the unit type: \ttfragment{unit}
We do not define the $\top$ type with an elimination rule.
Instead, we directly give it a uniqueness-rule (the $\eta$-equality) for convenience.
\end{defn}

\begin{defn}\label{def:top-sem}
We interpret the unit type in~\cref{def:top} in a contextual category (\cref{def:cxlcat}) $\CC$.
Consider $\lrbracket\Gamma\in\CC$, an \textit{interpretation of a context $\Gamma$},
\begin{enumerate}
\item The \textit{context $\Gamma$ extended by the type $\top$}, \fbox{$\Gamma,\top$},
is interpreted as $\lrbracket\Gamma\in\CC$ (same as the context being extended!),
denoted $\lrbracket{\Gamma,\top}\in\CC$, along with a display map (\cref{def:displaymap})
$\pi_{\lrbracket\top}\in\CC$ defined as the identity morphism $\id_{\lrbracket{\Gamma,\top}}$, denoted $\lrbracket\star$.
\item The introduction rule simply says that ``there is a term $\star$ of type $\top$''.
We interpret $\star$ as the identity morphism $\id_{\lrbracket{\Gamma,\top}}$. Apparently,
this is a section (\cref{def:section}) of the display map (see~\cref{def:ty-term-sem}), which is also the identity morphism.
\item The uniqueness rule of $\top$ is derived from the uniqueness of identity morphisms.
\end{enumerate}
% https://q.uiver.app/?q=WzAsMixbMiwwLCJcXEdhbW1hIl0sWzAsMCwiXFxscmJyYWNrZXR7XFxHYW1tYSxcXHRvcH0iXSxbMSwwLCJcXHBpX3tcXGxyYnJhY2tldCBcXHRvcH0iLDEseyJjdXJ2ZSI6MiwibGV2ZWwiOjIsInN0eWxlIjp7ImhlYWQiOnsibmFtZSI6Im5vbmUifX19XSxbMCwxLCJcXGxyYnJhY2tldFxcc3RhciIsMSx7ImN1cnZlIjoyLCJzdHlsZSI6eyJib2R5Ijp7Im5hbWUiOiJkYXNoZWQifX19XV0=&macro_url=https%3A%2F%2Fgist.githubusercontent.com%2Fice1000%2F47b7ea52f8c351607831f7f4afa9a79b%2Fraw%2F61947e4f504352e3a0363d5fa5edafe16fcc1ce4%2Fquiver-macros.tex
\[\begin{tikzcd}
	{\lrbracket{\Gamma,\top}} && \Gamma
	\arrow["{\pi_{\lrbracket \top}}"{description}, curve={height=12pt}, Rightarrow, no head, from=1-1, to=1-3]
	\arrow["\lrbracket\star"{description}, curve={height=12pt}, dashed, from=1-3, to=1-1]
\end{tikzcd}\]
The above diagram shows the interpretations of $\top$ and $\star$.
\end{defn}

\begin{remark}\label{rem:motivate-sem}
In~\cref{def:top-sem}, we interpret the formation, introduction, and elimination rules for a type theoretical construction.
Then, we \textit{prove} the uniqueness rule. If there are other equalities that we can prove in the categorical model
but cannot be derived from the type theory, we may add these equalities to make the type theory as they already make sense in another theory.
This is a benefit of categorical models of type theory.

Unfortunately, the $\top$ type is too simple to have anything interesting. In~\cref{sub:coprod},
we are going to see such an equality in~\cref{thm:prod-eta}.
\end{remark}

\subsection{Like a semiring: the (co)product type}\label{sub:coprod}
\begin{defn}[Product]\label{def:prod}
We extend~\cref{def:tt} with the (non-dependent) product type,
using the same notation as product objects (\cref{def:prodobj}): \ttfragment{prod}
\lessSpace{-1}
\end{defn}

\begin{lem}\label{lem:pullback-prodmor}
Consider the following pullback diagram, where $a,b$ are sections (\cref{def:section}) of $\pi_A,\pi_B$:
% https://q.uiver.app/?q=WzAsNCxbMSwxLCJcXEdhbW1hIl0sWzAsMSwiQSJdLFsxLDAsIkIiXSxbMCwwLCJBXFx0aW1lc19cXEdhbW1hIEIiXSxbMSwwLCJcXHBpX0EiXSxbMiwwLCJcXHBpX0IiLDJdLFszLDIsIlxccGlfMiIsMix7InN0eWxlIjp7ImJvZHkiOnsibmFtZSI6ImRhc2hlZCJ9fX1dLFszLDEsIlxccGlfMSIsMCx7InN0eWxlIjp7ImJvZHkiOnsibmFtZSI6ImRhc2hlZCJ9fX1dLFswLDEsImEiLDAseyJjdXJ2ZSI6LTF9XSxbMCwyLCJiIiwyLHsiY3VydmUiOjF9XV0=
\[\begin{tikzcd}
	{A\times_\Gamma B} & B \\
	A & \Gamma
	\arrow["{\pi_A}", from=2-1, to=2-2]
	\arrow["{\pi_B}"', from=1-2, to=2-2]
	\arrow["{\pi_2}"', dashed, from=1-1, to=1-2]
	\arrow["{\pi_1}", dashed, from=1-1, to=2-1]
	\arrow["a", curve={height=-6pt}, from=2-2, to=2-1]
	\arrow["b"', curve={height=6pt}, from=2-2, to=1-2]
\end{tikzcd}\]
Then, the product morphism (\cref{def:pullback}) of $a,b$ is a section of $\pi_A\circ\pi_1$ or $\pi_B\circ\pi_1$
(they are equivalent by commutativity).
\end{lem}
\begin{proof}
We rearrange the diagram for better intuition (where the product morphism is denoted $a\times_\Gamma b$):
% https://q.uiver.app/?q=WzAsNSxbMywxLCJcXEdhbW1hIl0sWzIsMSwiQSJdLFszLDAsIkIiXSxbMiwwLCJBXFx0aW1lc19cXEdhbW1hIEIiXSxbMCwwLCJcXEdhbW1hIl0sWzEsMCwiXFxwaV9BIl0sWzIsMCwiXFxwaV9CIiwyXSxbMywyLCJcXHBpXzIiLDIseyJzdHlsZSI6eyJib2R5Ijp7Im5hbWUiOiJkYXNoZWQifX19XSxbMywxLCJcXHBpXzEiLDAseyJzdHlsZSI6eyJib2R5Ijp7Im5hbWUiOiJkYXNoZWQifX19XSxbNCwxLCJhIiwxXSxbNCwyLCJiIiwxLHsiY3VydmUiOi0yfV0sWzQsMywiYVxcdGltZXNfXFxHYW1tYSBiIiwxLHsic3R5bGUiOnsiYm9keSI6eyJuYW1lIjoiZGFzaGVkIn19fV1d
\[\begin{tikzcd}
	\Gamma && {A\times_\Gamma B} & B \\
	&& A & \Gamma
	\arrow["{\pi_A}", from=2-3, to=2-4]
	\arrow["{\pi_B}"', from=1-4, to=2-4]
	\arrow["{\pi_2}"', dashed, from=1-3, to=1-4]
	\arrow["{\pi_1}", dashed, from=1-3, to=2-3]
	\arrow["a"{description}, from=1-1, to=2-3]
	\arrow["b"{description}, curve={height=-12pt}, from=1-1, to=1-4]
	\arrow["{a\times_\Gamma b}"{description}, dashed, from=1-1, to=1-3]
\end{tikzcd}\]
By definition of sections, $\pi_A\circ a=\id_\Gamma$. Then, by commutativity of the diagram,
$\pi_A\circ\pi_1\circ(a\times_\Gamma b)=\id_\Gamma$.
\end{proof}

\begin{defn}\label{def:prod-sem}
The \textit{context $\Gamma$ extended by the product type} (\cref{def:prod}) $A\times B$,
\fbox{$\Gamma,A\times B$}, is interpreted as the pullback (\cref{def:pullback})
$\lrbracket{\Gamma,A}\times_{\lrbracket\Gamma}\lrbracket{\Gamma,B}\in\CC$,
denoted $\lrbracket{\Gamma,A\times B}\in\CC$ in a contextual category (\cref{def:cxlcat}) $\CC$.

The interpretation of the formation rule is visualized below:
% https://q.uiver.app/?q=WzAsNSxbMiwxLCJcXGxyYnJhY2tldFxcR2FtbWEiXSxbMCwxLCJcXGxyYnJhY2tldHtcXEdhbW1hLEF9Il0sWzAsMCwiXFxscmJyYWNrZXR7XFxHYW1tYSxBXFx0aW1lcyBCfSJdLFsyLDAsIlxcbHJicmFja2V0e1xcR2FtbWEsIEJ9Il0sWzEsMV0sWzIsMCwiXFxwaV97XFxscmJyYWNrZXR7QVxcdGltZXMgQn19IiwxXSxbMiwxLCIiLDIseyJzdHlsZSI6eyJib2R5Ijp7Im5hbWUiOiJkYXNoZWQifX19XSxbMiwzLCIiLDEseyJzdHlsZSI6eyJib2R5Ijp7Im5hbWUiOiJkYXNoZWQifX19XSxbMywwLCJcXHBpX3tcXGxyYnJhY2tldCBCfSJdLFsxLDAsIlxccGlfe1xcbHJicmFja2V0IEF9IiwxXSxbMiw0LCIiLDEseyJzdHlsZSI6eyJuYW1lIjoiY29ybmVyIn19XV0=&macro_url=https%3A%2F%2Fgist.githubusercontent.com%2Fice1000%2F47b7ea52f8c351607831f7f4afa9a79b%2Fraw%2F61947e4f504352e3a0363d5fa5edafe16fcc1ce4%2Fquiver-macros.tex
\[\begin{tikzcd}
	{\lrbracket{\Gamma,A\times B}} && {\lrbracket{\Gamma, B}} \\
	{\lrbracket{\Gamma,A}} & {} & \lrbracket\Gamma
	\arrow["{\pi_{\lrbracket{A\times B}}}"{description}, from=1-1, to=2-3]
	\arrow[dashed, from=1-1, to=2-1]
	\arrow[dashed, from=1-1, to=1-3]
	\arrow["{\pi_{\lrbracket B}}", from=1-3, to=2-3]
	\arrow["{\pi_{\lrbracket A}}"{description}, from=2-1, to=2-3]
	\arrow["\lrcorner"{anchor=center, pos=0.125}, draw=none, from=1-1, to=2-2]
\end{tikzcd}\]
In the introduction rule, the inputs are $\lrbracket u$, a section (\cref{def:section}) of $\pi_{\lrbracket A}$,
and $\lrbracket v$, a section of $\pi_{\lrbracket B}$:
% https://q.uiver.app/?q=WzAsNCxbMSwxLCJcXGxyYnJhY2tldFxcR2FtbWEiXSxbMSwwLCJcXGxyYnJhY2tldFxcR2FtbWEsXFxscmJyYWNrZXQgQSJdLFswLDEsIlxcbHJicmFja2V0XFxHYW1tYSxcXGxyYnJhY2tldCBCIl0sWzAsMCwiXFxscmJyYWNrZXRcXEdhbW1hLFxcbHJicmFja2V0IHtBXFx0aW1lcyBCfSJdLFszLDEsIiIsMSx7InN0eWxlIjp7ImJvZHkiOnsibmFtZSI6ImRhc2hlZCJ9fX1dLFszLDIsIiIsMSx7InN0eWxlIjp7ImJvZHkiOnsibmFtZSI6ImRhc2hlZCJ9fX1dLFswLDIsIlxcbHJicmFja2V0IHYiLDFdLFszLDAsIiIsMix7InN0eWxlIjp7Im5hbWUiOiJjb3JuZXIifX1dLFswLDEsIlxcbHJicmFja2V0IHUiLDJdXQ==&macro_url=https%3A%2F%2Fgist.githubusercontent.com%2Fice1000%2F47b7ea52f8c351607831f7f4afa9a79b%2Fraw%2Fc368c27ee1f601d69c99ce4722eef5d32a3b644f%2Fquiver-macros.tex
\[\begin{tikzcd}
	{\lrbracket{\Gamma,A\times B}} & {\lrbracket{\Gamma,A}} \\
	{\lrbracket{\Gamma,B}} & \lrbracket\Gamma
	\arrow[dashed, from=1-1, to=1-2]
	\arrow[dashed, from=1-1, to=2-1]
	\arrow["{\lrbracket v}"{description}, from=2-2, to=2-1]
	\arrow["\lrcorner"{anchor=center, pos=0.125}, draw=none, from=1-1, to=2-2]
	\arrow["{\lrbracket u}"', from=2-2, to=1-2]
\end{tikzcd}\]
The introduction rule takes the product morphism of $\lrbracket u$ and $\lrbracket v$,
and it's a section of the display map by~\cref{lem:pullback-prodmor}.
The elimination rules are directly derived from the property of pullbacks.
% https://q.uiver.app/?q=WzAsNSxbMywxLCJcXGxyYnJhY2tldFxcR2FtbWEiXSxbMywwLCJcXGxyYnJhY2tldHtcXEdhbW1hLEF9Il0sWzAsMSwiXFxscmJyYWNrZXR7XFxHYW1tYSxCfSJdLFswLDAsIlxcbHJicmFja2V0e1xcR2FtbWEsQVxcdGltZXMgQn0iXSxbMSwxXSxbMywxLCJcXHBpXzIiLDEseyJzdHlsZSI6eyJib2R5Ijp7Im5hbWUiOiJkYXNoZWQifX19XSxbMywyLCJcXHBpXzEiLDIseyJzdHlsZSI6eyJib2R5Ijp7Im5hbWUiOiJkYXNoZWQifX19XSxbMCwyLCJcXGxyYnJhY2tldCB2IiwxXSxbMCwxLCJcXGxyYnJhY2tldCB1IiwyXSxbMCwzLCJcXGxyYnJhY2tldHtcXGxyYW5nbGV7dSx2fX0iLDFdLFszLDQsIiIsMCx7InN0eWxlIjp7Im5hbWUiOiJjb3JuZXIifX1dXQ==&macro_url=https%3A%2F%2Fgist.githubusercontent.com%2Fice1000%2F47b7ea52f8c351607831f7f4afa9a79b%2Fraw%2Fc368c27ee1f601d69c99ce4722eef5d32a3b644f%2Fquiver-macros.tex
\[\begin{tikzcd}
	{\lrbracket{\Gamma,A\times B}} &&& {\lrbracket{\Gamma,A}} \\
	{\lrbracket{\Gamma,B}} & {} && \lrbracket\Gamma
	\arrow["{\pi_2}"{description}, dashed, from=1-1, to=1-4]
	\arrow["{\pi_1}"', dashed, from=1-1, to=2-1]
	\arrow["{\lrbracket v}"{description}, from=2-4, to=2-1]
	\arrow["{\lrbracket u}"', from=2-4, to=1-4]
	\arrow["{\lrbracket{\lrangle{u,v}}}"{description}, from=2-4, to=1-1]
	\arrow["\lrcorner"{anchor=center, pos=0.125}, draw=none, from=1-1, to=2-2]
\end{tikzcd}\]
The computation rules hold by compositions.
\end{defn}

\begin{thm}[Uniqueness]\label{thm:prod-eta}
The interpretation in~\cref{def:prod-sem} satisfies the $\eta$-rule for the product type: \ttfragment{prod-eta}
\end{thm}
\begin{proof}
By doing some replacements of terms in the last diagram in~\cref{def:prod-sem}:
% https://q.uiver.app/?q=WzAsNSxbMywxLCJcXGxyYnJhY2tldFxcR2FtbWEiXSxbMywwLCJcXGxyYnJhY2tldHtcXEdhbW1hLCBBfSJdLFswLDEsIlxcbHJicmFja2V0e1xcR2FtbWEsQn0iXSxbMCwwLCJcXGxyYnJhY2tldHtcXEdhbW1hLEFcXHRpbWVzIEJ9Il0sWzEsMV0sWzMsMSwiXFxwaV8yIiwxLHsic3R5bGUiOnsiYm9keSI6eyJuYW1lIjoiZGFzaGVkIn19fV0sWzMsMiwiXFxwaV8xIiwyLHsic3R5bGUiOnsiYm9keSI6eyJuYW1lIjoiZGFzaGVkIn19fV0sWzAsMiwiXFxscmJyYWNrZXR7dC4yfSIsMV0sWzAsMSwiXFxscmJyYWNrZXR7dC4xfSIsMl0sWzAsMywiXFxscmJyYWNrZXR7dH0iLDFdLFszLDQsIiIsMSx7InN0eWxlIjp7Im5hbWUiOiJjb3JuZXIifX1dXQ==&macro_url=https%3A%2F%2Fgist.githubusercontent.com%2Fice1000%2F47b7ea52f8c351607831f7f4afa9a79b%2Fraw%2Fc368c27ee1f601d69c99ce4722eef5d32a3b644f%2Fquiver-macros.tex
\[\begin{tikzcd}
	{\lrbracket{\Gamma,A\times B}} &&& {\lrbracket{\Gamma, A}} \\
	{\lrbracket{\Gamma,B}} & {} && \lrbracket\Gamma
	\arrow["{\pi_2}"{description}, dashed, from=1-1, to=1-4]
	\arrow["{\pi_1}"', dashed, from=1-1, to=2-1]
	\arrow["{\lrbracket{t.2}}"{description}, from=2-4, to=2-1]
	\arrow["{\lrbracket{t.1}}"', from=2-4, to=1-4]
	\arrow["{\lrbracket{t}}"{description}, from=2-4, to=1-1]
	\arrow["\lrcorner"{anchor=center, pos=0.125}, draw=none, from=1-1, to=2-2]
\end{tikzcd}\]
\lessSpace{-1}
\end{proof}

\begin{defn}[Coproduct]\label{def:coprod}
We extend~\cref{def:tt} with the coproduct type: \ttfragment{coprod}
Note that in the two rules, \fbox{$ut$} and \fbox{$vt$} denote the operation of
``applying the singleton substitution object $t$'' to the term $u$ and $v$, respectively.
\end{defn}

\begin{remark}
Type theoretically, the coproduct type is dual to the product type.
Since product types are interpreted as pullbacks (see~\cref{def:prod-sem})
which are ``products in the overcategory'', the coproduct types should look like
``coproducts in the overcategory''.
\end{remark}

\begin{exercise}
Interpret the coproduct type (\cref{def:coprod}) in a contextual category (\cref{def:cxlcat}).
Beware of~\cref{warn:coprodpushout}.
\end{exercise}

% \begin{defn}\label{def:coprod-sem}
% The \textit{context $\Gamma$ extended by the coproduct type} (\cref{def:coprod}) $A\sqcup B$,
% $(\Gamma,A\sqcup B)$, is interpreted as the coproduct 
% $(\lrbracket{\Gamma,A})\sqcup_\Gamma(\lrbracket{\Gamma,B})\in\CC$,
% denoted $(\lrbracket{\Gamma,A\sqcup B})\in\CC$ in a contextual category (\cref{def:cxlcat}).

% The interpretation is exactly the same as product type as in~\cref{def:prod-sem} with all the arrows flipped
% and the names of the morphisms ``elim'', ``intro'' exchanged, so we omit the diagrams and the interpretation.
% \end{defn}

\subsection{Internal homs and the evaluation map}
\begin{defn}\label{def:smonoidalcat}
A category $\CC$ is \textit{strictly monoidal} if there is an operation (called the \textit{tensor product})
$\otimes$ such that for $A,B\in\CC$, $A\otimes B\in\CC$.
Apart from that, the following additional properties must hold:
\begin{enumerate}
\item For $A,B,C,D\in\CC$, $f\in\CC(A,C)$, and $g\in\CC(B,D)$,
there is a morphism $f\otimes g\in\CC(A\otimes B,C\otimes D)$.
\item There is an object $1\in\CC$, called the \textit{tensor unit},
satisfying that for every $A\in\CC$, $A\otimes 1 = 1\otimes A=A$.
\item The tensor product of objects and morphisms are strictly associative.
\end{enumerate}
\lessSpace{-0.2}
\end{defn}

\begin{defn}[Bifunctor]\label{def:bifunctor}
For (small) categories $\CC_1,\CC_2,\DD\in\CAT$, we take the product object (\cref{def:prodobj}) $\CC_1\times\CC_2$
and refer to a functor $\FF$ as a \textit{bifunctor} if it of form $\FF:\CC_1\times\CC_2\to\DD$.
We say that $\FF$ is a bifunctor \textit{from $\CC_1,\CC_2$ to $\DD$}.
\end{defn}

\begin{defn}\label{def:monoidalcat}
The~\cref{def:smonoidalcat} can be loosen into the general notion of \textit{monoidal categories}
in order to allow tensor products that are not strictly commutative (but up to isomorphism).
We define $\otimes$ as a bifunctor (\cref{def:bifunctor}) $\CC\otimes\CC\to\CC$,
also called the \textit{tensor product}, with the following additional structures:
\begin{enumerate}
\item The \textit{tensor unit} object $\textit1\in\CC$.
\item For $x,y,z\in\CC$, an isomorphism $\alpha_{x,y,z}\in\CC((x\otimes y)\otimes z,x\otimes(y\otimes z))$.
\item For $x\in\CC$, two isomorphisms $\rho_x\in\CC(x\otimes\textit1,x)$ and
$\lambda_x\in\CC(\textit1\otimes x,x)$.
\end{enumerate}
These structures commute the following diagrams:
% https://q.uiver.app/?q=WzAsOCxbMSwyLCIodyBcXG90aW1lcyB4KSBcXG90aW1lcyAoeSBcXG90aW1lcyB6KSJdLFswLDEsIigodyBcXG90aW1lcyB4ICkgXFxvdGltZXMgeSkgXFxvdGltZXMgeiJdLFsyLDEsIncgXFxvdGltZXMgKHggXFxvdGltZXMgKHkgXFxvdGltZXMgeikpIl0sWzAsMywiKHcgXFxvdGltZXMgKHggXFxvdGltZXMgeSkpIFxcb3RpbWVzIHoiXSxbMiwzLCJ3IFxcb3RpbWVzICggKHggXFxvdGltZXMgeSkgXFxvdGltZXMgeiApIl0sWzAsMCwiKHhcXG90aW1lcyBcXHRleHRpdDEpXFxvdGltZXMgeSJdLFsyLDAsInhcXG90aW1lcyAoXFx0ZXh0aXQxXFxvdGltZXMgeSkiXSxbMSwxLCJ4XFxvdGltZXMgeSJdLFsxLDAsIlxcYWxwaGFfe3cgXFxvdGltZXMgeCwgeSwgen0iLDFdLFswLDIsIlxcYWxwaGFfe3cseCx5IFxcb3RpbWVzIHp9IiwxXSxbMSwzLCJcXGFscGhhX3t3LHgseX0gXFxvdGltZXMgXFxpZF96IiwxXSxbMyw0LCJcXGFscGhhX3t3LHggXFxvdGltZXMgeSwgen0iLDFdLFs1LDYsIlxcYWxwaGFfe3gsXFx0ZXh0aXQxLHl9IiwxXSxbNSw3LCJcXHJob194XFxvdGltZXMgXFxpZF95IiwxXSxbNiw3LCJcXGlkX3hcXG90aW1lcyBcXGxhbWJkYV95IiwxXSxbNCwyLCJcXGlkX3cgXFxvdGltZXMgXFxhbHBoYV97eCx5LHp9IiwxXV0=
\[\begin{tikzcd}
	{(x\otimes \textit1)\otimes y} && {x\otimes (\textit1\otimes y)} \\
	{((w \otimes x ) \otimes y) \otimes z} & {x\otimes y} & {w \otimes (x \otimes (y \otimes z))} \\
	& {(w \otimes x) \otimes (y \otimes z)} \\
	{(w \otimes (x \otimes y)) \otimes z} && {w \otimes ( (x \otimes y) \otimes z )}
	\arrow["{\alpha_{w \otimes x, y, z}}"{description}, from=2-1, to=3-2]
	\arrow["{\alpha_{w,x,y \otimes z}}"{description}, from=3-2, to=2-3]
	\arrow["{\alpha_{w,x,y} \otimes \id_z}"{description}, from=2-1, to=4-1]
	\arrow["{\alpha_{w,x \otimes y, z}}"{description}, from=4-1, to=4-3]
	\arrow["{\alpha_{x,\textit1,y}}"{description}, from=1-1, to=1-3]
	\arrow["{\rho_x\otimes \id_y}"{description}, from=1-1, to=2-2]
	\arrow["{\id_x\otimes \lambda_y}"{description}, from=1-3, to=2-2]
	\arrow["{\id_w \otimes \alpha_{x,y,z}}"{description}, from=4-3, to=2-3]
\end{tikzcd}\]
\lessSpace{-1}
\end{defn}

\begin{defn}\label{def:cartmcat}
We say a monoidal category (\cref{def:monoidalcat}) to be \textit{cartesian} if its tensor products
are the product objects (\cref{def:prodobj}) and its tensor unit is the terminal object (\cref{def:termobj}).
\end{defn}

\begin{defn}\label{def:symmcat}
We say a monoidal category (\cref{def:monoidalcat}) to be \textit{symmetric} if its tensor product
is commutative up to isomorphism. This definition is slightly weaker than the conventional one,
but it is simpler.
\end{defn}

\begin{example}
In (many variations of) cubical type theory,
the category of cubes $\square_\otimes$ is symmetric monoidal (\cref{def:symmcat}).
A definition can be found in the doctoral thesis of Carlo Angiuli~\cite[\S 3.1]{CSCCTT}
(with geometric demonstrations).
\end{example}

\begin{defn}[EvalMap]\label{def:evalmap}
Consider a symmetric monoidal category $\CC$ (\cref{def:symmcat}) to be \textit{closed} if for objects $b,c\in\CC$,
there is an object $[b,c]\in\CC$ (called an \textit{internal hom}) together with a morphism
$\fs{eval}\in\CC([b,c]\otimes b, c)$ (called the \textit{evaluation map}) such that for any morphism
$f\in\CC(a\otimes b, c)$, there is a unique morphism $\lambda.f\in\CC(a, [b,c])$ that commutes the following diagram:
% https://q.uiver.app/?q=WzAsNSxbMiwwLCJhXFxvdGltZXMgYiJdLFsyLDEsIltiLGNdXFxvdGltZXMgYiJdLFs0LDEsImMiXSxbMCwwLCJhIl0sWzAsMSwiW2IsY10iXSxbMCwxLCJcXGxhbWJkYS4gZlxcb3RpbWVzIFxcaWRfYiIsMix7InN0eWxlIjp7ImJvZHkiOnsibmFtZSI6ImRhc2hlZCJ9fX1dLFsxLDIsIlxcdGV4dHtldmFsfSIsMV0sWzAsMiwiZiIsMV0sWzMsNCwiXFxsYW1iZGEuZiIsMix7InN0eWxlIjp7ImJvZHkiOnsibmFtZSI6ImRhc2hlZCJ9fX1dXQ==
\[\begin{tikzcd}
	a && {a\otimes b} \\
	{[b,c]} && {[b,c]\otimes b} && c
	\arrow["{\lambda. f\otimes \id_b}"', dashed, from=1-3, to=2-3]
	\arrow["{\text{eval}}"{description}, from=2-3, to=2-5]
	\arrow["f"{description}, from=1-3, to=2-5]
	\arrow["{\lambda.f}"', dashed, from=1-1, to=2-1]
\end{tikzcd}\]
The assignment of a unique $\lambda.f\in\CC(a, [b, c])$ for every $f\in\CC(a\otimes b, c)$
can be regarded as a hom set isomorphism $\lambda.\in\CC(a, [b, c])\simeq\CC(a\otimes b, c)$.
\end{defn}

\begin{defn}\label{def:closedsmcat}
A symmetric monoidal closed category (\cref{def:closedsmcat})
is said to be \textit{closed} if it has all internal homs and evaluation maps.
\end{defn}

\begin{lem}\label{lem:eval-functor}
Consider $\CC$ a symmetric monoidal closed category (\cref{def:closedsmcat}) and $a,b,e\in\CC$.
Every $t\in\CC(a, b)$ uniquely determines a functor on hom sets $\CC(e, [b, c])\to\CC(e, [a, c])$.
\end{lem}
\begin{proof}
Consider a morphism $f\in\CC(e, [b, c])$, it uniquely determines $\lambda. f\in\CC(e\otimes b, c)$,
which can be composed with $\id_e\otimes t$ to obtain a new, uniquely determined morphism, denoted $g\in\CC(e\otimes a, c)$.
Then we take $\lambda.^{-1}g\in\CC(e, [a, c])$.
% https://q.uiver.app/?q=WzAsOSxbMiwwLCJbYixjXSJdLFsyLDEsIlthLGNdIl0sWzAsMCwiYSJdLFswLDEsImIiXSxbMSwwLCJlIl0sWzMsMCwiZVxcb3RpbWVzIGIiXSxbNCwwLCJjIl0sWzMsMSwiZVxcb3RpbWVzIGEiXSxbMSwxLCJlIl0sWzIsMywidCIsMl0sWzQsMCwiZiJdLFs1LDYsIlxcbGFtYmRhLiBmIiwwLHsic3R5bGUiOnsiYm9keSI6eyJuYW1lIjoiZGFzaGVkIn19fV0sWzcsNSwiXFxpZF9lXFxvdGltZXMgdCJdLFs3LDYsImciLDJdLFs4LDEsIlxcbGFtYmRhLl57LTF9ZyIsMl1d
\[\begin{tikzcd}
	a & e & {[b,c]} & {e\otimes b} & c \\
	b & e & {[a,c]} & {e\otimes a}
	\arrow["t"', from=1-1, to=2-1]
	\arrow["f", from=1-2, to=1-3]
	\arrow["{\lambda. f}", dashed, from=1-4, to=1-5]
	\arrow["{\id_e\otimes t}", from=2-4, to=1-4]
	\arrow["g"', from=2-4, to=1-5]
	\arrow["{\lambda.^{-1}g}"', from=2-2, to=2-3]
\end{tikzcd}\]
We may also write the sequence of constructions as $\lambda.^{-1}(\lambda. f\circ(\id_e\otimes t))$.
\end{proof}

\begin{lem}\label{lem:tensor-hom-adjoint}
The following three pairs of hom presheaves (\cref{def:homfunc}) are naturally isomorphic
in a symmetric monoidal closed category (\cref{def:closedsmcat}).
\begin{align*}
&\CC(-\otimes b,c) &&\simeq &\CC(-,[b,c]) \\
&\CC(a\otimes -,c) &&\simeq &\CC(a,[-,c])
\end{align*}
\end{lem}
\begin{proof}
For every $u, v\in\CC$ and $f\in\CC(u, v)$, we validate the natural isomorphisms
by the following commutative diagrams:
% https://q.uiver.app/?q=WzAsOCxbMCwwLCJcXENDKHVcXG90aW1lcyBiLCBjKSJdLFswLDEsIlxcQ0Modlxcb3RpbWVzIGIsIGMpIl0sWzEsMCwiXFxDQyh1LCBbYixjXSkiXSxbMSwxLCJcXENDKHYsIFtiLGNdKSJdLFsyLDAsIlxcQ0MoYVxcb3RpbWVzIHUsIGMpIl0sWzIsMSwiXFxDQyhhXFxvdGltZXMgdiwgYykiXSxbMywwLCJcXENDKGEsIFt1LGNdKSJdLFszLDEsIlxcQ0MoYSwgW3YsY10pIl0sWzAsMSwiLVxcY2lyYyAoZlxcb3RpbWVzXFxpZF9iKSIsMl0sWzIsMywiLVxcY2lyYyBmIiwyXSxbMCwyLCJcXGxhbWJkYS4iLDAseyJzdHlsZSI6eyJ0YWlsIjp7Im5hbWUiOiJhcnJvd2hlYWQifX19XSxbMSwzLCJcXGxhbWJkYS4iLDAseyJzdHlsZSI6eyJ0YWlsIjp7Im5hbWUiOiJhcnJvd2hlYWQifX19XSxbNCw1LCItXFxjaXJjIChcXGlkX2FcXG90aW1lcyBmKSIsMl0sWzUsNywiXFxsYW1iZGEuIiwwLHsic3R5bGUiOnsidGFpbCI6eyJuYW1lIjoiYXJyb3doZWFkIn19fV0sWzQsNiwiXFxsYW1iZGEuIiwwLHsic3R5bGUiOnsidGFpbCI6eyJuYW1lIjoiYXJyb3doZWFkIn19fV0sWzYsNywiXFxjcmVme2xlbTpldmFsLWZ1bmN0b3J9IiwwLHsib2Zmc2V0Ijo1fV1d
\[\begin{tikzcd}
	{\CC(u\otimes b, c)} & {\CC(u, [b,c])} & {\CC(a\otimes u, c)} & {\CC(a, [u,c])} \\
	{\CC(v\otimes b, c)} & {\CC(v, [b,c])} & {\CC(a\otimes v, c)} & {\CC(a, [v,c])}
	\arrow["{-\circ (f\otimes\id_b)}"', from=1-1, to=2-1]
	\arrow["{-\circ f}"', from=1-2, to=2-2]
	\arrow["{\lambda.}", tail reversed, from=1-1, to=1-2]
	\arrow["{\lambda.}", tail reversed, from=2-1, to=2-2]
	\arrow["{-\circ (\id_a\otimes f)}"', from=1-3, to=2-3]
	\arrow["{\lambda.}", tail reversed, from=2-3, to=2-4]
	\arrow["{\lambda.}", tail reversed, from=1-3, to=1-4]
	\arrow["{\cref{lem:eval-functor}}", shift right=5, from=1-4, to=2-4]
\end{tikzcd}\]
\lessSpace{-1}
\end{proof}

\begin{lem}[Internalization]\label{lem:ccc-determine-lem}
In a symmetric monoidal closed category (\cref{def:closedsmcat}) $\CC$, its tensor unit (\cref{def:monoidalcat})
$\textit1\in\CC$, and objects $b,c\in\CC$, there is an isomorphism $\CC(b,c)\simeq\CC(\textit1,[b,c])$.
\end{lem}
\begin{proof}
By specializing the object $a$ in the equivalence in~\cref{lem:tensor-hom-adjoint} to \textit1:
$$\CC(\textit1,[b,c])\simeq\CC(\textit1\otimes b,c)\simeq\CC(b,c)$$
This reveals the intuition behind the name \textit{internal homs}.
\end{proof}

\begin{lem}\label{lem:ccc-determine-lem2}
The tersor unit used in~\cref{lem:ccc-determine-lem} can be replaced with any object $a\in\CC$ such that $a\otimes b\simeq b$. 
\end{lem}

\subsection{Fiber exponents}
\begin{glorious-defn}[CCC]\label{def:ccc}
We say a symmetric closed monoidal category (\cref{def:closedsmcat}) to be \textit{cartesian} if it
is also cartesian monoidal (\cref{def:cartmcat}).
In this case, the category is called a cartesian closed category, or CCC.
The internal homs of a CCC are also known as \textit{exponential objects} and we denote $[b,c]$ as $c^b$.
\end{glorious-defn}

\begin{example}\label{ex:funcset}
Since functions are sets too and they satisfy the natural isomorphisms in~\cref{def:closedsmcat},
$\SET$ is a CCC (\cref{def:ccc}). The evaluation map (\cref{def:evalmap}) in $\SET$
is just function application: for $A,B\in\SET$, the internal hom (\cref{def:closedsmcat}) $A\to B$
has the evaluation map $\fs{eval}\in\SET((A\to B)\times A,B)$.
\end{example}

\begin{defn}[LCCC]\label{def:lccc}
We say a category $\CC$ to be a \textit{locally cartesian closed} category or an LCCC if the overcategory $\CC_{/\Gamma}$
for every $\Gamma\in\CC$ is cartesian closed.
The relation between LCCC and type theories has been explored by Seely and Hofmann~\cite{LCCC,LCCC2}.
\end{defn}

\begin{lem}
An LCCC (\cref{def:lccc}) is a CCC (\cref{def:ccc}) if it has a terminal object.
\end{lem}
\begin{proof}
Consider an LCCC $\CC$ with a terminal object \textit1, we know that $\CC_{/\textit1}$ is a CCC.
By~\cref{lem:overcat-equiv}, $\CC$ is a CCC.
\end{proof}

\begin{lem}
In an LCCC (\cref{def:lccc}) $\CC$ and $\Gamma,A,B\in\CC$, the cartesian product in $\CC_\Gamma$ is given by
the pullback $A\times_\Gamma B\in\CC$. 
\end{lem}

\begin{defn}\label{def:fibexp}
In an LCCC (\cref{def:lccc}) $\CC$, for $\Gamma,A,B\in\CC$, $a\in\CC(A,\Gamma)$,
and $b\in\CC(B,\Gamma)$ (so that $a,b\in\CC_{/\Gamma}$),
the \textit{fiber exponent} $A\to_\Gamma B$ is the domain of the exponential object $b^a\in\CC_{/\Gamma}$.
The name is inspired from the alias ``fiber product'' of pullbacks (\cref{def:pullback})
and ``fiber coproduct'' of pushouts (\cref{def:pushout}).
% https://q.uiver.app/?q=WzAsNCxbMiwxLCJcXEdhbW1hIl0sWzIsMCwiQSJdLFswLDEsIkFcXHRvX1xcR2FtbWEgQiJdLFswLDAsIkIiXSxbMSwwLCJhIl0sWzMsMCwiYiIsMV0sWzIsMCwiYl5hIiwxXSxbMywxLCIiLDAseyJzdHlsZSI6eyJib2R5Ijp7Im5hbWUiOiJkYXNoZWQifX19XV0=&macro_url=https%3A%2F%2Fgist.githubusercontent.com%2Fice1000%2F47b7ea52f8c351607831f7f4afa9a79b%2Fraw%2Fc368c27ee1f601d69c99ce4722eef5d32a3b644f%2Fquiver-macros.tex
\[\begin{tikzcd}
	B && A \\
	{A\to_\Gamma B} && \Gamma
	\arrow["a", from=1-3, to=2-3]
	\arrow["b"{description}, from=1-1, to=2-3]
	\arrow["{b^a}"{description}, from=2-1, to=2-3]
	\arrow[dashed, from=1-1, to=1-3]
\end{tikzcd}\]
Note that by~\cref{lem:ccc-determine-lem}, there is an isomorphism
$\CC_{/\Gamma}(a,b)\simeq\CC_{/\Gamma}(\textit1_{\CC_{/\Gamma}}, b^a)$,
where $\textit1_{\CC_{/\Gamma}}=\id_\Gamma$.
\end{defn}

\subsection{Simple type theory: the function type}
\begin{defn}[Function]\label{def:func-tt}
We extend~\cref{def:tt} with the function type,
defined by the following rules: \ttfragment{function}
\end{defn}

\begin{defn}\label{def:func-sem}
In a contextual category (\cref{def:cxlcat}) $\CC$ which is also an LCCC, we can interpret \cref{def:func-tt}.

Consider types \fbox{$\Gvdash\isType A$,~$\Gvdash\isType B$,~$\Gvdash\isType C$}.
The formation of \textit{context $\Gamma$ extended by the function type $A \to B$}
corresponds to the fiber exponent (\cref{def:fibexp}) $\lrbracket{\Gamma,A} \to_{\lrbracket\Gamma} \lrbracket{\Gamma, B}$,
denoted $\lrbracket{\Gamma,A\to B}\in\CC$, whose display map (\cref{def:displaymap}) is given by the
exponential object $\pi_{\lrbracket{A\to B}}=\pi_{\lrbracket B}^{\pi_{\lrbracket A}}\in\CC_{/\lrbracket\Gamma}$.
% https://q.uiver.app/?q=WzAsNCxbMiwxLCJcXGxyYnJhY2tldFxcR2FtbWEiXSxbMiwwLCJcXGxyYnJhY2tldFxcR2FtbWEsXFxscmJyYWNrZXQge0F9Il0sWzAsMSwiXFxscmJyYWNrZXRcXEdhbW1hLFxcbHJicmFja2V0e0FcXHRvIEJ9Il0sWzAsMCwiXFxscmJyYWNrZXRcXEdhbW1hLFxcbHJicmFja2V0IHtCfSJdLFsxLDAsIlxccGlfe1xcbHJicmFja2V0IEF9Il0sWzMsMCwiXFxwaV97XFxscmJyYWNrZXQgQn0iLDFdLFsyLDAsIlxccGlfe1xcbHJicmFja2V0e0FcXHRvIEJ9fSIsMV1d&macro_url=https%3A%2F%2Fgist.githubusercontent.com%2Fice1000%2F47b7ea52f8c351607831f7f4afa9a79b%2Fraw%2Fc368c27ee1f601d69c99ce4722eef5d32a3b644f%2Fquiver-macros.tex
\[\begin{tikzcd}
	{\lrbracket{\Gamma,B}} && {\lrbracket{\Gamma,A}} \\
	{\lrbracket{\Gamma,A\to B}} && \lrbracket\Gamma
	\arrow["{\pi_{\lrbracket A}}", from=1-3, to=2-3]
	\arrow["{\pi_{\lrbracket B}}"{description}, from=1-1, to=2-3]
	\arrow["{\pi_{\lrbracket{A\to B}}}"{description}, from=2-1, to=2-3]
\end{tikzcd}\]
The introduction rule takes a morphism $\lrbracket u\in\CC(\lrbracket{\Gamma,A}, \lrbracket{\Gamma,B})$
such that $\pi_{\lrbracket B}\circ\lrbracket u=\pi_{\lrbracket A}$ to the morphism
$\lrbracket{\lambda. u}\in\CC(\lrbracket\Gamma,\lrbracket{\Gamma,A\to B})$ by the isomorphism in~\cref{def:fibexp}:
% https://q.uiver.app/?q=WzAsNCxbMiwxLCJcXGxyYnJhY2tldFxcR2FtbWEiXSxbMiwwLCJcXGxyYnJhY2tldFxcR2FtbWEsXFxscmJyYWNrZXQge0F9Il0sWzAsMSwiXFxscmJyYWNrZXRcXEdhbW1hLFxcbHJicmFja2V0e0FcXHRvIEJ9Il0sWzAsMCwiXFxscmJyYWNrZXRcXEdhbW1hLFxcbHJicmFja2V0IHtCfSJdLFsxLDAsIlxccGlfe1xcbHJicmFja2V0IEF9Il0sWzMsMCwiXFxwaV97XFxscmJyYWNrZXQgQn0iLDFdLFswLDIsIlxcbHJicmFja2V0e1xcbGFtYmRhLnV9IiwxXSxbMSwzLCJcXGxyYnJhY2tldCB1IiwxXV0=&macro_url=https%3A%2F%2Fgist.githubusercontent.com%2Fice1000%2F47b7ea52f8c351607831f7f4afa9a79b%2Fraw%2Fc368c27ee1f601d69c99ce4722eef5d32a3b644f%2Fquiver-macros.tex
\[\begin{tikzcd}
	{\lrbracket{\Gamma,B}} && {\lrbracket{\Gamma,A}} \\
	{\lrbracket{\Gamma,A\to B}} && \lrbracket\Gamma
	\arrow["{\pi_{\lrbracket A}}", from=1-3, to=2-3]
	\arrow["{\pi_{\lrbracket B}}"{description}, from=1-1, to=2-3]
	\arrow["{\lrbracket{\lambda.u}}"{description}, from=2-3, to=2-1]
	\arrow["{\lrbracket u}"{description}, from=1-3, to=1-1]
\end{tikzcd}\]
The rest of the rules involve the fiber product $\lrbracket{\Gamma,A\to B}\times_{\lrbracket \Gamma}
\lrbracket{\Gamma,A}\in\CC$, denoted $\lrbracket{\Gamma,A\times(A\to B)}$
(the $\times$ notation corresponds to the fact that the fiber product corresponds to the product type former, see~\cref{def:prod-sem}).
For convenience, we denote $\id_{\lrbracket{\Gamma,A\to B}} \times_{\lrbracket \Gamma} (f\circ \pi_{\lrbracket{A\to B}})$
as $\text{data}(f)$ for morphism $f\in\CC(\lrbracket\Gamma,\lrbracket{\Gamma,A})$.

The elimination rule is interpreted by morphism composition and the evaluation map (\cref{def:evalmap}),
visualized as the following diagram which commutes everywhere:
% https://q.uiver.app/?q=WzAsNixbMywxLCJcXGxyYnJhY2tldFxcR2FtbWEiXSxbMywwLCJcXGxyYnJhY2tldHtcXEdhbW1hLEF9Il0sWzAsMSwiXFxscmJyYWNrZXR7XFxHYW1tYSxBXFx0byBCfSJdLFsxLDAsIlxcbHJicmFja2V0e1xcR2FtbWEsQn0iXSxbMCwwLCJcXGxyYnJhY2tldHtcXEdhbW1hLEFcXHRpbWVzKEFcXHRvIEIpfSJdLFsxLDFdLFswLDIsIlxcbHJicmFja2V0IHUiLDFdLFswLDEsIlxcbHJicmFja2V0IHYiLDJdLFs0LDIsIiIsMix7ImN1cnZlIjotMSwic3R5bGUiOnsiYm9keSI6eyJuYW1lIjoiZGFzaGVkIn19fV0sWzQsMSwiIiwxLHsiY3VydmUiOi0yLCJzdHlsZSI6eyJib2R5Ijp7Im5hbWUiOiJkYXNoZWQifX19XSxbNCw1LCIiLDEseyJzdHlsZSI6eyJuYW1lIjoiY29ybmVyIn19XSxbNCwzLCJcXHJte2V2YWx9IiwyXSxbMiw0LCJcXHRleHR7ZGF0YX0oXFxscmJyYWNrZXQgdikiLDAseyJjdXJ2ZSI6LTF9XSxbMCwzLCJcXGxyYnJhY2tldHtcXGFwcGx5IHUgdn0iLDFdXQ==&macro_url=https%3A%2F%2Fgist.githubusercontent.com%2Fice1000%2F47b7ea52f8c351607831f7f4afa9a79b%2Fraw%2Fc368c27ee1f601d69c99ce4722eef5d32a3b644f%2Fquiver-macros.tex
\[\begin{tikzcd}
	{\lrbracket{\Gamma,A\times(A\to B)}} & {\lrbracket{\Gamma,B}} && {\lrbracket{\Gamma,A}} \\
	{\lrbracket{\Gamma,A\to B}} & {} && \lrbracket\Gamma
	\arrow["{\lrbracket u}"{description}, from=2-4, to=2-1]
	\arrow["{\lrbracket v}"', from=2-4, to=1-4]
	\arrow[curve={height=-6pt}, dashed, from=1-1, to=2-1]
	\arrow[curve={height=-12pt}, dashed, from=1-1, to=1-4]
	\arrow["\lrcorner"{anchor=center, pos=0.125}, draw=none, from=1-1, to=2-2]
	\arrow["{\rm{eval}}"', from=1-1, to=1-2]
	\arrow["{\text{data}(\lrbracket v)}", curve={height=-6pt}, from=2-1, to=1-1]
	\arrow["{\lrbracket{\apply u v}}"{description}, from=2-4, to=1-2]
\end{tikzcd}\]
Composing with the evaluation map (\cref{def:evalmap}) interprets the $\beta$-equality:
% https://q.uiver.app/?q=WzAsNixbMywxLCJcXGxyYnJhY2tldFxcR2FtbWEiXSxbMywwLCJcXGxyYnJhY2tldHtcXEdhbW1hLEF9Il0sWzAsMSwiXFxscmJyYWNrZXR7XFxHYW1tYSxBXFx0byBCfSJdLFsxLDAsIlxcbHJicmFja2V0e1xcR2FtbWEsQn0iXSxbMCwwLCJcXGxyYnJhY2tldHtcXEdhbW1hLEFcXHRpbWVzKEFcXHRvIEIpfSJdLFsxLDFdLFswLDIsIlxcbHJicmFja2V0e1xcbGFtYmRhLnV9IiwxXSxbMCwxLCJcXGxyYnJhY2tldCB2IiwyXSxbNCwyLCIiLDIseyJjdXJ2ZSI6LTEsInN0eWxlIjp7ImJvZHkiOnsibmFtZSI6ImRhc2hlZCJ9fX1dLFs0LDEsIiIsMSx7ImN1cnZlIjotMiwic3R5bGUiOnsiYm9keSI6eyJuYW1lIjoiZGFzaGVkIn19fV0sWzQsNSwiIiwxLHsic3R5bGUiOnsibmFtZSI6ImNvcm5lciJ9fV0sWzQsMywiXFxybXtldmFsfSIsMl0sWzIsNCwiXFx0ZXh0e2RhdGF9KFxcbHJicmFja2V0IHYpIiwwLHsiY3VydmUiOi0xfV0sWzAsMywiXFxscmJyYWNrZXQgdVxcY2lyYyBcXGxyYnJhY2tldCB2IiwxXSxbMSwzLCJcXGxyYnJhY2tldCB1IiwxXV0=&macro_url=https%3A%2F%2Fgist.githubusercontent.com%2Fice1000%2F47b7ea52f8c351607831f7f4afa9a79b%2Fraw%2Fc368c27ee1f601d69c99ce4722eef5d32a3b644f%2Fquiver-macros.tex
\[\begin{tikzcd}
	{\lrbracket{\Gamma,A\times(A\to B)}} & {\lrbracket{\Gamma,B}} && {\lrbracket{\Gamma,A}} \\
	{\lrbracket{\Gamma,A\to B}} & {} && \lrbracket\Gamma
	\arrow["{\lrbracket{\lambda.u}}"{description}, from=2-4, to=2-1]
	\arrow["{\lrbracket v}"', from=2-4, to=1-4]
	\arrow[curve={height=-6pt}, dashed, from=1-1, to=2-1]
	\arrow[curve={height=-12pt}, dashed, from=1-1, to=1-4]
	\arrow["\lrcorner"{anchor=center, pos=0.125}, draw=none, from=1-1, to=2-2]
	\arrow["{\rm{eval}}"', from=1-1, to=1-2]
	\arrow["{\text{data}(\lrbracket v)}", curve={height=-6pt}, from=2-1, to=1-1]
	\arrow["{\lrbracket u\circ \lrbracket v}"{description}, from=2-4, to=1-2]
	\arrow["{\lrbracket u}"{description}, from=1-4, to=1-2]
\end{tikzcd}\]
Note that we may also define the display map of $\lrbracket{\Gamma,A\times(A\to B)}$,
given by $\pi_{\lrbracket B}\circ \fs{eval}$. But we don't have to.
\end{defn}

\section{Dependent type theory}\label{sec:more-type}
\begin{defn}[Monic]\label{def:monic}
We say a morphism $f\in\CC(A,B)$ to be a \textit{monomorphism}, a \textit{mono},
or \textit{monic} if for every $C\in\CC$ and $g_1,g_2\in\CC(C,A)$ such that:
\begin{center}
$f\circ g_1=f\circ g_2 \implies g_1=g_2$
\end{center}
A monomorphism is also known as a \textit{left-cancallative} morphism.
The above equality is visualized below:
% https://q.uiver.app/?q=WzAsMyxbMCwwLCJDIl0sWzMsMCwiQSJdLFs1LDAsIkIiXSxbMCwxLCJnXzEiLDEseyJjdXJ2ZSI6LTF9XSxbMCwxLCJnXzIiLDEseyJjdXJ2ZSI6MX1dLFsxLDIsImYiLDFdXQ==
\[\begin{tikzcd}
	C &&& A && B
	\arrow["{g_1}"{description}, curve={height=-6pt}, from=1-1, to=1-4]
	\arrow["{g_2}"{description}, curve={height=6pt}, from=1-1, to=1-4]
	\arrow["f"{description}, from=1-4, to=1-6]
\end{tikzcd}\]
\lessSpace{-1}
\end{defn}

\begin{example}[Injection]\label{ex:setmono}
In $\SET$, a morphism is monic (\cref{def:monic}) if and only if it is an injective function.
\end{example}

% \begin{defn}[Monic]\label{def:monic2}
% This is an alternative to \cref{def:monic}.
% In a category $\CC$, if for every $C\in\CC$ the hom functor (\cref{def:homfunc}) $\CC(C,-)$
% takes $f\in\CC(A,B)$ to injective functions $\CC(C,f):\CC(C,A)\to\CC(C,B)$, we say $f$ to be \textit{monic}.
% \end{defn}

\begin{lem}\label{lem:monoiso}
For a mono (\cref{def:monic}) $f$, if there exist a morphism $g$ such that $f\circ g=\id_{cod(f)}$,
then $f$ is an isomorphism.
\end{lem}
\begin{proof}
$f\circ g\circ f=\id_{cod(f)}\circ f=f=f\circ \id_{dom(f)}$ and by the definition of mono and the associativity of composition
$g\circ f=\id_{dom(f)}$. Thus $f$ an isomorphism.
\end{proof}

\begin{remark}
There is an intuitive justification of~\cref{lem:monoiso}: in $\SET$,
invertible and injective (see~\cref{ex:setmono}) functions are isomorphisms.
\end{remark}

\begin{defn}[Epic]\label{def:epic}
We say a morphism $f\in\CC(A,B)$ to be an \textit{epimorphism}, an \textit{epi},
or \textit{epic} if for every $C\in\CC$ and $g_1,g_2\in\CC(B,C)$ such that:
\begin{center}
$g_1\circ f=g_2\circ f \implies g_1=g_2$
\end{center}
An epimorphism is also known as a \textit{right-cancallative} morphism.
The above equality is visualized below:
% https://q.uiver.app/?q=WzAsMyxbMCwwLCJDIl0sWzMsMCwiQiJdLFs1LDAsIkEiXSxbMSwwLCJnXzEiLDEseyJjdXJ2ZSI6MX1dLFsxLDAsImdfMiIsMSx7ImN1cnZlIjotMX1dLFsyLDEsImYiLDFdXQ==
\[\begin{tikzcd}
	C &&& B && A
	\arrow["{g_1}"{description}, curve={height=6pt}, from=1-4, to=1-1]
	\arrow["{g_2}"{description}, curve={height=-6pt}, from=1-4, to=1-1]
	\arrow["f"{description}, from=1-6, to=1-4]
\end{tikzcd}\]
\lessSpace{-1}
\end{defn}

\begin{example}[Surjection]\label{ex:setepi}
In $\SET$, a morphism is epic (\cref{def:epic}) if and only if it is a surjective function.
\end{example}

\begin{lem}\label{lem:isoepic}
Isomorphisms are epic (\cref{def:epic}).
\end{lem}
\begin{proof}
For an isomorphism $f$, we show that if there exist morphisms
$g_1,g_2$ such that $g_1\circ f=g_2\circ f$, then $g_1=g_2$:
$$g_1 = g_1 \circ f \circ f^{-1} = g_2 \circ f \circ f^{-1} = g_2$$
\end{proof}

\begin{remark}
A monomorphism (\cref{def:monic}) $f\in\CC(A, B)$ is epic (\cref{def:epic}) in the opposite category
(\cref{def:oppocat}) and vice versa. In other words, monic and epic are dual (\cref{conv:duality}) concepts.
\end{remark}

\subsection{Free lunch: the sigma type}
\begin{defn}[Sigma]\label{def:sigmatt}
We extend~\cref{def:tt} with the following type:\ttfragment{desum}
This type is known as the \textit{dependent sum type} or the \textit{sigma type}.
\end{defn}

\begin{defn}\label{def:sigma-sem}
We interpret sigma types (\cref{def:sigmatt}) using the compositions of display maps (\cref{def:displaymap}).
Consider a contextual category (\cref{def:cxlcat}) $\CC$, the formation of \textit{context $\Gamma$ extended by the sigma type $\Sigma_A B$},
$\lrbracket{\Gamma,\Sigma_A B}\in\CC$, is defined as $\lrbracket{\Gamma,A,B}\in\CC$ (the context $\Gamma$ extended by type $A$ then by type $B$)
whose display map is simply the composition of the display maps of type $A$ and $B$
($\pi_{\lrbracket{\Sigma_A B}}=\pi_{\lrbracket A}\circ\pi_{\lrbracket B}$).
% https://q.uiver.app/?q=WzAsNCxbMCwxLCJcXGxyYnJhY2tldFxcR2FtbWEiXSxbMiwxLCJcXGxyYnJhY2tldHtcXEdhbW1hLEF9Il0sWzQsMSwiXFxscmJyYWNrZXR7XFxHYW1tYSxBLEJ9Il0sWzQsMCwiXFxscmJyYWNrZXR7XFxHYW1tYSxcXFNpZ21hX0EgQn0iXSxbMSwwLCJcXHBpX3tcXGxyYnJhY2tldCBBfSJdLFsyLDEsIlxccGlfe1xcbHJicmFja2V0IEJ9Il0sWzMsMiwiIiwwLHsibGV2ZWwiOjIsInN0eWxlIjp7ImhlYWQiOnsibmFtZSI6Im5vbmUifX19XSxbMywwLCJcXHBpX3tcXGxyYnJhY2tldHtcXFNpZ21hX0EgQn19IiwyXV0=&macro_url=https%3A%2F%2Fgist.githubusercontent.com%2Fice1000%2F47b7ea52f8c351607831f7f4afa9a79b%2Fraw%2Ff5bd21d4bd401a9f3672269a2816c728e53bd9c0%2Fquiver-macros.tex
\[\begin{tikzcd}
	&&&& {\lrbracket{\Gamma,\Sigma_A B}} \\
	\lrbracket\Gamma && {\lrbracket{\Gamma,A}} && {\lrbracket{\Gamma,A,B}}
	\arrow["{\pi_{\lrbracket A}}", from=2-3, to=2-1]
	\arrow["{\pi_{\lrbracket B}}", from=2-5, to=2-3]
	\arrow[Rightarrow, no head, from=1-5, to=2-5]
	\arrow["{\pi_{\lrbracket{\Sigma_A B}}}"', from=1-5, to=2-1]
\end{tikzcd}\]
The introduction rule takes two morphisms.
One is $\lrbracket u\in\CC(\lrbracket{\Gamma}, \lrbracket{\Gamma,A})$,
and by the reindexing functor (\cref{def:reindex}) we can obtain the following pullback
(the fact that the diagram is a pullback is by~\cref{lem:reindex-pullback})
with a unique morphism $\pi_{\lrbracket B}^*(\lrbracket u)$ (denoted $\lrbracket{u}'$), an object $\lrbracket{\Gamma,Bu}$
(which corresponds to the context $\Gamma$ extended by the substituted type $Bu$),
and its display map $\pi_{\lrbracket{\Gamma,Bu}}\in\CC(\lrbracket\Gamma,\lrbracket{\Gamma,Bu})$:
% https://q.uiver.app/?q=WzAsNixbMCwxLCJcXGxyYnJhY2tldFxcR2FtbWEiXSxbMiwxLCJcXGxyYnJhY2tldHtcXEdhbW1hLEF9Il0sWzIsMCwiXFxscmJyYWNrZXR7XFxHYW1tYSxBLEJ9Il0sWzAsMCwiXFxscmJyYWNrZXR7XFxHYW1tYSxCdX0iXSxbMSwxXSxbMywxLCJcXGxyYnJhY2tldHtcXEdhbW1hLFxcU2lnbWFfQSBCfSJdLFswLDEsIlxcbHJicmFja2V0e3t1fX0iLDFdLFsyLDEsIlxccGlfe1xcbHJicmFja2V0IEJ9Il0sWzMsMiwiXFxscmJyYWNrZXR7dX0nIiwxLHsic3R5bGUiOnsiYm9keSI6eyJuYW1lIjoiZGFzaGVkIn19fV0sWzMsNCwiIiwyLHsic3R5bGUiOnsibmFtZSI6ImNvcm5lciJ9fV0sWzMsMCwiXFxwaV97XFxscmJyYWNrZXR7QnV9fSIsMix7InN0eWxlIjp7ImJvZHkiOnsibmFtZSI6ImRhc2hlZCJ9fX1dLFs1LDIsIiIsMCx7ImxldmVsIjoyLCJzdHlsZSI6eyJoZWFkIjp7Im5hbWUiOiJub25lIn19fV0sWzUsMCwiXFxwaV97XFxscmJyYWNrZXR7XFxTaWdtYV9BIEJ9fSIsMSx7ImN1cnZlIjotMn1dXQ==&macro_url=https%3A%2F%2Fgist.githubusercontent.com%2Fice1000%2F47b7ea52f8c351607831f7f4afa9a79b%2Fraw%2Ff5bd21d4bd401a9f3672269a2816c728e53bd9c0%2Fquiver-macros.tex
\[\begin{tikzcd}
	{\lrbracket{\Gamma,Bu}} && {\lrbracket{\Gamma,A,B}} \\
	\lrbracket\Gamma & {} & {\lrbracket{\Gamma,A}} & {\lrbracket{\Gamma,\Sigma_A B}}
	\arrow["{\lrbracket{{u}}}"{description}, from=2-1, to=2-3]
	\arrow["{\pi_{\lrbracket B}}", from=1-3, to=2-3]
	\arrow["{\lrbracket{u}'}"{description}, dashed, from=1-1, to=1-3]
	\arrow["\lrcorner"{anchor=center, pos=0.125}, draw=none, from=1-1, to=2-2]
	\arrow["{\pi_{\lrbracket{Bu}}}"', dashed, from=1-1, to=2-1]
	\arrow[Rightarrow, no head, from=2-4, to=1-3]
	\arrow["{\pi_{\lrbracket{\Sigma_A B}}}"{description}, curve={height=-12pt}, from=2-4, to=2-1]
\end{tikzcd}\]
The other input morphism is an inverse of $\pi_{\lrbracket{\Gamma,Bu}}$, denoted
$\lrbracket v\in\CC(\lrbracket{\Gamma}, \lrbracket{\Gamma,Bu})$.
We define the introduction rule as the composition $\lrbracket{u}'\circ\lrbracket{v}$,
which is inverse to $\pi_{\lrbracket{\Sigma_A B}}$ due to the commutativity of the following diagram:
% https://q.uiver.app/?q=WzAsNixbMCwxLCJcXGxyYnJhY2tldFxcR2FtbWEiXSxbMiwxLCJcXGxyYnJhY2tldHtcXEdhbW1hLEF9Il0sWzIsMCwiXFxscmJyYWNrZXR7XFxHYW1tYSxBLEJ9Il0sWzMsMSwiXFxscmJyYWNrZXR7XFxHYW1tYSxcXFNpZ21hX0EgQn0iXSxbMCwwLCJcXGxyYnJhY2tldHtcXEdhbW1hLEJ1fSJdLFsxLDFdLFswLDEsIlxcbHJicmFja2V0e3V9IiwxXSxbMywyLCIiLDAseyJsZXZlbCI6Miwic3R5bGUiOnsiaGVhZCI6eyJuYW1lIjoibm9uZSJ9fX1dLFswLDQsIlxcbHJicmFja2V0e3Z9Il0sWzQsMiwiXFxscmJyYWNrZXR7dX0nIiwxLHsic3R5bGUiOnsiYm9keSI6eyJuYW1lIjoiZGFzaGVkIn19fV0sWzQsNSwiIiwyLHsic3R5bGUiOnsibmFtZSI6ImNvcm5lciJ9fV0sWzAsMiwiXFxscmJyYWNrZXR7XFxsbHJyYW5nbGV7dSx2fX0iLDFdLFszLDAsIlxccGlfe1xcbHJicmFja2V0e1xcU2lnbWFfQSBCfX0iLDEseyJjdXJ2ZSI6LTJ9XV0=&macro_url=https%3A%2F%2Fgist.githubusercontent.com%2Fice1000%2F47b7ea52f8c351607831f7f4afa9a79b%2Fraw%2Ff5bd21d4bd401a9f3672269a2816c728e53bd9c0%2Fquiver-macros.tex
\[\begin{tikzcd}
	{\lrbracket{\Gamma,Bu}} && {\lrbracket{\Gamma,A,B}} \\
	\lrbracket\Gamma & {} & {\lrbracket{\Gamma,A}} & {\lrbracket{\Gamma,\Sigma_A B}}
	\arrow["{\lrbracket{u}}"{description}, from=2-1, to=2-3]
	\arrow[Rightarrow, no head, from=2-4, to=1-3]
	\arrow["{\lrbracket{v}}", from=2-1, to=1-1]
	\arrow["{\lrbracket{u}'}"{description}, dashed, from=1-1, to=1-3]
	\arrow["\lrcorner"{anchor=center, pos=0.125}, draw=none, from=1-1, to=2-2]
	\arrow["{\lrbracket{\llrrangle{u,v}}}"{description}, from=2-1, to=1-3]
	\arrow["{\pi_{\lrbracket{\Sigma_A B}}}"{description}, curve={height=-12pt}, from=2-4, to=2-1]
\end{tikzcd}\]
The elimination rules are interpreted as well by compositions.
Consider a term $\lrbracket s\in\CC(\lrbracket\Gamma, \lrbracket{\Gamma,\Sigma_A B})$:
\begin{enumerate}
\item The first projection $\lrbracket{s.\fstProj}\in\CC(\lrbracket\Gamma,\lrbracket{\Gamma,A})$
is defined as $\pi_{\lrbracket B}\circ \lrbracket s$.
\item The second projection $\lrbracket{s.\sndProj}\in\CC(\lrbracket\Gamma,\lrbracket{\Gamma,B{s.\fstProj}})$
is defined by taking pullback of $\lrbracket s$ along $\lrbracket{s.\fstProj}'$.
\end{enumerate}
Diagrammatically:
% https://q.uiver.app/?q=WzAsNixbMCwxLCJcXGxyYnJhY2tldFxcR2FtbWEiXSxbMiwyLCJcXGxyYnJhY2tldHtcXEdhbW1hLEF9Il0sWzIsMCwiXFxscmJyYWNrZXR7XFxHYW1tYSxBLEJ9Il0sWzIsMSwiXFxscmJyYWNrZXR7XFxHYW1tYSxcXFNpZ21hX0EgQn0iXSxbMCwwLCJcXGxyYnJhY2tldHtcXEdhbW1hLEIocy5cXGZzdFByb2opfSJdLFsxLDFdLFswLDEsIlxcbHJicmFja2V0e3MuXFxmc3RQcm9qfSIsMSx7InN0eWxlIjp7ImJvZHkiOnsibmFtZSI6ImRhc2hlZCJ9fX1dLFszLDIsIiIsMCx7ImxldmVsIjoyLCJzdHlsZSI6eyJoZWFkIjp7Im5hbWUiOiJub25lIn19fV0sWzQsMiwiXFxscmJyYWNrZXR7cy5cXGZzdFByb2p9JyIsMV0sWzQsNSwiIiwyLHsic3R5bGUiOnsibmFtZSI6ImNvcm5lciJ9fV0sWzAsNCwiXFxscmJyYWNrZXR7cy5cXHNuZFByb2p9Il0sWzMsMSwiXFxwaV97XFxscmJyYWNrZXQgQn0iXSxbMCwzLCJcXGxyYnJhY2tldCBzIiwxXV0=&macro_url=https%3A%2F%2Fgist.githubusercontent.com%2Fice1000%2F47b7ea52f8c351607831f7f4afa9a79b%2Fraw%2Ff5bd21d4bd401a9f3672269a2816c728e53bd9c0%2Fquiver-macros.tex
\[\begin{tikzcd}
	{\lrbracket{\Gamma,B(s.\fstProj)}} && {\lrbracket{\Gamma,A,B}} \\
	\lrbracket\Gamma & {} & {\lrbracket{\Gamma,\Sigma_A B}} \\
	&& {\lrbracket{\Gamma,A}}
	\arrow["{\lrbracket{s.\fstProj}}"{description}, dashed, from=2-1, to=3-3]
	\arrow[Rightarrow, no head, from=2-3, to=1-3]
	\arrow["{\lrbracket{s.\fstProj}'}"{description}, from=1-1, to=1-3]
	\arrow["\lrcorner"{anchor=center, pos=0.125}, draw=none, from=1-1, to=2-2]
	\arrow["{\lrbracket{s.\sndProj}}", from=2-1, to=1-1]
	\arrow["{\pi_{\lrbracket B}}", from=2-3, to=3-3]
	\arrow["{\lrbracket s}"{description}, from=2-1, to=2-3]
\end{tikzcd}\]
The computation rules hold by compositions.
This definition is inspired from~\cite{Guest0x0}.
\end{defn}

\begin{remark}
The sigma type is the easiest dependent type to interpret, because it only requires context extension.
\end{remark}

\begin{cor}\label{cor:nondepsigma}
Suppose in a contextual category $\CC$ where we have interpretations of two types $A, B'$ in context $\Gamma$.
Consider an object $\lrbracket{\Gamma,A,B}$ given by the pullback of the following square:
% https://q.uiver.app/?q=WzAsNCxbMCwxLCJcXGxyYnJhY2tldFxcR2FtbWEiXSxbMSwxLCJcXGxyYnJhY2tldHtcXEdhbW1hLEF9Il0sWzAsMCwiXFxscmJyYWNrZXR7XFxHYW1tYSxCJ30iXSxbMSwwLCJcXGxyYnJhY2tldHtcXEdhbW1hLEEsQn0iXSxbMSwwLCJcXHBpX3tcXGxyYnJhY2tldCBBfSJdLFsyLDAsIlxccGlfe1xcbHJicmFja2V0e0InfX0iLDJdLFszLDIsIiIsMCx7InN0eWxlIjp7ImJvZHkiOnsibmFtZSI6ImRhc2hlZCJ9fX1dLFszLDEsIiIsMSx7InN0eWxlIjp7ImJvZHkiOnsibmFtZSI6ImRhc2hlZCJ9fX1dLFszLDAsIiIsMSx7InN0eWxlIjp7Im5hbWUiOiJjb3JuZXIifX1dXQ==&macro_url=https%3A%2F%2Fgist.githubusercontent.com%2Fice1000%2F47b7ea52f8c351607831f7f4afa9a79b%2Fraw%2Ff5bd21d4bd401a9f3672269a2816c728e53bd9c0%2Fquiver-macros.tex
\[\begin{tikzcd}
	{\lrbracket{\Gamma,B'}} & {\lrbracket{\Gamma,A,B}} \\
	\lrbracket\Gamma & {\lrbracket{\Gamma,A}}
	\arrow["{\pi_{\lrbracket A}}", from=2-2, to=2-1]
	\arrow["{\pi_{\lrbracket{B'}}}"', from=1-1, to=2-1]
	\arrow[dashed, from=1-2, to=1-1]
	\arrow[dashed, from=1-2, to=2-2]
	\arrow["\lrcorner"{anchor=center, pos=0.125, rotate=-90}, draw=none, from=1-2, to=2-1]
\end{tikzcd}\]
Then, the interpretation of the sigma type $\Sigma_A B$ (see~\cref{def:sigma-sem}) is equivalent to the
interpretation of the product type $A\times B'$ (\cref{def:prod-sem}).
\end{cor}
\begin{proof}
The diagrams are the same (both are pullbacks).
\end{proof}

\begin{remark}
Note that in~\cref{cor:nondepsigma}, $B$ is simply $B'$ in a weakened context,
so this corollary shows that (type theoretically), when $B$ does not depend on $A$,
the sigma type $\Sigma_A B$ is essentially the product type $A \times B$.

Consider $\lrbracket u\in\CC(\lrbracket{\Gamma}, \lrbracket{\Gamma,A})$, we diagram the sigma type:
% https://q.uiver.app/?q=WzAsNixbMCwxLCJcXGxyYnJhY2tldFxcR2FtbWEiXSxbMSwxLCJcXGxyYnJhY2tldHtcXEdhbW1hLEF9Il0sWzAsMCwiXFxscmJyYWNrZXR7XFxHYW1tYSxCJ30iXSxbMSwwLCJcXGxyYnJhY2tldHtcXEdhbW1hLEEsQn0iXSxbMiwwLCJcXGxyYnJhY2tldHtcXEdhbW1hLEJ1fSJdLFsyLDEsIlxcbHJicmFja2V0XFxHYW1tYSJdLFsxLDAsIlxccGlfe1xcbHJicmFja2V0IEF9IiwyXSxbMiwwLCJcXHBpX3tcXGxyYnJhY2tldHtCJ319IiwyXSxbMywyLCIiLDAseyJzdHlsZSI6eyJib2R5Ijp7Im5hbWUiOiJkYXNoZWQifX19XSxbMywxLCIiLDEseyJzdHlsZSI6eyJib2R5Ijp7Im5hbWUiOiJkYXNoZWQifX19XSxbMywwLCIiLDEseyJzdHlsZSI6eyJuYW1lIjoiY29ybmVyIn19XSxbNCwzLCIiLDAseyJzdHlsZSI6eyJib2R5Ijp7Im5hbWUiOiJkYXNoZWQifX19XSxbNCw1LCIiLDIseyJzdHlsZSI6eyJib2R5Ijp7Im5hbWUiOiJkYXNoZWQifX19XSxbNSwxLCJcXGxyYnJhY2tldCB1IiwyXSxbNCwxLCIiLDIseyJzdHlsZSI6eyJuYW1lIjoiY29ybmVyIn19XSxbNSwwLCJcXGlkX3tcXGxyYnJhY2tldFxcR2FtbWF9IiwxLHsiY3VydmUiOi0yfV1d&macro_url=https%3A%2F%2Fgist.githubusercontent.com%2Fice1000%2F47b7ea52f8c351607831f7f4afa9a79b%2Fraw%2Ff5bd21d4bd401a9f3672269a2816c728e53bd9c0%2Fquiver-macros.tex
\[\begin{tikzcd}
	{\lrbracket{\Gamma,B'}} & {\lrbracket{\Gamma,A,B}} & {\lrbracket{\Gamma,Bu}} \\
	\lrbracket\Gamma & {\lrbracket{\Gamma,A}} & \lrbracket\Gamma
	\arrow["{\pi_{\lrbracket A}}"', from=2-2, to=2-1]
	\arrow["{\pi_{\lrbracket{B'}}}"', from=1-1, to=2-1]
	\arrow[dashed, from=1-2, to=1-1]
	\arrow[dashed, from=1-2, to=2-2]
	\arrow["\lrcorner"{anchor=center, pos=0.125, rotate=-90}, draw=none, from=1-2, to=2-1]
	\arrow[dashed, from=1-3, to=1-2]
	\arrow[dashed, from=1-3, to=2-3]
	\arrow["{\lrbracket u}"', from=2-3, to=2-2]
	\arrow["\lrcorner"{anchor=center, pos=0.125, rotate=-90}, draw=none, from=1-3, to=2-2]
	\arrow["{\id_{\lrbracket\Gamma}}"{description}, curve={height=-12pt}, from=2-3, to=2-1]
\end{tikzcd}\]
If we simplify the identity morphism, we get:
% https://q.uiver.app/?q=WzAsNCxbMCwxLCJcXGxyYnJhY2tldFxcR2FtbWEiXSxbMCwwLCJcXGxyYnJhY2tldHtcXEdhbW1hLEInfSJdLFsxLDAsIlxcbHJicmFja2V0e1xcR2FtbWEsQnV9Il0sWzEsMSwiXFxscmJyYWNrZXRcXEdhbW1hIl0sWzEsMCwiXFxwaV97XFxscmJyYWNrZXR7Qid9fSIsMl0sWzIsMywiIiwyLHsic3R5bGUiOnsiYm9keSI6eyJuYW1lIjoiZGFzaGVkIn19fV0sWzMsMCwiXFxpZF97XFxscmJyYWNrZXRcXEdhbW1hfSIsMl0sWzIsMSwiIiwxLHsic3R5bGUiOnsiYm9keSI6eyJuYW1lIjoiZGFzaGVkIn19fV0sWzIsMCwiIiwyLHsic3R5bGUiOnsibmFtZSI6ImNvcm5lciJ9fV1d&macro_url=https%3A%2F%2Fgist.githubusercontent.com%2Fice1000%2F47b7ea52f8c351607831f7f4afa9a79b%2Fraw%2Ff5bd21d4bd401a9f3672269a2816c728e53bd9c0%2Fquiver-macros.tex
\[\begin{tikzcd}
	{\lrbracket{\Gamma,B'}} & {\lrbracket{\Gamma,Bu}} \\
	\lrbracket\Gamma & \lrbracket\Gamma
	\arrow["{\pi_{\lrbracket{B'}}}"', from=1-1, to=2-1]
	\arrow[dashed, from=1-2, to=2-2]
	\arrow["{\id_{\lrbracket\Gamma}}"', from=2-2, to=2-1]
	\arrow[dashed, from=1-2, to=1-1]
	\arrow["\lrcorner"{anchor=center, pos=0.125, rotate=-90}, draw=none, from=1-2, to=2-1]
\end{tikzcd}\]
By~\cref{lem:pullback-along-id}, $\lrbracket{\Gamma,Bu}\cong\lrbracket{\Gamma,B'}$ (up to unique isomorphism).
It is left as an exercise to make sense of this isomorphism type theoretically.
\end{remark}

\begin{cor}
A contextual category (\cref{def:cxlcat}) is democratic (\cref{def:democ}).
\end{cor}
\begin{proof}
For any $\Gamma\in\CC$, by~\cref{def:ctxly} we know that it's generated by
the terminal object \textit1 and context extension. We replace the context
extension operation with taking sigma types, and we get a sigma type in the empty context.
This fulfills the object needed for democracy.
\end{proof}

\subsection{Equalizers: the extensional equality type}
\begin{remark}[McBride]
We begin this section with a quotation by Conor McBride from his suspended Twitter account%
\footnote{\url{https://twitter.com/pigworker}}:
\begin{quotation}
Never trust a type theorist who has not changed their mind about equality.
\end{quotation}
\lessSpace{-0.2}
\end{remark}

\begin{defn}[Id]\label{def:id-tt}
We extend~\cref{def:tt} with the following type:\ttfragment{extid}
This type is known as the \textit{extensional equality type}, in contrast to the \textit{intensional equality type}.
We will only talk about the extensional one for convenience.
\end{defn}

\begin{terminology}
The introduction rule in~\cref{def:id-tt} is known as the rule for \textit{reflexivity},
and the elimination rule is known as \textit{equality reflection}.
\end{terminology}

\begin{defn}[Equalizer]\label{def:equalizer}
In a category $\CC$ and for $f,g\in\CC(X,Y)$ (we refer to pairs of morphisms like this as \textit{parallel morphisms}),
there might be some objects $E\in\CC$ together with morphisms of form $e\in\CC(E,X)$ commuting the following diagram
(so that $f\circ e=g\circ e\in\CC(E,Y)$):
% https://q.uiver.app/?q=WzAsMyxbMiwwLCJYIl0sWzUsMCwiWSJdLFswLDAsIkUiXSxbMCwxLCJmIiwxLHsiY3VydmUiOi0xfV0sWzAsMSwiZyIsMSx7ImN1cnZlIjoxfV0sWzIsMCwiZSIsMV1d
\[\begin{tikzcd}
	E && X &&& Y
	\arrow["f"{description}, curve={height=-6pt}, from=1-3, to=1-6]
	\arrow["g"{description}, curve={height=6pt}, from=1-3, to=1-6]
	\arrow["e"{description}, from=1-1, to=1-3]
\end{tikzcd}\]
We take the terminal object of the full subcategory (\cref{def:fullsub,def:fullsub2}) of $\CC_{/X}$
commuting the above diagram which consists of morphisms like $e$, and refer to the object $E$ (denoted $\Eqlz f g$)
and the morphism $e$ (denoted $\eqlz f g$) as the \textit{equalizer} of $f$ and $g$.
We bring these notations into the diagram above:
% https://q.uiver.app/?q=WzAsMyxbMiwwLCJYIl0sWzUsMCwiWSJdLFswLDAsIlxcRXFseiBmIGciXSxbMCwxLCJmIiwxLHsiY3VydmUiOi0xfV0sWzAsMSwiZyIsMSx7ImN1cnZlIjoxfV0sWzIsMCwiXFxlcWx6IGYgZyIsMSx7InN0eWxlIjp7ImJvZHkiOnsibmFtZSI6ImRhc2hlZCJ9fX1dXQ==&macro_url=https%3A%2F%2Fgist.githubusercontent.com%2Fice1000%2F47b7ea52f8c351607831f7f4afa9a79b%2Fraw%2Ff5bd21d4bd401a9f3672269a2816c728e53bd9c0%2Fquiver-macros.tex
\[\begin{tikzcd}
	{\Eqlz f g} && X &&& Y
	\arrow["f"{description}, curve={height=-6pt}, from=1-3, to=1-6]
	\arrow["g"{description}, curve={height=6pt}, from=1-3, to=1-6]
	\arrow["{\eqlz f g}"{description}, dashed, from=1-1, to=1-3]
\end{tikzcd}\]
In case each parallel morphisms in $\CC$ has an equalizer, we say that $\CC$ \textit{has all equalizers}.
The definition of equalizers is from~\cite{Equalizer}.
\end{defn}

\begin{lem}\label{lem:eqlz-pullbacks}
A category $\CC$ has all equalizers (\cref{def:equalizer}) if it has a terminal object
and all pullbacks (\cref{def:pullback}).
\end{lem}
\begin{proof}
By~\cref{lem:prod-from-pullback} we know $\CC$ has products.
We construct the equalizer for arbitrary parallel morphisms $f,g\in\CC(X,Y)$:
% https://q.uiver.app/?q=WzAsNSxbMCwwLCJcXEVxbHogZiBnIl0sWzAsMSwiWSJdLFsyLDEsIllcXHRpbWVzIFkiXSxbMiwwLCJYIl0sWzEsMV0sWzAsMSwiIiwyLHsic3R5bGUiOnsiYm9keSI6eyJuYW1lIjoiZGFzaGVkIn19fV0sWzEsMiwiKFxcaWRfWSxcXGlkX1kpIiwxXSxbMCwzLCJcXGVxbHogZiBnIiwxLHsic3R5bGUiOnsiYm9keSI6eyJuYW1lIjoiZGFzaGVkIn19fV0sWzMsMiwiKGYsZykiXSxbMCw0LCIiLDIseyJzdHlsZSI6eyJuYW1lIjoiY29ybmVyIn19XV0=&macro_url=https%3A%2F%2Fgist.githubusercontent.com%2Fice1000%2F47b7ea52f8c351607831f7f4afa9a79b%2Fraw%2Ff5bd21d4bd401a9f3672269a2816c728e53bd9c0%2Fquiver-macros.tex
\[\begin{tikzcd}
	{\Eqlz f g} && X \\
	Y & {} & {Y\times Y}
	\arrow[dashed, from=1-1, to=2-1]
	\arrow["{(\id_Y,\id_Y)}"{description}, from=2-1, to=2-3]
	\arrow["{\eqlz f g}"{description}, dashed, from=1-1, to=1-3]
	\arrow["{(f,g)}", from=1-3, to=2-3]
	\arrow["\lrcorner"{anchor=center, pos=0.125}, draw=none, from=1-1, to=2-2]
\end{tikzcd}\]
\lessSpace{-1}
\end{proof}

\begin{lem}[{\cite[Proposition 1.3]{Equalizer}}]\label{lem:eqlz-mono}
For a category $\CC$ and $f,g\in\CC(X,Y)$, $\eqlz f g$ is monic (\cref{def:monic}).
\end{lem}

\begin{lem}\label{lem:eqlzid}
In a category $\CC$ and a morphism $f\in\CC(X,Y)$, the equalizer of $f$ and $f$ is $\id_X$.
\end{lem}

\begin{defn}\label{def:id-sem}
We interpret the extensional identity type in a contextual category (\cref{def:cxlcat}) $\CC$ as equalizers (\cref{def:equalizer}).
The formation of the \textit{context $\Gamma$ extended by $\Idty A a b$} (denoted $\lrbracket{\Gamma,\Idty A a b}\in\CC$)
is by taking the equalizer $\Eqlz{\lrbracket a}{\lrbracket b}$, where the display map (\cref{def:displaymap})
is given by the equalizer: $\pi_{\lrbracket{\Idty A a b}}=\eqlz{\lrbracket a}{\lrbracket b}$:
% https://q.uiver.app/?q=WzAsMyxbMiwwLCJcXGxyYnJhY2tldFxcR2FtbWEiXSxbMCwwLCJcXGxyYnJhY2tldFxcR2FtbWEsXFxscmJyYWNrZXQgQSJdLFs1LDAsIlxcbHJicmFja2V0XFxHYW1tYSxcXGxyYnJhY2tldHtcXElkdHkgQSBhIGJ9Il0sWzAsMSwiXFxscmJyYWNrZXQgYSIsMSx7ImN1cnZlIjoxfV0sWzAsMSwiXFxscmJyYWNrZXQgYiIsMSx7ImN1cnZlIjotMX1dLFsyLDAsIlxcZXFsentcXGxyYnJhY2tldCBhfXtcXGxyYnJhY2tldCBifSIsMSx7InN0eWxlIjp7ImJvZHkiOnsibmFtZSI6ImRhc2hlZCJ9fX1dXQ==&macro_url=https%3A%2F%2Fgist.githubusercontent.com%2Fice1000%2F47b7ea52f8c351607831f7f4afa9a79b%2Fraw%2Ff5bd21d4bd401a9f3672269a2816c728e53bd9c0%2Fquiver-macros.tex
\[\begin{tikzcd}
	{\lrbracket{\Gamma,A}} && \lrbracket\Gamma &&& {\lrbracket{\Gamma,\Idty A a b}}
	\arrow["{\lrbracket a}"{description}, curve={height=6pt}, from=1-3, to=1-1]
	\arrow["{\lrbracket b}"{description}, curve={height=-6pt}, from=1-3, to=1-1]
	\arrow["{\eqlz{\lrbracket a}{\lrbracket b}}"{description}, dashed, from=1-6, to=1-3]
\end{tikzcd}\]
By~\cref{lem:eqlzid}, $\eqlz{\lrbracket a}{\lrbracket a}=\id_{\lrbracket\Gamma}$.
So, we define $\lrbracket{\reflcon a}=\id_{\lrbracket\Gamma}$:
% https://q.uiver.app/?q=WzAsMyxbMiwwLCJcXGxyYnJhY2tldFxcR2FtbWEiXSxbMCwwLCJcXGxyYnJhY2tldFxcR2FtbWEsXFxscmJyYWNrZXQgQSJdLFs0LDAsIlxcbHJicmFja2V0XFxHYW1tYSxcXGxyYnJhY2tldHtcXElkdHkgQSBhIGF9Il0sWzAsMSwiXFxscmJyYWNrZXQgYSIsMV0sWzIsMCwiXFxscmJyYWNrZXR7XFxyZWZsY29uIGF9IiwxLHsibGV2ZWwiOjIsInN0eWxlIjp7ImhlYWQiOnsibmFtZSI6Im5vbmUifX19XV0=&macro_url=https%3A%2F%2Fgist.githubusercontent.com%2Fice1000%2F47b7ea52f8c351607831f7f4afa9a79b%2Fraw%2Ff5bd21d4bd401a9f3672269a2816c728e53bd9c0%2Fquiver-macros.tex
\[\begin{tikzcd}
	{\lrbracket{\Gamma,A}} && \lrbracket\Gamma && {\lrbracket{\Gamma,\Idty A a a}}
	\arrow["{\lrbracket a}"{description}, from=1-3, to=1-1]
	\arrow["{\lrbracket{\reflcon a}}"{description}, Rightarrow, no head, from=1-5, to=1-3]
\end{tikzcd}\]
In case we have a morphism $f$ corresponding to an instance of the identity type, it is an inverse to the equalizer
($\eqlz{\lrbracket a}{\lrbracket b}\circ f=\id_{\lrbracket{\Gamma,\Idty A a b}}$),
and by~\cref{lem:monoiso} $f$ is an isomorphism.
Then, by~\cref{lem:isoepic} $\lrbracket a=\lrbracket b$ holds, and that makes sense of the elimination rule:
% https://q.uiver.app/?q=WzAsMyxbMiwwLCJcXGxyYnJhY2tldFxcR2FtbWEiXSxbMCwwLCJcXGxyYnJhY2tldFxcR2FtbWEsXFxscmJyYWNrZXQgQSJdLFs0LDAsIlxcbHJicmFja2V0XFxHYW1tYSxcXGxyYnJhY2tldHtcXElkdHkgQSBhIGJ9Il0sWzAsMSwiXFxscmJyYWNrZXQgYSIsMSx7ImN1cnZlIjoyfV0sWzAsMSwiXFxscmJyYWNrZXQgYiIsMSx7ImN1cnZlIjotMn1dLFswLDIsImYiLDIseyJzdHlsZSI6eyJib2R5Ijp7Im5hbWUiOiJkYXNoZWQifSwiaGVhZCI6eyJuYW1lIjoiaGFycG9vbiIsInNpZGUiOiJib3R0b20ifX19XSxbMiwwLCJcXGVxbHp7XFxscmJyYWNrZXQgYX17XFxscmJyYWNrZXQgYn0iLDIseyJzdHlsZSI6eyJib2R5Ijp7Im5hbWUiOiJkYXNoZWQifSwiaGVhZCI6eyJuYW1lIjoiaGFycG9vbiIsInNpZGUiOiJib3R0b20ifX19XSxbMSwwLCJcXHBpX3tcXGxyYnJhY2tldCBBfSIsMV1d&macro_url=https%3A%2F%2Fgist.githubusercontent.com%2Fice1000%2F47b7ea52f8c351607831f7f4afa9a79b%2Fraw%2Ff5bd21d4bd401a9f3672269a2816c728e53bd9c0%2Fquiver-macros.tex
\[\begin{tikzcd}
	{\lrbracket{\Gamma,A}} && \lrbracket\Gamma && {\lrbracket{\Gamma,\Idty A a b}}
	\arrow["{\lrbracket a}"{description}, curve={height=12pt}, from=1-3, to=1-1]
	\arrow["{\lrbracket b}"{description}, curve={height=-12pt}, from=1-3, to=1-1]
	\arrow["f"', dashed, harpoon', from=1-3, to=1-5]
	\arrow["{\eqlz{\lrbracket a}{\lrbracket b}}"', dashed, harpoon', from=1-5, to=1-3]
	\arrow["{\pi_{\lrbracket A}}"{description}, from=1-1, to=1-3]
\end{tikzcd}\]
\lessSpace{-1}
\end{defn}

\begin{lem}[UIP]\label{lem:uip}
The interpretation in~\cref{def:id-sem} gives rise to the following
``uniqueness of the identity proof'' rule: \ttfragment{uip2}
\end{lem}
\begin{proof}
By~\cref{lem:eqlzid} and the uniqueness of the identity morphism.
For visualization, see the second diagram in~\cref{def:id-sem}.
\end{proof}

\begin{thm}[Uniqueness]\label{thm:uip}
The interpretation in~\cref{def:id-sem} gives rise to the following ``uniqueness rule''
of the identity type: \ttfragment{uip}
\end{thm}
\begin{proof} (Type theory perspective)
By equality reflection, the existence of $p$ and $q$ implies $a=b$, so $p$ and $q$ are also
instances of $\Idty A a a$, and by~\cref{lem:uip} they are both equal to $\reflcon a$.
\end{proof}
\begin{proof} (Categorical perspective)
$\lrbracket p\circ\eqlz{\lrbracket a}{\lrbracket b}=\lrbracket q\circ\eqlz{\lrbracket a}{\lrbracket b}$,
and then $\lrbracket p = \lrbracket q$ since $\eqlz{\lrbracket a}{\lrbracket b}$ is monic.
\end{proof}

\subsection{Adjoint functors and counits}\label{sec:adjoint}
\begin{defn}[Adjoint]\label{def:adjoint}
Consider two categories $\CC,\DD$ and two functors $L:\CC\to\DD$ and $R:\DD\to\CC$.
We say $L$ and $R$ constitute \textit{a pair of adjoint functors}
(also known as \textit{an adjoint pair of functors}), denoted $L\dashv R:\CC\rightleftarrows\DD$,
if for any $C\in\CC, D\in\DD$, we have the following hom set isomorphism:
$$\DD(L(C), D)\simeq\CC(C, R(D))$$
In other words, the following functors are naturally isomorphic:
\begin{align*}
&\DD(L(-), D) &&\simeq &\CC(-, R(D)) \\
&\DD(L(C), -) &&\simeq &\CC(C, R(-))
\end{align*}
In commutative diagrams, we denote adjoint functors like this:
% https://q.uiver.app/?q=WzAsMixbMiwwLCJcXENDIl0sWzAsMCwiXFxERCJdLFswLDEsIkwiLDIseyJvZmZzZXQiOjN9XSxbMSwwLCJSIiwyLHsib2Zmc2V0IjozfV0sWzIsMywiIiwwLHsibGV2ZWwiOjEsInN0eWxlIjp7Im5hbWUiOiJhZGp1bmN0aW9uIn19XV0=
\[\begin{tikzcd}
	\DD && \CC
	\arrow[""{name=0, anchor=center, inner sep=0}, "L"', shift right=3, from=1-3, to=1-1]
	\arrow[""{name=1, anchor=center, inner sep=0}, "R"', shift right=3, from=1-1, to=1-3]
	\arrow["\dashv"{anchor=center, rotate=-90}, draw=none, from=0, to=1]
\end{tikzcd}\]
The relationship between adjoint functors is called \textit{adjunction}.
\end{defn}

\begin{example}\label{ex:tensor-hom}
Consider a category with internal homs (\cref{def:closedsmcat}) $\CC$.
For any $A\in\CC$, there is the adjoint pair (\cref{def:adjoint})
$(-\otimes A)\dashv[A, -]$ (both functors are endofunctors on $\CC$)
proved as a part of~\cref{lem:tensor-hom-adjoint}. In other words,
$$\CC((-\otimes A), -)\simeq\CC(-, [A, -])$$
This adjoint pair is known as the \textit{tensor-hom adjunction}.
\end{example}

\begin{remark}
Recall the intuition of \textit{internal homs} as of~\cref{lem:ccc-determine-lem},
we may think of $\CC((B\otimes A), C)$ as $\CC(\textit1, [B\otimes A, C])$.
Similarly we may think of $\CC(B, [A, C])$ as $\CC(\textit1, [B, [A, C]])$.

The equivalence between the two reveals a relationship between $[B\otimes A, C]$ and $[B, [A, C]]$.
This is related to the idea of \textit{currying} in functional programming and lambda calculus.

Rewriting the above formulae with the notation in~\cref{def:ccc} and assume the relationship to be an equivalence,
we get $C^{(B\times A)}=(C^A)^B$. This is a known fact in elementary algebra (and is true for at least $A, B, C\in\mathbb Q$).
\end{remark}

\begin{example}[Left]\label{ex:leftadj}
Consider a contextual category (\cref{def:cxlcat}) $\CC$ with objects $\Gamma, A$ and
a display map (\cref{def:displaymap}) $\pi_A\in\CC(A,\Gamma)$ which can also be written as $\pi_A\in\CC_{/\Gamma}$.
Recall the notation in~\cref{def:reindex}, that $\pi_A^*:\CC_{/\Gamma}\to\CC_{/A}$
is the ``weakening'' functor (see~\cref{ex:weakening}) that ``adds'' $A$ to the context without using it.

Every display map $\pi_B\in\CC_{/A}$ (of an object $B\in\CC$) corresponds to an object
$\Sigma_A B$ with display map $\pi_{\Sigma_A B}\in\CC_{/\Gamma}$ by the construction in~\cref{def:sigma-sem},
so we can think of $\Sigma_A:\CC_{/A}\to\CC_{/\Gamma}$ as a functor that sends any
display map $\pi_B\in\CC(B, A)$ to the display map $\pi_{\Sigma_A B}\in\CC(\Sigma_A B, \Gamma)$,
that ``packs'' $A$ together with the input object.

Fix a display map $\pi_A$, we show $\Sigma_A\dashv\pi_A^*$ by showing that for any display map $\pi_B, \pi_C$
there is isomorphic hom sets $\CC_{/\Gamma}(\pi_{\Sigma_A B}, \pi_C)\simeq\CC_{/A}(\pi_B, \pi_A^*(\pi_C))$.
The input can be organized in this way (recall the notation for cartesian lifting in~\cref{def:fibr}):
% https://q.uiver.app/?q=WzAsNixbMCwwLCJCIixbNDYsNjYsMzAsMV1dLFswLDEsIlxcU2lnbWFfQSBCIixbMjQ2LDczLDQ0LDFdXSxbMSwxLCJcXEdhbW1hIl0sWzEsMCwiQSIsWzQ2LDY2LDMwLDFdXSxbMiwxLCJDIl0sWzIsMCwiXFxwaV9BXiooQykiXSxbMCwzLCJcXHBpX0IiLDAseyJjb2xvdXIiOls0Niw2NiwzMF19LFs0Niw2NiwzMCwxXV0sWzQsMiwiXFxwaV9DIl0sWzMsMiwiXFxwaV9BIiwyLHsiY29sb3VyIjpbNDYsNjYsMzBdfSxbNDYsNjYsMzAsMV1dLFs1LDQsIlxcb3ZlcmxpbmV7XFxwaV9BfShDKSJdLFs1LDMsIlxccGlfQV4qKFxccGlfQykiLDJdLFs1LDIsIiIsMix7InN0eWxlIjp7Im5hbWUiOiJjb3JuZXIifX1dLFswLDEsIiIsMix7ImxldmVsIjoyLCJzdHlsZSI6eyJoZWFkIjp7Im5hbWUiOiJub25lIn19fV0sWzEsMiwiXFxwaV97XFxTaWdtYV9BIEJ9IiwyLHsiY29sb3VyIjpbMjQ2LDczLDQ0XX0sWzI0Niw3Myw0NCwxXV1d
\[\begin{tikzcd}
	\textcolor{rgb,255:red,127;green,103;blue,26}{B} & \textcolor{rgb,255:red,127;green,103;blue,26}{A} & {\pi_A^*(C)} \\
	\textcolor{rgb,255:red,47;green,30;blue,194}{\Sigma_A B} & \Gamma & C
	\arrow["{\pi_B}", color={rgb,255:red,127;green,103;blue,26}, from=1-1, to=1-2]
	\arrow["{\pi_C}", from=2-3, to=2-2]
	\arrow["{\pi_A}"', color={rgb,255:red,127;green,103;blue,26}, from=1-2, to=2-2]
	\arrow["{\overline{\pi_A}(C)}", from=1-3, to=2-3]
	\arrow["{\pi_A^*(\pi_C)}"', from=1-3, to=1-2]
	\arrow["\lrcorner"{anchor=center, pos=0.125, rotate=-90}, draw=none, from=1-3, to=2-2]
	\arrow[Rightarrow, no head, from=1-1, to=2-1]
	\arrow["{\pi_{\Sigma_A B}}"', color={rgb,255:red,47;green,30;blue,194}, from=2-1, to=2-2]
\end{tikzcd}\]
Note that $\pi_A^*(C)$ is just $C$ in a weakened context.
Observe that $\pi_A^*(C)$ is the pullback of $\pi_A$ and $\pi_C$,
so it coincides with the interpretation of the product type (\cref{def:prod-sem}).
The hom sets that we need to show to be isomorphic are $l, r$ shown below:
% https://q.uiver.app/?q=WzAsNixbMCwwLCJCIixbNDYsNjYsMzAsMV1dLFswLDEsIlxcU2lnbWFfQSBCIixbMjQ2LDczLDQ0LDFdXSxbMSwxLCJcXEdhbW1hIl0sWzEsMCwiQSIsWzQ2LDY2LDMwLDFdXSxbMiwxLCJDIl0sWzIsMCwiQVxcdGltZXNfXFxHYW1tYSBDIl0sWzAsMywiIiwwLHsiY29sb3VyIjpbNDYsNjYsMzBdfV0sWzQsMl0sWzMsMiwiIiwyLHsiY29sb3VyIjpbNDYsNjYsMzBdfV0sWzUsNCwiXFxwaV8yIiwwLHsic3R5bGUiOnsiYm9keSI6eyJuYW1lIjoiZGFzaGVkIn19fV0sWzUsMywiXFxwaV8xIiwwLHsic3R5bGUiOnsiYm9keSI6eyJuYW1lIjoiZGFzaGVkIn19fV0sWzUsMiwiIiwyLHsic3R5bGUiOnsibmFtZSI6ImNvcm5lciJ9fV0sWzAsMSwiIiwyLHsibGV2ZWwiOjIsInN0eWxlIjp7ImhlYWQiOnsibmFtZSI6Im5vbmUifX19XSxbMSwyLCIiLDIseyJjb2xvdXIiOlsyNDYsNzMsNDRdfV0sWzEsNCwibCIsMSx7ImN1cnZlIjoyfV0sWzAsNSwiciIsMSx7ImN1cnZlIjotMn1dXQ==
\[\begin{tikzcd}
	\textcolor{rgb,255:red,127;green,103;blue,26}{B} & \textcolor{rgb,255:red,127;green,103;blue,26}{A} & {A\times_\Gamma C} \\
	\textcolor{rgb,255:red,47;green,30;blue,194}{\Sigma_A B} & \Gamma & C
	\arrow[draw={rgb,255:red,127;green,103;blue,26}, from=1-1, to=1-2]
	\arrow[from=2-3, to=2-2]
	\arrow[draw={rgb,255:red,127;green,103;blue,26}, from=1-2, to=2-2]
	\arrow["{\pi_2}", dashed, from=1-3, to=2-3]
	\arrow["{\pi_1}", dashed, from=1-3, to=1-2]
	\arrow["\lrcorner"{anchor=center, pos=0.125, rotate=-90}, draw=none, from=1-3, to=2-2]
	\arrow[Rightarrow, no head, from=1-1, to=2-1]
	\arrow[draw={rgb,255:red,47;green,30;blue,194}, from=2-1, to=2-2]
	\arrow["l"{description}, curve={height=12pt}, from=2-1, to=2-3]
	\arrow["r"{description}, curve={height=-12pt}, from=1-1, to=1-3]
\end{tikzcd}\]
To go from $r$ to $l$, we take the composition with $\pi_2$. The other direction is obtained
by the definition of pullback (\cref{def:pullback}):
% https://q.uiver.app/?q=WzAsNSxbMiwwLCJcXFNpZ21hX0EgQiJdLFswLDEsIlxcR2FtbWEiXSxbMCwwLCJBIixbNDYsNjYsMzAsMV1dLFsxLDEsIkMiXSxbMSwwLCJBXFx0aW1lc19cXEdhbW1hIEMiXSxbMywxXSxbMiwxLCIiLDIseyJjb2xvdXIiOls0Niw2NiwzMF19XSxbNCwzLCJcXHBpXzIiLDAseyJzdHlsZSI6eyJib2R5Ijp7Im5hbWUiOiJkYXNoZWQifX19XSxbNCwyLCJcXHBpXzEiLDAseyJzdHlsZSI6eyJib2R5Ijp7Im5hbWUiOiJkYXNoZWQifX19XSxbNCwxLCIiLDIseyJzdHlsZSI6eyJuYW1lIjoiY29ybmVyIn19XSxbMCwzLCJsIiwxLHsiY3VydmUiOi0yfV0sWzAsNCwiciIsMCx7InN0eWxlIjp7ImJvZHkiOnsibmFtZSI6ImRhc2hlZCJ9fX1dLFswLDIsIlxccGlfQiIsMSx7ImN1cnZlIjoyfV1d
\[\begin{tikzcd}
	\textcolor{rgb,255:red,127;green,103;blue,26}{A} & {A\times_\Gamma C} & {\Sigma_A B} \\
	\Gamma & C
	\arrow[from=2-2, to=2-1]
	\arrow[draw={rgb,255:red,127;green,103;blue,26}, from=1-1, to=2-1]
	\arrow["{\pi_2}", dashed, from=1-2, to=2-2]
	\arrow["{\pi_1}", dashed, from=1-2, to=1-1]
	\arrow["\lrcorner"{anchor=center, pos=0.125, rotate=-90}, draw=none, from=1-2, to=2-1]
	\arrow["l"{description}, curve={height=-12pt}, from=1-3, to=2-2]
	\arrow["r", dashed, from=1-3, to=1-2]
	\arrow["{\pi_B}"{description}, curve={height=12pt}, from=1-3, to=1-1]
\end{tikzcd}\]
The two directions are inverse to each other by the uniqueness of morphisms induced by pullbacks.
\end{example}

\begin{terminology}
The construction $\Sigma_A$ in~\cref{ex:leftadj} is also known as the
(categorical sense of) \textit{dependent sums}.
\end{terminology}

\begin{defn}[Counit]\label{def:counit}
Consider an adjunction (\cref{def:adjoint}) $L\dashv R:\CC\rightleftarrows\DD$,
we may obtain two endofunctors $L(R(-)):\DD\to\DD$ and $R(L(-)):\CC\to\CC$
that commute the following diagram:
% https://q.uiver.app/?q=WzAsNSxbMCwwLCJcXENDIl0sWzEsMCwiXFxERCJdLFsyLDAsIlxcQ0MiXSxbMywwLCJcXEREIl0sWzQsMCwiXFxDQyJdLFswLDEsIkwiXSxbMSwyLCJSIl0sWzIsMywiTCJdLFswLDMsIkwiLDEseyJsYWJlbF9wb3NpdGlvbiI6MzAsIm9mZnNldCI6LTIsImN1cnZlIjotM31dLFszLDQsIlIiXSxbMSw0LCJSIiwxLHsibGFiZWxfcG9zaXRpb24iOjcwLCJvZmZzZXQiOi0yLCJjdXJ2ZSI6LTN9XV0=
\[\begin{tikzcd}
	\CC & \DD & \CC & \DD & \CC
	\arrow["L", from=1-1, to=1-2]
	\arrow["R", from=1-2, to=1-3]
	\arrow["L", from=1-3, to=1-4]
	\arrow["L"{description, pos=0.3}, shift left=2, curve={height=-18pt}, from=1-1, to=1-4]
	\arrow["R", from=1-4, to=1-5]
	\arrow["R"{description, pos=0.7}, shift left=2, curve={height=-18pt}, from=1-2, to=1-5]
\end{tikzcd}\]
The commutativity is known as \textit{triangle identities}.
The natural transformation $\epsilon:L(R(-))\to\id_\DD$ is called the \textit{counit}
of the adjunction, and $\eta:\id_\CC\to R(L(-))$ is called the \textit{unit}.
\end{defn}

\begin{example}\label{ex:tensor-hom-counit}
Consider the tensor-hom adjunction (\cref{ex:tensor-hom}) in a suitable category $\CC$
and objects $A, B\in\CC$. The component of the counit $\epsilon_B\in\CC(L(R(B)), B)$,
which expands to $\epsilon_B\in\CC([A,B]\otimes A, B)$,
is the evaluation map (\cref{def:evalmap}).
\end{example}

\begin{exercise}
Verify that using the $\epsilon_B$ morphism in~\cref{ex:tensor-hom-counit}
to replace the eval morphism in the diagram in~\cref{def:evalmap}
still makes the diagram commutative.
\end{exercise}

\subsection{Another adjoint: the pi type}\label{sec:pi}
\begin{defn}[Pi]\label{def:pi-tt}
We extend~\cref{def:tt} with the \textit{pi type} (also known as the \textit{dependent product type}),
defined by the following typing rules: \ttfragment{deprod}
\end{defn}

\begin{defn}[Right]\label{def:deprod}
Similar to~\cref{ex:leftadj}, we may take the \textit{right adjoint} of a weakening functor
(called the \textit{dependent product}).
Note that unlike left adjoints, right adjoints of weakening functors do not necessarily exist.
Here we assume their existence and discuss their properties.

Consider $\Gamma,A\in\CC$ with a display map $\pi_A\in\CC(A,\Gamma)$,
we have the weakening functor $\pi_A^*:\CC_{/\Gamma}\to\CC_{/A}$.
The right adjoint functor, denoted $\Pi_{A}:\CC_{/A}\to\CC_{/\Gamma}$, sends any display map
$\pi_{B}\in\CC_{/A}$ to a morphism $\pi_{\Pi_A B}\in\CC_{/\Gamma}$
which we need to show to be a display map.
So we have the following familiar setup, where we replace $\pi_A^*(C)$ with $A\times_\Gamma C$
and $B$ with $\Sigma_A B$ to offer more type theoretical intuition:
% https://q.uiver.app/?q=WzAsNixbMCwwLCJcXFNpZ21hX0EgQiIsWzQ2LDY2LDMwLDFdXSxbMCwxLCJcXFBpX0EgQiIsWzI0Niw3Myw0NCwxXV0sWzEsMSwiXFxHYW1tYSIsWzI0Niw3Myw0NCwxXV0sWzEsMCwiQSIsWzQ2LDY2LDMwLDFdXSxbMiwxLCJDIixbMjQ2LDczLDQ0LDFdXSxbMiwwLCJBXFx0aW1lc19cXEdhbW1hIEMiLFs0Niw2NiwzMCwxXV0sWzAsMywiXFxwaV9CIiwwLHsiY29sb3VyIjpbNDYsNjYsMzBdfSxbNDYsNjYsMzAsMV1dLFs0LDIsIlxccGlfQyIsMCx7ImNvbG91ciI6WzI0Niw3Myw0NF19LFsyNDYsNzMsNDQsMV1dLFszLDIsIlxccGlfQSIsMl0sWzUsNCwiIiwwLHsic3R5bGUiOnsiYm9keSI6eyJuYW1lIjoiZGFzaGVkIn19fV0sWzUsMywiIiwyLHsiY29sb3VyIjpbNDYsNjYsMzBdLCJzdHlsZSI6eyJib2R5Ijp7Im5hbWUiOiJkYXNoZWQifX19XSxbNSwyLCIiLDIseyJzdHlsZSI6eyJuYW1lIjoiY29ybmVyIn19XSxbMSwyLCJcXHBpX3tcXFBpX0EgQn0iLDIseyJjb2xvdXIiOlsyNDYsNzMsNDRdfSxbMjQ2LDczLDQ0LDFdXV0=
\[\begin{tikzcd}
	\textcolor{rgb,255:red,127;green,103;blue,26}{\Sigma_A B} & \textcolor{rgb,255:red,127;green,103;blue,26}{A} & \textcolor{rgb,255:red,127;green,103;blue,26}{A\times_\Gamma C} \\
	\textcolor{rgb,255:red,47;green,30;blue,194}{\Pi_A B} & \textcolor{rgb,255:red,47;green,30;blue,194}{\Gamma} & \textcolor{rgb,255:red,47;green,30;blue,194}{C}
	\arrow["{\pi_B}", color={rgb,255:red,127;green,103;blue,26}, from=1-1, to=1-2]
	\arrow["{\pi_C}", color={rgb,255:red,47;green,30;blue,194}, from=2-3, to=2-2]
	\arrow["{\pi_A}"', from=1-2, to=2-2]
	\arrow[dashed, from=1-3, to=2-3]
	\arrow[draw={rgb,255:red,127;green,103;blue,26}, dashed, from=1-3, to=1-2]
	\arrow["\lrcorner"{anchor=center, pos=0.125, rotate=-90}, draw=none, from=1-3, to=2-2]
	\arrow["{\pi_{\Pi_A B}}"', color={rgb,255:red,47;green,30;blue,194}, from=2-1, to=2-2]
\end{tikzcd}\]
The adjointedness says that for any display map $\pi_B$,
the following set isomorphism holds:
$$\CC_{/A}(\pi_A^*(\pi_C), \pi_B)\simeq\CC_{/\Gamma}(\pi_C, \pi_{\Pi_A B})$$
Equivalently,
$$\CC_{/A}(\pi_1, \pi_B)\simeq\CC_{/\Gamma}(\pi_C, \pi_{\Pi_A B})$$
Diagrammatically, we have $l$ and $r$ uniquely determine each other:
% https://q.uiver.app/?q=WzAsNixbMCwwLCJcXFNpZ21hX0EgQiIsWzQ2LDY2LDMwLDFdXSxbMCwxLCJcXFBpX0EgQiIsWzI0Niw3Myw0NCwxXV0sWzEsMSwiXFxHYW1tYSIsWzI0Niw3Myw0NCwxXV0sWzEsMCwiQSIsWzQ2LDY2LDMwLDFdXSxbMiwxLCJDIixbMjQ2LDczLDQ0LDFdXSxbMiwwLCJBXFx0aW1lc19cXEdhbW1hIEMiLFs0Niw2NiwzMCwxXV0sWzAsMywiIiwwLHsiY29sb3VyIjpbNDYsNjYsMzBdfV0sWzQsMiwiIiwwLHsiY29sb3VyIjpbMjQ2LDczLDQ0XX1dLFszLDJdLFs1LDQsIiIsMCx7InN0eWxlIjp7ImJvZHkiOnsibmFtZSI6ImRhc2hlZCJ9fX1dLFs1LDMsIiIsMix7ImNvbG91ciI6WzQ2LDY2LDMwXSwic3R5bGUiOnsiYm9keSI6eyJuYW1lIjoiZGFzaGVkIn19fV0sWzUsMiwiIiwyLHsic3R5bGUiOnsibmFtZSI6ImNvcm5lciJ9fV0sWzEsMiwiIiwyLHsiY29sb3VyIjpbMjQ2LDczLDQ0XX1dLFs1LDAsImwiLDEseyJjdXJ2ZSI6MiwiY29sb3VyIjpbNDYsNjYsMzBdfSxbNDYsNjYsMzAsMV1dLFs0LDEsInIiLDEseyJjdXJ2ZSI6LTIsImNvbG91ciI6WzI0Niw3Myw0NF19LFsyNDYsNzMsNDQsMV1dXQ==
\[\begin{tikzcd}
	\textcolor{rgb,255:red,127;green,103;blue,26}{\Sigma_A B} & \textcolor{rgb,255:red,127;green,103;blue,26}{A} & \textcolor{rgb,255:red,127;green,103;blue,26}{A\times_\Gamma C} \\
	\textcolor{rgb,255:red,47;green,30;blue,194}{\Pi_A B} & \textcolor{rgb,255:red,47;green,30;blue,194}{\Gamma} & \textcolor{rgb,255:red,47;green,30;blue,194}{C}
	\arrow[draw={rgb,255:red,127;green,103;blue,26}, from=1-1, to=1-2]
	\arrow[draw={rgb,255:red,47;green,30;blue,194}, from=2-3, to=2-2]
	\arrow[from=1-2, to=2-2]
	\arrow[dashed, from=1-3, to=2-3]
	\arrow[draw={rgb,255:red,127;green,103;blue,26}, dashed, from=1-3, to=1-2]
	\arrow["\lrcorner"{anchor=center, pos=0.125, rotate=-90}, draw=none, from=1-3, to=2-2]
	\arrow[draw={rgb,255:red,47;green,30;blue,194}, from=2-1, to=2-2]
	\arrow["l"{description}, color={rgb,255:red,127;green,103;blue,26}, curve={height=12pt}, from=1-3, to=1-1]
	\arrow["r"{description}, color={rgb,255:red,47;green,30;blue,194}, curve={height=-12pt}, from=2-3, to=2-1]
\end{tikzcd}\]
Suppose $C=\Gamma$ and thus $\pi_C=\id_\Gamma$, by~\cref{lem:pullback-along-id} there is $A\times_\Gamma \Gamma=A$.
By that we can reassemble the above diagram as:
% https://q.uiver.app/?q=WzAsNixbMywxLCJcXFNpZ21hX0EgQiIsWzQ2LDY2LDMwLDFdXSxbMCwwLCJcXFBpX0EgQiIsWzI0Niw3Myw0NCwxXV0sWzEsMCwiXFxHYW1tYSIsWzI0Niw3Myw0NCwxXV0sWzIsMCwiQSIsWzQ2LDY2LDMwLDFdXSxbMSwxLCJcXEdhbW1hIixbMjQ2LDczLDQ0LDFdXSxbMiwxLCJBIixbNDYsNjYsMzAsMV1dLFs0LDIsIiIsMCx7ImxldmVsIjoyLCJjb2xvdXIiOlsyNDYsNzMsNDRdLCJzdHlsZSI6eyJoZWFkIjp7Im5hbWUiOiJub25lIn19fV0sWzMsMl0sWzUsNF0sWzUsMywiIiwyLHsibGV2ZWwiOjIsImNvbG91ciI6WzQ2LDY2LDMwXSwic3R5bGUiOnsiaGVhZCI6eyJuYW1lIjoibm9uZSJ9fX1dLFs1LDIsIiIsMix7InN0eWxlIjp7Im5hbWUiOiJjb3JuZXIifX1dLFsyLDEsIlxcTGFtYmRhLmwiLDIseyJjb2xvdXIiOlsyNDYsNzMsNDRdLCJzdHlsZSI6eyJib2R5Ijp7Im5hbWUiOiJkYXNoZWQifX19LFsyNDYsNzMsNDQsMV1dLFs1LDAsImwiLDIseyJjb2xvdXIiOls0Niw2NiwzMF19LFs0Niw2NiwzMCwxXV0sWzMsMCwibCIsMCx7ImNvbG91ciI6WzQ2LDY2LDMwXX0sWzQ2LDY2LDMwLDFdXSxbNCwxLCJcXExhbWJkYS5sIiwwLHsiY29sb3VyIjpbMjQ2LDczLDQ0XSwic3R5bGUiOnsiYm9keSI6eyJuYW1lIjoiZGFzaGVkIn19fSxbMjQ2LDczLDQ0LDFdXV0=
\[\begin{tikzcd}
	\textcolor{rgb,255:red,47;green,30;blue,194}{\Pi_A B} & \textcolor{rgb,255:red,47;green,30;blue,194}{\Gamma} & \textcolor{rgb,255:red,127;green,103;blue,26}{A} \\
	& \textcolor{rgb,255:red,47;green,30;blue,194}{\Gamma} & \textcolor{rgb,255:red,127;green,103;blue,26}{A} & \textcolor{rgb,255:red,127;green,103;blue,26}{\Sigma_A B}
	\arrow[draw={rgb,255:red,47;green,30;blue,194}, Rightarrow, no head, from=2-2, to=1-2]
	\arrow[from=1-3, to=1-2]
	\arrow[from=2-3, to=2-2]
	\arrow[draw={rgb,255:red,127;green,103;blue,26}, Rightarrow, no head, from=2-3, to=1-3]
	\arrow["\lrcorner"{anchor=center, pos=0.125, rotate=180}, draw=none, from=2-3, to=1-2]
	\arrow["{\Lambda.l}"', color={rgb,255:red,47;green,30;blue,194}, dashed, from=1-2, to=1-1]
	\arrow["l"', color={rgb,255:red,127;green,103;blue,26}, from=2-3, to=2-4]
	\arrow["l", color={rgb,255:red,127;green,103;blue,26}, from=1-3, to=2-4]
	\arrow["{\Lambda.l}", color={rgb,255:red,47;green,30;blue,194}, dashed, from=2-2, to=1-1]
\end{tikzcd}\]
Now, for every $l\in\CC(A, \Sigma_A B)$ uniquely determines a morphism denoted $\Lambda.l\in\CC(\Gamma,\Pi_A B)$.
This corresponds to the introduction rule of the pi type.

Now, suppose $C=\Pi_A B$ and thus $\pi_C=\id_{\Pi_A B}$. We pass $\id_{\Pi_A B}$ to the adjunction
and we uniquely obtain a morphism $\fs{Eval}\in\CC(A\otimes\Pi_A B, B)$.
This morphism corresponds to the evaluation map but for dependent products.
% https://q.uiver.app/?q=WzAsNixbMCwwLCJcXFNpZ21hX0EgQiIsWzQ2LDY2LDMwLDFdXSxbMCwxLCJcXFBpX0EgQiIsWzI0Niw3Myw0NCwxXV0sWzEsMSwiXFxHYW1tYSIsWzI0Niw3Myw0NCwxXV0sWzEsMCwiQSIsWzQ2LDY2LDMwLDFdXSxbMiwxLCJcXFBpX0EgQiIsWzI0Niw3Myw0NCwxXV0sWzIsMCwiQVxcdGltZXNfXFxHYW1tYSBcXFBpX0EgQiIsWzQ2LDY2LDMwLDFdXSxbMCwzLCIiLDAseyJjb2xvdXIiOls0Niw2NiwzMF19XSxbNCwyLCIiLDAseyJjb2xvdXIiOlsyNDYsNzMsNDRdfV0sWzMsMl0sWzUsNCwiIiwwLHsic3R5bGUiOnsiYm9keSI6eyJuYW1lIjoiZGFzaGVkIn19fV0sWzUsMywiIiwyLHsiY29sb3VyIjpbNDYsNjYsMzBdLCJzdHlsZSI6eyJib2R5Ijp7Im5hbWUiOiJkYXNoZWQifX19XSxbNSwyLCIiLDIseyJzdHlsZSI6eyJuYW1lIjoiY29ybmVyIn19XSxbMSwyLCIiLDIseyJjb2xvdXIiOlsyNDYsNzMsNDRdfV0sWzUsMCwiXFxybXtFdmFsfSIsMSx7ImN1cnZlIjoyLCJjb2xvdXIiOls0Niw2NiwzMF19LFs0Niw2NiwzMCwxXV0sWzQsMSwiXFxpZF97XFxQaV9BIEJ9IiwxLHsiY3VydmUiOi0yLCJsZXZlbCI6MiwiY29sb3VyIjpbMjQ2LDczLDQ0XSwic3R5bGUiOnsiaGVhZCI6eyJuYW1lIjoibm9uZSJ9fX0sWzI0Niw3Myw0NCwxXV1d
\[\begin{tikzcd}
	\textcolor{rgb,255:red,127;green,103;blue,26}{\Sigma_A B} & \textcolor{rgb,255:red,127;green,103;blue,26}{A} & \textcolor{rgb,255:red,127;green,103;blue,26}{A\times_\Gamma \Pi_A B} \\
	\textcolor{rgb,255:red,47;green,30;blue,194}{\Pi_A B} & \textcolor{rgb,255:red,47;green,30;blue,194}{\Gamma} & \textcolor{rgb,255:red,47;green,30;blue,194}{\Pi_A B}
	\arrow[draw={rgb,255:red,127;green,103;blue,26}, from=1-1, to=1-2]
	\arrow[draw={rgb,255:red,47;green,30;blue,194}, from=2-3, to=2-2]
	\arrow[from=1-2, to=2-2]
	\arrow[dashed, from=1-3, to=2-3]
	\arrow[draw={rgb,255:red,127;green,103;blue,26}, dashed, from=1-3, to=1-2]
	\arrow["\lrcorner"{anchor=center, pos=0.125, rotate=-90}, draw=none, from=1-3, to=2-2]
	\arrow[draw={rgb,255:red,47;green,30;blue,194}, from=2-1, to=2-2]
	\arrow["{\rm{Eval}}"{description}, color={rgb,255:red,127;green,103;blue,26}, curve={height=12pt}, from=1-3, to=1-1]
	\arrow["{\id_{\Pi_A B}}"{description}, color={rgb,255:red,47;green,30;blue,194}, curve={height=-12pt}, Rightarrow, no head, from=2-3, to=2-1]
\end{tikzcd}\]
Consider an inverse of $\pi_B$ and an inverse of $\pi_A$, there is
$\fs{Eval}\circ(a\times \Lambda. f)=f\circ a$.
This corresponds to the elimination and computation rule of the pi type.
\end{defn}

\begin{exercise}
Verify that the morphism $\Lambda.l$ in~\cref{def:deprod} is a left-inverse of $\pi_{\Pi_A B}$
(Hint: both morphisms are given by the same adjunction).
\end{exercise}

\begin{exercise}
Reinterpret Eval in~\cref{def:deprod} using counit (\cref{def:counit}).
\end{exercise}

\begin{remark}
The $B$ in diagrams in~\cref{def:deprod} are equivalent to $\Sigma_A B$.
This somehow reflects a vague relationship between dependent sums and dependent products.
\end{remark}

\begin{defn}\label{def:pi-sem}
We interpret the pi type in a contextual category (\cref{def:cxlcat}) $\CC$
with right adjoints (\cref{def:deprod}) of weakening functors (\cref{ex:weakening}).
Consider interpretations $\lrbracket\Gamma,\lrbracket{\Gamma,A},\lrbracket{\Gamma,\Sigma_A B}\in\CC$
with corresponding display maps, and we take the dependent product of $\pi_{\lrbracket A}$,
denoted $\Pi_{\lrbracket A}$. The \textit{context $\Gamma$ extended by the pi type $\Pi_A B$}
is interpreted as $\Pi_{\lrbracket A}(\lrbracket{\Gamma,\Sigma_A B})$, denoted $\lrbracket{\Gamma,\Pi_A B}$,
with its display map $\Pi_{\lrbracket A}(\pi_{\lrbracket B})$, denoted $\pi_{\lrbracket{\Pi_A B}}$:
% https://q.uiver.app/?q=WzAsNCxbMCwwLCJcXGxyYnJhY2tldHtcXFNpZ21hX0EgQn0iXSxbMCwxLCJcXGxyYnJhY2tldHtcXEdhbW1hLFxcUGlfQSBCfSIsWzI0Niw3Myw0NCwxXV0sWzEsMSwiXFxscmJyYWNrZXRcXEdhbW1hIixbMjQ2LDczLDQ0LDFdXSxbMSwwLCJcXGxyYnJhY2tldCBBIl0sWzAsMywiXFxwaV97XFxscmJyYWNrZXQgQn0iXSxbMywyLCJcXHBpX3tcXGxyYnJhY2tldCBBfSIsMl0sWzEsMiwiXFxwaV97XFxscmJyYWNrZXR7XFxQaV9BIEJ9fSIsMix7ImNvbG91ciI6WzI0Niw3Myw0NF19LFsyNDYsNzMsNDQsMV1dXQ==
\[\begin{tikzcd}
	{\lrbracket{\Sigma_A B}} & {\lrbracket A} \\
	\textcolor{rgb,255:red,47;green,30;blue,194}{\lrbracket{\Gamma,\Pi_A B}} & \textcolor{rgb,255:red,47;green,30;blue,194}{\lrbracket\Gamma}
	\arrow["{\pi_{\lrbracket B}}", from=1-1, to=1-2]
	\arrow["{\pi_{\lrbracket A}}"', from=1-2, to=2-2]
	\arrow["{\pi_{\lrbracket{\Pi_A B}}}"', color={rgb,255:red,47;green,30;blue,194}, from=2-1, to=2-2]
\end{tikzcd}\]
In the introduction rule, for (the interpretation of) a term $\lrbracket u\in\CC(\lrbracket{\Gamma,A},\lrbracket{\Gamma,\Sigma_A B})$
we define $\lrbracket{\Lambda.u}$ as $\Lambda.\lrbracket{u}$ as in~\cref{def:deprod}.
The diagram is omitted because it's exactly the same as the one in~\cref{def:deprod}
except that the name of the morphism $l$ is replaced with $\lrbracket u$.

In the elimination rule, we also take a similar construction as in~\cref{def:deprod},
which also justifies the computation rule.
\end{defn}

\begin{thm}
In a contextual category $\CC$ with dependent products (\cref{def:deprod}), for $A\in\CC$:
$$\Sigma_A\dashv\pi_A^*\dashv\Pi_A$$
is an ``adjoint triple''.
\end{thm}

\subsection{Subobject classifiers: the universe of all propositions}
\begin{defn}[Prop]\label{def:prop-tt}
We extend~\cref{def:tt} with a type $\Prop$,
defined by the following typing rules: \ttfragment{prop}
We can think of $R$ as an injective function that maps the unique instance (when exists) of a type into an instance of $\Prop$.
\end{defn}

\begin{example}[Truth]\label{ex:truth}
The unit type (\cref{def:top}), $\top$, has a corresponding proposition
(\cref{def:prop-tt}) $R(\top, \Lambda.\Lambda.\reflcon\star):\Prop$.
\end{example}

\begin{terminology}\label{conv:inhabit}
When talking about the instances of a type, we say that the type is \textit{inhabited}
if it has at least one instance, and these instances are called \textit{inhabitants}.

If the type has no instance, we say it is \textit{uninhabited}.
\end{terminology}

\begin{remark}
The \cref{def:prop-tt} is an imitation of the $\Prop$ universe in
the (extended) calculus of constructions \cite{CoC,LUTT,LUTT2},
which can be regarded as the type of all types (called \textit{propositions}) that
``have either one single inhabitant (\cref{conv:inhabit}) or have no instance at all''.
The second last rule in~\cref{def:prop-tt} captures such \textit{uniquely inhabited} property of propositions.

By having a universe for \textit{some} types (propositions),
we partially addresses~\cref{warn:univ}.
\end{remark}

\begin{defn}[SubObj]\label{def:subobjclfir}
For a category $\CC$ with a terminal object (\cref{def:termobj}) \textit1,
all pullbacks (\cref{def:pullback}),
a selected object $\Omega\in\CC$ called the \textit{truth value object},
and a mono (\cref{def:monic}) $\truecon\in\CC(\textit1,\Omega)$, called the \textit{subobject classifier}, such that:

Every mono $\pi_U\in\CC(U,\Gamma)$ (for some $U,\Gamma\in\CC$) uniquely determines
a morphism $\chi_U\in\CC(\Gamma,\Omega)$ such that the following square is a pullback square:
% https://q.uiver.app/?q=WzAsNSxbMCwwLCJVIl0sWzAsMSwiXFx0ZXh0aXQgMSJdLFsyLDEsIlxcT21lZ2EiXSxbMiwwLCJcXEdhbW1hIl0sWzEsMV0sWzAsMSwiXFx0ZXh0aXQgMSIsMl0sWzEsMiwiXFx0ZXh0e3RydWV9IiwxXSxbMCwzLCJcXHBpX1UiLDFdLFszLDIsIlxcY2hpX1UiLDAseyJzdHlsZSI6eyJib2R5Ijp7Im5hbWUiOiJkYXNoZWQifX19XSxbMCw0LCIiLDIseyJzdHlsZSI6eyJuYW1lIjoiY29ybmVyIn19XV0=
\[\begin{tikzcd}
	U && \Gamma \\
	{\textit 1} & {} & \Omega
	\arrow["{\textit 1}"', from=1-1, to=2-1]
	\arrow["{\truecon}"{description}, from=2-1, to=2-3]
	\arrow["{\pi_U}"{description}, from=1-1, to=1-3]
	\arrow["{\chi_U}", dashed, from=1-3, to=2-3]
	\arrow["\lrcorner"{anchor=center, pos=0.125}, draw=none, from=1-1, to=2-2]
\end{tikzcd}\]
The morphism $\chi_U$ is also known as the \textit{characteristic map} or \textit{classifying map}
of the \textit{subobject} $\pi_U$.
\end{defn}

\begin{example}[Set]\label{ex:set-subobj}
In $\SET$, the truth value object (\cref{def:subobjclfir}) $\Omega$ is the 2-valued set $\textbf2:=\{t,f\}$,
and $\truecon\in\SET(\textit1,\textbf2), \truecon(x):=t$.
For every $A\in\SET$ and $B\subseteq A\in\SET$ a subobject (that $\pi_B$ is the natural inclusion (\cref{not:set})),
the characteristic map is defined as
$\chi_B(a):=\left\{\begin{array}{ll}
	t &\text{if } a\in B \\
	f &\text{otherwise}
\end{array}\right\}$. For each natural inclusion morphism in $\SET$
can we find a subobject classifier for it. We specialize the diagram in~\cref{def:subobjclfir} into $\SET$:
% https://q.uiver.app/?q=WzAsNSxbMCwwLCJCIl0sWzAsMSwiXFx0ZXh0aXQgMSJdLFsyLDEsIlxce3QsZlxcfSJdLFsyLDAsIkEiXSxbMSwxXSxbMCwxLCJcXHRleHRpdCAxIiwyXSxbMSwyLCJ0IiwxXSxbMCwzLCJcXHN1YnNldGVxIiwxXSxbMywyLCJcXGNoaV9CIiwwLHsic3R5bGUiOnsiYm9keSI6eyJuYW1lIjoiZGFzaGVkIn19fV0sWzAsNCwiIiwyLHsic3R5bGUiOnsibmFtZSI6ImNvcm5lciJ9fV1d
\[\begin{tikzcd}
	B && A \\
	{\textit 1} & {} & {\{t,f\}}
	\arrow["{\textit 1}"', from=1-1, to=2-1]
	\arrow["t"{description}, from=2-1, to=2-3]
	\arrow["\subseteq"{description}, from=1-1, to=1-3]
	\arrow["{\chi_B}", dashed, from=1-3, to=2-3]
	\arrow["\lrcorner"{anchor=center, pos=0.125}, draw=none, from=1-1, to=2-2]
\end{tikzcd}\]
The pullback square characterizes $\chi_B$ to be a function that
``returns $f$ as much as possible'' because it is terminal in $\SET(B,A)$.
\end{example}

\begin{defn}\label{def:prop-ct}
We interpret the universe of propositions (\cref{def:prop-tt}) in a contextual category
(\cref{def:cxlcat}) $\CC$ with a subobject classifier (\cref{def:subobjclfir}) defined as~\cref{ex:truth}.

Consider a type \fbox{$\Gvdash\isType A$}, it is a proposition when uniquely inhabited
(\cref{conv:inhabit}) by $\Gvdash u:A$ or uninhabited.
In the following diagram, \textit{the context $\Gamma$ extended by the proposition $A$} is the pullback
$\lrbracket{\Gamma,A}\in\CC$, and the interpretation of the $\Prop$ universe is the truth object,
where $\top_\Prop$ denotes any randomly selected proposition like
${R(\top,\Lambda.\Lambda.\reflcon\star)}$:
% https://q.uiver.app/?q=WzAsNSxbMCwwLCJcXGxyYnJhY2tldFxcR2FtbWEsXFxscmJyYWNrZXQgQSJdLFsyLDAsIlxcbHJicmFja2V0XFxHYW1tYSJdLFsyLDEsIlxcbHJicmFja2V0XFxQcm9wIl0sWzAsMSwiXFx0ZXh0aXQxIl0sWzEsMV0sWzAsMSwiXFxwaV97XFxscmJyYWNrZXQgQX0iLDFdLFsxLDIsIlxcbHJicmFja2V0e1IoQSxwKX0iLDAseyJzdHlsZSI6eyJib2R5Ijp7Im5hbWUiOiJkYXNoZWQifX19XSxbMywyLCJcXGxyYnJhY2tldHtSKFxcdG9wLFxcTGFtYmRhLlxcTGFtYmRhLlxcc3Rhcil9IiwxXSxbMCwzLCIiLDIseyJzdHlsZSI6eyJib2R5Ijp7Im5hbWUiOiJkYXNoZWQifX19XSxbMCw0LCIiLDIseyJzdHlsZSI6eyJuYW1lIjoiY29ybmVyIn19XV0=&macro_url=https%3A%2F%2Fgist.githubusercontent.com%2Fice1000%2F47b7ea52f8c351607831f7f4afa9a79b%2Fraw%2Ff5bd21d4bd401a9f3672269a2816c728e53bd9c0%2Fquiver-macros.tex
\[\begin{tikzcd}
	{\lrbracket{\Gamma,A}} && \lrbracket\Gamma \\
	\textit1 & {} & \lrbracket\Prop
	\arrow["{\pi_{\lrbracket A}}"{description}, from=1-1, to=1-3]
	\arrow["{\lrbracket{R(A,p)}}", dashed, from=1-3, to=2-3]
	\arrow["{\lrbracket{\top_\Prop}}"{description}, from=2-1, to=2-3]
	\arrow[dashed, from=1-1, to=2-1]
	\arrow["\lrcorner"{anchor=center, pos=0.125}, draw=none, from=1-1, to=2-2]
\end{tikzcd}\]
\begin{enumerate}
\item Since $\pi_{\lrbracket A}$ is mono, its inverse $\lrbracket u\in\CC(\lrbracket\Gamma,\lrbracket{\Gamma,A})$
(if exists) must be unique. This uniqueness justifies the uniqueness rule of propositions and is guaranteed by the proof $p$.
\item For every proposition $A$, we define a (unique) term $R(A,p)\in\Prop$ that corresponds to $A$.
The uniqueness justifies the extensionality (see~\cref{thm:propext}) of $\Prop$ and the existence justifies the completeness.
\item The $\El-$ operation takes a morphism $A\in\CC(\lrbracket\Gamma,\lrbracket\Prop)$ (which is an instance of $\Prop$)
on the left of the above diagram and returns the pullback. To some extent, $\El-$ and $R(-,p)$ are inverse to each other.
\end{enumerate}
\lessSpace{-1}
\end{defn}

\begin{thm}[Extensionality]\label{thm:propext}
The extensionality of propositions holds in the interpretation in~\cref{def:prop-ct}: \ttfragment{propext}
\end{thm}

\begin{thm}\label{lem:id-isprop}
For any type $A$ and its instances $x:A,y:A$, \fbox{$\Idty A x y$} is a proposition. In other words,
the following type is inhabited: $$\Pi_{x:A}\Pi_{y:A}\Pi_{i:\Idty A x y}\Pi_{j:\Idty A x y}\Idty{\Idty A x y}i j$$
\end{thm}
\begin{proof}
By~\cref{thm:uip} we can prove it by reflexivity: $\textsf{uip}:=\Lambda x.\Lambda.\Lambda.\Lambda.\reflcon{\reflcon x}$
(alternatively we can write $\textsf{uip}:=\Lambda.\Lambda y.\Lambda.\Lambda.\reflcon{\reflcon y}$).
\end{proof}

\begin{remark}
There is a more type theoretical way to think about the subobject classifier diagram in~\cref{def:prop-ct},
which may help type theorists to intuite the definition. We present the following informal diagram,
which uses sigma types (\cref{def:sigmatt}):
% https://q.uiver.app/?q=WzAsNSxbMCwwLCJcXEdhbW1hLEEiXSxbMCwxLCJcXFNpZ21hX3tQOlxcUHJvcH0gUCJdLFsyLDEsIlxcUHJvcCJdLFsyLDAsIlxcR2FtbWEiXSxbMSwxXSxbMCwxLCJcXGxyYW5nbGV7XFx0b3BfXFxQcm9wLCBcXHN0YXJ9IiwyXSxbMSwyLCIuMSIsMV0sWzAsMywiXFxwaV9BIiwxXSxbMywyLCJSKEEsIHApIiwwLHsic3R5bGUiOnsiYm9keSI6eyJuYW1lIjoiZGFzaGVkIn19fV0sWzAsNCwiIiwyLHsic3R5bGUiOnsibmFtZSI6ImNvcm5lciJ9fV1d&macro_url=https%3A%2F%2Fgist.githubusercontent.com%2Fice1000%2F47b7ea52f8c351607831f7f4afa9a79b%2Fraw%2Ff5bd21d4bd401a9f3672269a2816c728e53bd9c0%2Fquiver-macros.tex
\[\begin{tikzcd}
	{\Gamma,A} && \Gamma \\
	{\Sigma_{P:\Prop} P} & {} & \Prop
	\arrow["{\lrangle{\top_\Prop, \star}}"', from=1-1, to=2-1]
	\arrow["{.1}"{description}, from=2-1, to=2-3]
	\arrow["{\pi_A}"{description}, from=1-1, to=1-3]
	\arrow["{R(A, p)}", dashed, from=1-3, to=2-3]
	\arrow["\lrcorner"{anchor=center, pos=0.125}, draw=none, from=1-1, to=2-2]
\end{tikzcd}\]
It is obvious that $\Sigma_{P:\Prop} P$ is a terminal object, but to show that it's terminal,
we need to define $\Prop$ first. This is why the diagram is only introduced informally.
\begin{enumerate}
\item The uniqueness of $R(A, p)$ (corresponds to $\chi_A$ which is by definition unique)
  corresponds to propositional extensionality.
\item The fact that $A$ is uniquely inhabited makes $\pi_A$ a mono,
  because the inverse of $\pi_A$ is unique.
\item For all uniquely inhabited types, we can find the morphism $R(A, p)$.
  This corresponds to the fact that ``uniquely inhabited types are propositions''. 
\end{enumerate}
\end{remark}

\subsection{Internalized constructions}
\begin{defn}[Bool]\label{rem:bool-tt}
We define a type Bool in the type theory in~\cref{def:tt} with the extensions (\cref{def:top,def:coprod}) as:
\begin{enumerate}
\item $\text{Bool}:=\top\sqcup\top$.
\item $\truecon:=\textsf{inl}(\star)$.
\item $\falsecon:=\textsf{inr}(\star)$.
\item $\lrangle{t~?~u:v}:=\textsf{match}(t,u,v)$.
\end{enumerate}
From the above definition, we can derive the following typing rules for Bool: \ttfragment{bool}
\end{defn}

\begin{defn}[PropTrunc]\label{def:proptrunc}
For every well-formed type \fbox{$\Gvdash\isType A$}, we define a type
\fbox{$\Gvdash\isType{||A||}$}, called the \textit{propositional truncation of $A$}:
\begin{enumerate}
\item $||A||:=\prod_{P:\Prop}(A\to \El P) \to \El P$ (using~\cref{def:func-tt,def:pi-tt,def:prop-tt}).
\item For $\Gvdash u:A$, $\Gvdash|u|:||A||$
(unfolds to $\Gvdash|u|:\prod_{P:\Prop} (A\to\El P) \to \El P$),
defined as $|u|:=\Lambda P.\lambda f.\apply f u$.
\item The elimination, that for $\Gvdash u:||A||$ we can eliminate it to another proposition $\Gvdash P:\Prop$
with the map $\Gvdash f:A\to\El P$ by $\apply{\apply u P}f$.
\end{enumerate}
From the above definition, we can derive the following typing rules for $||-||$: \ttfragment{proptrunc}
\end{defn}

\begin{defn}\label{def:idprop}
We construct an alternative definition of Id that lives in $\Prop$ (see~\cref{lem:id-isprop}) by
$\Idprop A a b:=R(\Idty A a b, \apply{\apply{\textsf{uip}}a}b)$. We derive the following judgment: \ttfragment{idprop}
\end{defn}

\begin{defn}\label{def:void}
We define the empty type $\bot:=\prod_{P:\Prop}P$ using~\cref{def:pi-tt,def:prop-tt}.
\end{defn}

\section{Alternative constructions}
\begin{defn}[DiscreteCat]\label{def:discat}
We say a category $\CC$ to be \textit{discrete} if it has no morphism other than identity morphisms.
\end{defn}

\begin{prop}
$\CC$ is discrete $\iff$ $\CC \simeq \text{Ob}(\CC)$.
\end{prop}

\begin{defn}[Discreteness]\label{def:discfib}
We say a fibration $p:\EE\to\CC$ to be \textit{discrete} if each fiber (\cref{def:fiber}) is discrete (\cref{def:discat}).
\end{defn}

\begin{remark}
In the literature, we sometimes ask the fibration $p$ of a CwA (\cref{def:cwa})
to be a discrete fibration (\cref{def:discfib}). This is what Pitts~\cite[\S 6]{catl}
has done implicitly (without mentioning the notion of fibrations (\cref{def:fibr})
and comprehension categories (\cref{def:compcat})).
\end{remark}

\begin{defn}\label{def:fam}
The category of indexed families of sets, $\FAM$, consists of the following:
\begin{enumerate}
\item Objects $(U_x)_{x\in X}$, where $X$ is a set and for each $x\in X$, $U_x$ is a set.
\item Morphisms $(f,(g_x)_{x\in X}) \in\FAM((U_x)_{x\in X}, (V_y)_{y\in Y})$,
where $f:X\to Y$ is a function (called the \textit{reindexing function}, similar to~\cref{def:reindex})
and for each $x\in X$, $g_x:U_x\to V_{f(x)}$ is a function.
\end{enumerate}
\lessSpace{-1.5}
\end{defn}

\subsection{Type categories}
\begin{glorious-defn}[Pitts' TypeCat]\label{def:tycat}
A \textit{type category} consists of:
\begin{enumerate}
\item A category $\CC$, in analogy to the category of contexts (\cref{def:cat-ctx}).
\item A presheaf (\cref{def:presh}) $\text{Ty}:\OpCat\CC\to\SET$.
  We refer to $\Ty\Gamma$ as the set of \textit{dependent types indexed by $\Gamma$}.
\item A \textit{context extension} operation, constructing $(\Gamma,A)\in\CC$ from $A\in\Ty\Gamma$.
\item A \textit{projection morphism} $\pi_A\in\CC((\Gamma,A),\Gamma)$ for every $A$, the inverse of context extension.
\end{enumerate}
We also require the object $(\Delta,\sigma^*A)\in\CC$ (where $\sigma^*A\in\Ty\Delta$
is called the \textit{pullback of $A$ along $\Delta$})
and the morphism $\sigma^A:\CC((\Delta,\sigma^*A),(\Gamma,A))$ to exist
(assuming $\Gamma\in\CC,\Delta\in\CC$, $A\in\Ty\Gamma$) by the following pullback square:
% https://q.uiver.app/?q=WzAsNCxbMCwwLCJcXERlbHRhLFxcc2lnbWFeKkEiXSxbMCwxLCJcXERlbHRhIl0sWzEsMCwiXFxHYW1tYSxBIl0sWzEsMSwiXFxHYW1tYSJdLFswLDEsIlxccGlfe1xcc2lnbWFeKkF9IiwyXSxbMCwyLCJcXHNpZ21hX0EiXSxbMiwzLCJcXHBpX0EiXSxbMSwzLCJcXHNpZ21hIiwyXSxbMCwzLCIiLDEseyJzdHlsZSI6eyJuYW1lIjoiY29ybmVyIn19XV0=
\[\begin{tikzcd}
  {\Delta,\sigma^*A} & {\Gamma,A} \\
  \Delta & \Gamma
  \arrow["{\pi_{\sigma^*A}}"', from=1-1, to=2-1]
  \arrow["{\sigma^A}", from=1-1, to=1-2]
  \arrow["{\pi_A}", from=1-2, to=2-2]
  \arrow["\sigma"', from=2-1, to=2-2]
  \arrow["\lrcorner"{anchor=center, pos=0.125}, draw=none, from=1-1, to=2-2]
\end{tikzcd}\]
The following \textit{strictness condition} must hold in the pullback square:
\begin{align*}
\id_\Gamma^*A&= A & \id_\Gamma^A&=\id_{\Gamma,A} \\
\gamma^*(\sigma^*A) &= (\sigma\circ\gamma)^* A & \sigma^A\circ \gamma^{\sigma^*A}&=(\sigma\circ \gamma)^A
\end{align*}
This definition is due to~\cite[definition 6.3.3]{catl}.
\end{glorious-defn}

\begin{remark}
The ``projection morphism'' in~\cref{def:tycat} is similar to the ``display maps'' in~\cref{def:displaymap},
and the pullback square in~\cref{def:tycat} is similar to the pullback square in~\cref{rem:ty-subst}.
\end{remark}

\begin{lem}\label{lem:tycat-embed}
For $\Gamma\in\CC$, the set $\Ty\Gamma$ has an embedding in the objects in the
overcategory (\cref{def:slice}) $\CC_{/\Gamma}$.
\end{lem}
\begin{proof}
By mapping $A\in\Ty\Gamma$ to $\pi_A\in\CC((\Gamma,A),\Gamma)$.
\end{proof}

\begin{remark}\label{rem:want-tycat}
By~\cref{lem:tycat-embed}, we may add morphisms between the elements of $\Ty\Gamma$ to form a category
in order to put more categorical structures into it. This will make Ty into a functor $\text{Ty}:\OpCat\CC\to\CAT$,
similar to the functor $\Psi:\OpCat\CC\to\CAT$ (\cref{lem:fake-funct}) in a split (\cref{def:split,term:fullsplit})
comprehension category (since we have the strictness conditions in~\cref{def:tycat}).
\end{remark}

\begin{notation}
We define two syntactical shorthands for type categories (\cref{def:tycat}) for convenience.
\begin{itemize}
\item For $A\in\Ty{\textit1}$, we write $A\in\CC$ for $(\textit1, A)\in\CC$.
\item $\Ty[\Gamma]{A, B}$ for the hom set $\Ty\Gamma(A, B)$.
\end{itemize}
\lessSpace{-1.5}
\end{notation}

\begin{defn}[TyMor]\label{def:tymor}
The hom set $\Ty[\Gamma]{A, B}$ is defined as the hom set $\CC_{/\Gamma}(\pi_A,\pi_B)$.
Identities and compositions are inherited from $\CC$.
This realizes~\cref{rem:want-tycat}.
\end{defn}

\begin{remark}\label{rem:diff-cwa-compcat}
The original definition of a CwA \textit{is} discrete (and is almost the same as a type category),
which means that the morphisms in the fibers in $\EE$ are all identity morphisms.
In~\cref{def:cwa}, on the other hand, we allow arbitrary morphisms in $\EE$ as long as
$\FF$ preserves cartesianness and $p$ is a full split fibration.

The benefit of defining a CwA based on a comprehension category is that comprehension categories are
appreciated by mathematicians since they have their own well-established fibrations.
Curious readers may refer to~\cite[example 4.10]{CompCat}.
\end{remark}

\subsection{Categories with families}
\begin{glorious-defn}[Dybjer's CwF]\label{def:cwf}
A \textit{category with families} (CwF)~\cite{CwF2,CwF-Dybjer,CwF-Hofmann} is a structure consisting of:
\begin{enumerate}
\item A category interpreting contexts (\cref{def:cat-ctx}) $\CC$.
\item A $\FAM$-valued presheaf $T:\OpCat\CC\to\FAM$ (\cref{def:fam,def:presh}).
\item A context comprehension operation defined in~\cref{def:cwf-ctxcomp}.
It is similar to~\cref{def:ctxext,def:tycat}.
\end{enumerate}
\lessSpace{-1.5}
\end{glorious-defn}

\begin{remark}
We can also have a contextual (\cref{def:ctxly}) CwF by requiring an $\ell$ operation on its $\CC$ category,
or a democratic (\cref{def:democ}) CwF by requiring such types to exists.
\end{remark}

\begin{notation}\label{not:subst-mor}
We have the following syntactical shorthands:
\begin{enumerate}
\item $\Ty\Gamma=X$ for $\Gamma\in\CC$ if $T(\Gamma)=(U_x)_{x\in X}$. In fact, Ty is a presheaf.
\item $\Tm\Gamma A=U_A$ for $A\in\Ty\Gamma$ if $T(\Gamma)=(U_x)_{x\in X}$.
\item For a substitution object $\sigma\in\CC(\Gamma,\Delta)$, there exist the following function:
$T(\sigma):(\Tm\Delta A)_{A\in\Ty\Delta}\to(\Tm\Gamma B)_{B\in\Ty\Gamma}$.
\end{enumerate}
\lessSpace{-1.5}
\end{notation}

\begin{defn}[Substitution]\label{def:subst-cwf}
The morphism $T(\sigma)$ in~\cref{not:subst-mor} consists of two maps:
\begin{enumerate}
\item A reindexing map $\sigma^*:\Ty\Delta\to\Ty\Gamma$ called the \textit{substitution on types}, similar to~\cref{def:reindex}.
\item A map $\sigma^*:\Tm\Delta A\to\Tm\Gamma{\sigma^*(A)}$ called the \textit{substitution on terms}.
\end{enumerate}
Note that the names of these two maps are overloaded. These maps constitute the following commutative square:
% https://q.uiver.app/?q=WzAsNixbMCwxLCJcXERlbHRhIl0sWzAsMCwiXFxHYW1tYSJdLFsxLDAsIlxcVHlcXEdhbW1hIl0sWzEsMSwiXFxUeVxcRGVsdGEiXSxbMiwwLCJcXFRtXFxHYW1tYXtcXHNpZ21hXiooQSl9Il0sWzIsMSwiXFxUbVxcRGVsdGEgQSJdLFsxLDAsIlxcc2lnbWEiXSxbMywyLCJcXHNpZ21hXioiLDJdLFs1LDQsIlxcc2lnbWFeKiIsMl1d
\[\begin{tikzcd}
	\Gamma & \Ty\Gamma & {\Tm\Gamma{\sigma^*(A)}} \\
	\Delta & \Ty\Delta & {\Tm\Delta A}
	\arrow["\sigma", from=1-1, to=2-1]
	\arrow["{\sigma^*}"', from=2-2, to=1-2]
	\arrow["{\sigma^*}"', from=2-3, to=1-3]
\end{tikzcd}\]
Beware of the directions of the arrows. A substitution $\sigma\in\CC(\Gamma,\Delta)$ brings types
formed in $\Delta$ to $\Gamma$, as discussed in~\cref{rem:spoiler2}.
\end{defn}

\begin{defn}[Comprehension]\label{def:cwf-ctxcomp}
For $\Gamma\in\CC$ and $A\in\Ty\Gamma$, the \textit{context comprehension} of a CwF
consists of an assigned context $\Gamma\cdot A$ and two functions $p_{\Gamma,A}:\Gamma\cdot A \to \Gamma$
(which is also a substitution (\cref{def:subst-cwf}))
and $q_{\Gamma,A}\in\Tm{\Gamma\cdot A}{(p_{\Gamma,A})^*(A)}$, such that the following property holds:

For every $\sigma\in\CC(\Gamma,\Delta)$, $A\in\Ty\Delta$, and $a\in\Tm\Gamma{\sigma^*(A)}$,
a substitution object $\sigma\cdot a\in\CC(\Gamma,\Delta\cdot A)$ is uniquely identified such that
$p_{\Delta,A} \circ (\sigma\cdot a)=\sigma$ and $(\sigma\cdot a)^*q_{\Delta,A}=a$.
The characterization is visualized in the following commutative parallelogram:
% https://q.uiver.app/?q=WzAsNCxbMiwwLCJcXERlbHRhXFxjZG90IEEiXSxbMiwxLCJcXEdhbW1hIl0sWzQsMCwiXFxEZWx0YSJdLFswLDEsIlxcc2lnbWFeKihBKSJdLFsxLDAsIlxcc2lnbWFcXGNkb3QgYSIsMCx7InN0eWxlIjp7ImJvZHkiOnsibmFtZSI6ImRhc2hlZCJ9fX1dLFsxLDIsIlxcc2lnbWEiLDJdLFswLDIsInBfe1xcRGVsdGEsQX0iXSxbMCwzLCJxX3tcXERlbHRhLEF9IiwyXSxbMSwzLCJhIl1d
\[\begin{tikzcd}
	&& {\Delta\cdot A} && \Delta \\
	{\sigma^*(A)} && \Gamma
	\arrow["{\sigma\cdot a}", dashed, from=2-3, to=1-3]
	\arrow["\sigma"', from=2-3, to=1-5]
	\arrow["{p_{\Delta,A}}", from=1-3, to=1-5]
	\arrow["{q_{\Delta,A}}"', from=1-3, to=2-1]
	\arrow["a", from=2-3, to=2-1]
\end{tikzcd}\]
\lessSpace{-1}
\end{defn}

% \begin{thm}\label{thm:cwf-tycat}
% A CwF is equivalent to a type category.
% \end{thm}
% \begin{proof}
% We try to equate a CwF with a type category by identifying:
% \begin{enumerate}
% \item The category interpreting contexts $\CC$. Both of them have such categories.
% \item The (discrete) category of $\Gamma$-indexed dependent types for $\Gamma\in\CC$.
% They are called $\Ty\Gamma$ in both categories, both are sets.
% \item The set of terms of type $A$ in a CwF is $\Tm\Gamma A$, while in a type category it is
% just the set of left inverses of $\pi_A\in\CC_{/\Gamma}$.
% \end{enumerate}
% \lessSpace{-1}
% \end{proof}

% \begin{exercise}
% Prove that the context comprehension operation in a CwF (\cref{def:cwf-ctxcomp})
% and a type category (\cref{def:tycat}) are equivalent.
% \end{exercise}

% \begin{cor}
% A CwF is equivalent to a CwA.
% \end{cor}
% \begin{proof}
% By~\cref{thm:cwf-tycat,thm:cwa-tycat}.
% \end{proof}

% \begin{remark}
% We consider a CwF (\cref{def:cwf}) as a CwA (\cref{def:cwa}) or a type category
% (\cref{def:tycat}) with additional but unnecessary structures.
% \end{remark}

\begin{remark}
The structure of a CwF is \textit{very} close to the syntax of type theory --
both the formation of contexts and types and the typing of substitutions and terms
have a direct correspondence to a ``belong to'' relation in a CwF.
In~\cite{CwF2}, Castellan even used the notation of substitution from type theory directly
($\_[\gamma]$ instead of $\gamma^*$), but we avoid their notation for consistency with the rest of this document.
\end{remark}

\begin{terminology}
The model of type theory based on a CwF (\cref{def:cwf}) is known as
the \textit{presheaf} model since the notion of ``dependent types defined in a context'' is
interpreted as a presheaf.
\end{terminology}

\section{More category theory}\label{sec:more-cat}
\begin{defn}[Pushout]\label{def:pushout}
The notion dual to pullback (\cref{def:pullback}) is \textit{pushout} (or \textit{pushforward}),
characterizing objects by taking pullbacks in the opposite category (\cref{def:oppocat}), denoted and diagramed below:
% https://q.uiver.app/?q=WzAsNCxbMCwxLCJBIl0sWzEsMCwiQiJdLFsxLDEsIkFcXHNxY3VwX0MgQiJdLFswLDAsIkMiXSxbMCwyXSxbMSwyXSxbMywwXSxbMywxXSxbMiwzLCIiLDEseyJzdHlsZSI6eyJuYW1lIjoiY29ybmVyIn19XV0=
\[\begin{tikzcd}
	C & B \\
	A & {A\sqcup_C B}
	\arrow[from=2-1, to=2-2]
	\arrow[from=1-2, to=2-2]
	\arrow[from=1-1, to=2-1]
	\arrow[from=1-1, to=1-2]
	\arrow["\lrcorner"{anchor=center, pos=0.125, rotate=180}, draw=none, from=2-2, to=1-1]
\end{tikzcd}\]
The pushout $(A\sqcup_C B)\in\CC$ is also known as the \textit{fiber coproduct} of $A$ and $B$.
\end{defn}

\begin{defn}[Undercategory]\label{def:coslice}
For a category $\CC$ and $X\in\CC$, the \textit{undercategory} $\prescript{}{/X}\CC$
(also known as the \textit{coslice category under} $X$) is the overcategory ${\OpCat\CC}_{/X}$.
The objects of an undercategory are certain morphisms \textit{from} $X$ in $\CC$
commuting the (reversed version of the) diagrams in~\cref{def:slice}.
\end{defn}

\begin{defn}[Cocartesianness]\label{def:cocart-mor}
Similar to~\cref{def:cart-mor}, where we have a functor $p:\EE\to\CC$ and objects $\Gamma\in\CC$, $E,D\in\EE$.
We say a morphism $g\in\EE(D,E)$ to be \textit{Grothendieck cocartesian} if for all $g'\in\EE(D,E')$ (where $E'\in\EE$) there is a unique
morphism $f\in\EE(E',E)$ such that $g'\circ f=g$. This universal property is visualized below:
% https://q.uiver.app/?q=WzAsNSxbMiwwLCJwKEUpIl0sWzIsMSwiXFxHYW1tYSJdLFswLDEsIkQiXSxbMCwwLCJFJyJdLFsxLDAsIkUiXSxbMSwwLCJwKGcpIiwyXSxbMiwzLCJnJyJdLFszLDQsImYiLDAseyJzdHlsZSI6eyJib2R5Ijp7Im5hbWUiOiJkYXNoZWQifX19XSxbMiw0LCJnIiwxXSxbOCw1LCJwIiwyLHsic2hvcnRlbiI6eyJzb3VyY2UiOjIwLCJ0YXJnZXQiOjIwfX1dXQ==
\[\begin{tikzcd}
	{E'} & E & {p(E)} \\
	D && \Gamma
	\arrow[""{name=0, anchor=center, inner sep=0}, "{p(g)}"', from=2-3, to=1-3]
	\arrow["{g'}", from=2-1, to=1-1]
	\arrow["f", dashed, from=1-1, to=1-2]
	\arrow[""{name=1, anchor=center, inner sep=0}, "g"{description}, from=2-1, to=1-2]
\end{tikzcd}\]
The diagram shows why $g$ is sometimes known as an \textit{initial lifting}.
This definition is due to Jacobs~\cite[definition 2.1 (i)]{CompCat} due to Grothendieck~\cite{Fibration}.
\end{defn}

\begin{defn}[Opfibration]\label{def:opfibr}
Similar to~\cref{def:fibr}, we define a functor $p:\EE\to\CC$ to be an \textit{opfibration}
if composition preserves Grothendieck cocartesianness and for every $\sigma\in\CC(p(D), \Gamma)$
there is a Grothendieck cocartesian morphism $g\in\EE(D,E)$ such that $p(g)=\sigma$.
The visualization is almost identical to the one in~\cref{def:fibr}.
\end{defn}

\begin{cor}
$p:\EE\to\CC$ is an opfibration $\iff$ $\OpCat p:\OpCat\EE\to\OpCat\CC$ is a fibration.
\end{cor}

\begin{defn}[Bifibration]\label{def:bifibr}
A functor that is a fibration (\cref{def:fibr}) and a opfibration (\cref{def:opfibr}) is
called a \textit{bifibration}.
\end{defn}

\begin{example}
Extending~\cref{ex:codproj,ex:domproj}:
\begin{itemize}
\item The domain projection fibration $dom:\CC^\to\to\CC$ is a bifibration if it has all pushouts (\cref{def:pushout}).
\item The codomain projection functor $cod:\CC^\to\to\CC$ is an opfibration (\cref{def:opfibr}).
\end{itemize}
We can also replace the word ``fibration'' with ``bifibration'' in \cref{ex:codproj}.
\end{example}

% \begin{defn}
% For a functor $p:\EE\to\CC$, objects $E,D\in\EE$, and morphism $\sigma\in\CC(p(E),p(D))$,
% we write $\EE_\sigma(D,E)$ for the set of morphisms $f\in\EE(D,E)$ such that $p(f)=\sigma$.
% In other words, $EE_\sigma(D,E):=p^{-1}(\sigma)$.
% If $p$ is a fibration, then the fiber (\cref{def:fiber}) $\EE_{p(D)}(D,\sigma^*(E))$ is equivalent
% to the set $\EE$
% \end{defn}

\begin{defn}[Cartesianness]\label{def:gen-cart}
For a functor $p:\EE\to\CC$ and $f\in\EE(D,E)$, we say that $f$ is (strongly, see~\cref{his:cart}) \textit{$p$-cartesian} if for every
$f'\in\EE(D',E)$ there is a commutative triangle $p(f)\circ \overline g=p(f')$ for some $\overline g$ that
can uniquely determine a commutative triangle $f\circ g=f'$ for some $g$. The visualization is:
% https://q.uiver.app/?q=WzAsNixbNSwwLCJwKEUpIl0sWzMsMSwicChEJykiXSxbMiwxLCJEIl0sWzAsMSwiRCciXSxbMiwwLCJFIl0sWzUsMSwicChEKSJdLFsxLDAsInAoZicpIiwxXSxbMywyLCIiLDIseyJzdHlsZSI6eyJib2R5Ijp7Im5hbWUiOiJkYXNoZWQifX19XSxbMyw0LCJmJyJdLFsyLDQsImYiXSxbNSwwLCJwKGYpIiwyXSxbMSw1XSxbOSw2LCJwIiwxLHsic2hvcnRlbiI6eyJzb3VyY2UiOjIwLCJ0YXJnZXQiOjIwfX1dXQ==
\[\begin{tikzcd}
	&& E &&& {p(E)} \\
	{D'} && D & {p(D')} && {p(D)}
	\arrow[""{name=0, anchor=center, inner sep=0}, "{p(f')}"{description}, from=2-4, to=1-6]
	\arrow[dashed, from=2-1, to=2-3]
	\arrow["{f'}", from=2-1, to=1-3]
	\arrow[""{name=1, anchor=center, inner sep=0}, "f", from=2-3, to=1-3]
	\arrow["{p(f)}"', from=2-6, to=1-6]
	\arrow[from=2-4, to=2-6]
	\arrow["p"{description}, shorten <=13pt, shorten >=13pt, Rightarrow, from=1, to=0]
\end{tikzcd}\]
In the diagram, the triangle on the right can uniquely determine (called \textit{lift to}) the one on the left.
This definition is taken from~\cite[\href{https://kerodon.net/tag/01T1}{\S 01T1}]{kerodon}.
\end{defn}

\begin{defn}[Enrichment]\label{def:enrich}
A category $C$ \textit{enriched over} a monoidal (\cref{def:monoidalcat}) category $K$ is a category
that replaces its hom sets with objects in $K$ (called \textit{hom objects} or \textit{objects of morphisms}).
This means we can have morphisms between the hom objects. Formally, $C$ has the following structures:
\begin{enumerate}
\item Objects and morphisms, just like other categories.
Morphisms are objects in $K$, so we will denote objects in $K$ as $C(a,b)$ for $a,b\in C$.
\item For $a,b,c\in C$, there is a morphism $\circ_{a,b,c}\in K(C(a,b)\otimes C(b,c),C(a,c))$,
called the \textit{composition morphism}.
\item For the tensor unit $1\in K$ and $a\in C$, there is an \textit{identity morphism} $1_a\in K(1,C(a,a))$.
\end{enumerate}
Such that the following diagrams commute:

\lessSpace{-1.5}
% https://q.uiver.app/?q=WzAsMTAsWzAsMCwiQyhiLGIpXFxvdGltZXMgQyhhLGIpIl0sWzEsMCwiQyhhLGIpIl0sWzIsMCwiQyhhLGIpXFxvdGltZXMgQyhhLGEpIl0sWzAsMSwiMVxcb3RpbWVzIEMoYSxiKSJdLFsyLDEsIkMoYSxiKVxcb3RpbWVzMSJdLFsxLDEsIkMoYSxkKSJdLFsyLDIsIkMoYyxkKSBcXG90aW1lcyBDKGEsYykiXSxbMCwyLCJDKGIsZClcXG90aW1lcyBDKGEsYikiXSxbMCwzLCIoQyhjLGQpXFxvdGltZXMgQyhiLGMpKSBcXG90aW1lcyBDKGEsYikiXSxbMiwzLCJDKGMsZCkgXFxvdGltZXMgKEMoYixjKSBcXG90aW1lcyBDKGEsYikpIl0sWzAsMSwiXFxjaXJjX3thLGIsYn0iLDFdLFsyLDEsIlxcY2lyY197YSxhLGJ9IiwxXSxbMywwLCIxX2FcXG90aW1lc1xcaWRfe0MoYSxiKX0iXSxbNCwyLCJcXGlkX3tDKGEsYil9XFxvdGltZXMxX2EiLDJdLFszLDEsIlxcbGFtYmRhX3tDKGEsYil9IiwxXSxbNCwxLCJcXHJob197QyhhLGIpfSIsMV0sWzcsNSwiXFxjaXJjX3thLGIsZH0iLDFdLFs2LDUsIlxcY2lyY197YSxjLGR9IiwxXSxbOCw5LCJcXGFscGhhX3tDKGMsZCksQyhiLGMpLCBDKGEsYil9Il0sWzgsNywiXFxjaXJjX3tiLGMsZH1cXG90aW1lcyBcXGlkX3tDKGEsYil9Il0sWzksNiwiXFxpZF97QyhjLGQpfVxcb3RpbWVzIFxcY2lyY197YSxiLGN9IiwyXV0=&macro_url=https%3A%2F%2Fgist.githubusercontent.com%2Fice1000%2F47b7ea52f8c351607831f7f4afa9a79b%2Fraw%2Fbed9ddea8cd95fb03de082dc6b3fc55b9d754c90%2Fquiver-macros.tex
\[\begin{tikzcd}
	{C(b,b)\otimes C(a,b)} & {C(a,b)} & {C(a,b)\otimes C(a,a)} \\
	{1\otimes C(a,b)} & {C(a,d)} & {C(a,b)\otimes1} \\
	{C(b,d)\otimes C(a,b)} && {C(c,d) \otimes C(a,c)} \\
	{(C(c,d)\otimes C(b,c)) \otimes C(a,b)} && {C(c,d) \otimes (C(b,c) \otimes C(a,b))}
	\arrow["{\circ_{a,b,b}}"{description}, from=1-1, to=1-2]
	\arrow["{\circ_{a,a,b}}"{description}, from=1-3, to=1-2]
	\arrow["{1_a\otimes\id_{C(a,b)}}", from=2-1, to=1-1]
	\arrow["{\id_{C(a,b)}\otimes1_a}"', from=2-3, to=1-3]
	\arrow["{\lambda_{C(a,b)}}"{description}, from=2-1, to=1-2]
	\arrow["{\rho_{C(a,b)}}"{description}, from=2-3, to=1-2]
	\arrow["{\circ_{a,b,d}}"{description}, from=3-1, to=2-2]
	\arrow["{\circ_{a,c,d}}"{description}, from=3-3, to=2-2]
	\arrow["{\alpha_{C(c,d),C(b,c), C(a,b)}}", from=4-1, to=4-3]
	\arrow["{\circ_{b,c,d}\otimes \id_{C(a,b)}}", from=4-1, to=3-1]
	\arrow["{\id_{C(c,d)}\otimes \circ_{a,b,c}}"', from=4-3, to=3-3]
\end{tikzcd}\]
\lessSpace{-0.5}
\end{defn}

\begin{example}\label{ex:enrichcat}
A category enriched in $\CAT$ is a strict 2-category (\cref{def:bicat}).
This is a realization of \cref{rem:bicat-enrich}.
\end{example}

\subsection{Potential mistakes and counterexamples}
\begin{warning}\label{warn:prodnotterm}
For a category $\CC$ and objects $A,B\in\CC$, we take the full subcategory (\cref{def:fullsub,def:fullsub2})
of $\CC$ by selecting objects $X\in\CC$ such that both $\CC(X,A)$ and $\CC(X,B)$ are nonempty.

The terminal object in such full subcategory is \textit{not} the product object $A\times B\in\CC$.
The reason is that the terminal object in such subcategory \textit{does not commute}
(although the morphisms exist) the diagram in~\cref{def:prodobj}.
\end{warning}

\begin{example}
In $\SET$, for sufficiently large sets (say, size larger than 2) $A,B\in\SET$, the full subcategory of $\SET$
like in~\cref{warn:prodnotterm} is equivalent to $\SET$, whose terminal object is the trivial set $\{\emptyset\}$.
This is definitely not the product $A\times B$.
\end{example}

\begin{remark}
Similar to~\cref{warn:prodnotterm}, pullbacks cannot be characterized as a terminal object in a subcategory either.
\end{remark}

\begin{warning}\label{warn:coprodpushout}
Coproduct types are \textit{not} interpreted as pushouts (\cref{def:pushout}) which are
coproducts in the undercategory (\cref{def:coslice}). Instead, they correspond to coproduct
objects in the overcategory (\cref{def:slice}).
\end{warning}

\begin{warning}
We \textit{cannot} turn isomorphisms of hom sets in the overcategory into the original category.
In other words, given a category $\CC$, objects $A,B,C,D,X\in\CC$, and the following morphisms:
% https://q.uiver.app/?q=WzAsNSxbMiwxLCJYIl0sWzAsMSwiQSJdLFswLDAsIkIiXSxbNCwwLCJEIl0sWzQsMSwiQyJdLFsxLDAsImEiLDFdLFsyLDAsImIiLDFdLFszLDAsImQiLDFdLFs0LDAsImMiLDFdXQ==
\[\begin{tikzcd}
	B &&&& D \\
	A && X && C
	\arrow["a"{description}, from=2-1, to=2-3]
	\arrow["b"{description}, from=1-1, to=2-3]
	\arrow["d"{description}, from=1-5, to=2-3]
	\arrow["c"{description}, from=2-5, to=2-3]
\end{tikzcd}\]
There is $\CC_{/X}(a,b)\simeq\CC_{/X}(c,d)\centernot\implies\CC(A,B)\simeq\CC(C,D)$ because $\CC(A,B)$ may contain
more morphisms than $\CC_{/X}(a,b)$, and similarly for $\CC(C,D)$ and $\CC_{/X}(c,d)$.
\end{warning}

\begin{example}
In~\cref{def:fibexp}, even $\CC_{/\Gamma}(a,b)\simeq\CC_{/\Gamma}(\textit1_{\CC_{/\Gamma}}, b^a)$,
we cannot conclude $\CC(B,A)\simeq\CC(A\to_\Gamma B,\Gamma)$.
\end{example}

\begin{warning}
Speaking of internalized definitions, we \textit{cannot yet} define types by large elimination
because there is not a type for all types (\cref{warn:univ}) (although we have a type for all propositions).
\end{warning}

\begin{warning}
For a category $\CC$, its terminal object $\textit1\in\CC$, an arbitrary object $A\in\CC$,
and a morphism $f\in\CC(\textit1,A)$, $f$ is \textit{not} the terminal object in the overcategory $\CC_{/X}$.

Terminal objects in overcategories are identity morphisms.
\end{warning}

\begin{warning}
Consider a CwA (\cref{def:cwa}) $\FF:\EE\to\CC^\to$ that is contextual.
It is not the case that $\text{Ob}(\EE_\Gamma)\simeq\text{Ob}(\CC_{/\Gamma})$ (disregarding~\cref{conv:dontcare})
a set isomorphism. It is true that $\text{Ob}(\EE_\Gamma)$ is isomorphic to the set of
display maps of codomain $\Gamma$ as mentioned in~\cref{conv:dontcare},
but there can be more objects in $\CC_{/\Gamma}$, because not all morphisms are display maps.
\end{warning}

\section*{Acknowledgement}
% \begin{description}
% \item[In person]
We are grateful to (in random order) Yuchen Wu, Niels van der Weide, Ende Jin, Anqur Lu, Jonathan Sterling,
Parker Liu, Aoyang Yu, Xu Huang, Jack Xu,
and GitHub users \textsf{@dramforever}, \textsf{@AliceLogos}, \textsf{@niltok}, and \textsf{@Guest0x0}
for their useful suggestions on improving this document.
We thank \textsf{@AliceLogos} for carefully checking the definitions in the draft version of this document,
and Valery Isaev for sharing the basic ideas on categorical semantics of dependent type theories.
We also thank Jack Xu for checking the technical details in the sophisticated categorical proofs.
% \item[At large] We thank the authors of:

The development of this document is based on the following artifacts:
\begin{itemize}
\item The \LaTeX{} system, its \TeX{Live} distribution\footnote{\url{https://www.tug.org/texlive}},
the \hologo{BibTeX} tool, and the mathpazo\footnote{\url{https://www.ctan.org/pkg/mathpazo}} font.
\item The ``quiver'' tool\footnote{\url{https://q.uiver.app}},
an interactive commutative-diagramming tool.
\item ``Detexify''\footnote{\url{https://detexify.kirelabs.org/classify.html}},
a tool that suggests \LaTeX{} commands from a drawing.
\item ``\hologo{BibTeX} Tidy''\footnote{\url{https://flamingtempura.github.io/bibtex-tidy}},
a smart formatter for \hologo{BibTeX} entries.
\item The Emacs operating system and the
AUC\TeX{}\footnote{\url{https://www.gnu.org/software/auctex}} major-mode of it.
\item The ``ncatlab'' website\footnote{\url{https://ncatlab.org/nlab/show/HomePage}},
the ``kerodon'' website~\cite{kerodon}, the \TeX{} Q\&A website\footnote{\url{https://tex.stackexchange.com}} and the Wikipedia.
\item Several social networks connecting researchers together.
\end{itemize}
% We also thank the mathematicians who invented category theory and the type theorists
% who tried to interpret dependent type theories in categorical constructions.
% We are also grateful for them because they did not explain some structures too well
% that the author can understand immediately, so there is still a space for this
% diagram-rich introduction.
% \end{description}

\printbibliography
\end{document}